\documentclass[12pt]{article}
\pdfoutput=1
\usepackage{amsfonts,hyperref}
\usepackage{color}
\usepackage{appendix}
\usepackage{amsmath}
\usepackage{graphicx}
\usepackage[latin1]{inputenc}
\usepackage[french,english]{babel}
\usepackage{geometry}
\usepackage{etoolbox}
\usepackage{fancyhdr}
\usepackage{url}
\patchcmd{\thebibliography}{\section*}{\section}{}{}

\addto\captionsfrench{}
\addto\captionsenglish{}

\newcommand\encadremath[1]{\vbox{\hrule\hbox{\vrule\kern8pt
\vbox{\kern8pt \hbox{$\displaystyle #1$}\kern8pt}
\kern8pt\vrule}\hrule}}
\def\enca#1{\vbox{\hrule\hbox{
\vrule\kern8pt\vbox{\kern8pt \hbox{$\displaystyle #1$}
\kern8pt} \kern8pt\vrule}\hrule}}

\newcommand\figureframex[3]{
\begin{figure}[bth]
\hrule\hbox{\vrule\kern8pt
\vbox{\kern8pt \vbox{
\begin{center}
{\mbox{\epsfxsize=#1.truecm\epsfbox{#2}}}
\end{center}
\caption{#3}
}\kern8pt}
\kern8pt\vrule}\hrule
\end{figure}
}

\makeatletter
\@addtoreset{equation}{section}
\makeatother
\newtheorem{theorem}{Theorem}[section]
\newtheorem{conjecture}{Conjecture}[section]
\newtheorem{remark}{Remark}[section]
\newtheorem{proposition}{Proposition}[section]
\newtheorem{lemma}{Lemma}[section]
\newtheorem{corollary}{Corollary}[section]
\newtheorem{definition}{Definition}[section]

\def\br{\begin{remark}\rm\small}
\def\er{\end{remark}}
\def\bt{\begin{theorem}}
\def\et{\end{theorem}}
\def\bd{\begin{definition}}
\def\ed{\end{definition}}
\def\bp{\begin{proposition}}
\def\ep{\end{proposition}}
\def\bl{\begin{lemma}}
\def\el{\end{lemma}}
\def\bc{\begin{corollary}}
\def\ec{\end{corollary}}
\def\beaq{\begin{eqnarray}}
\def\eeaq{\end{eqnarray}}

\newcommand{\beq}{\begin{equation}}
\newcommand{\eeq}{\end{equation}}
\newcommand{\bea}{\begin{eqnarray}}
\newcommand{\eea}{\end{eqnarray}}
\newcommand{\beqq}{\begin{equation*}}
\newcommand{\eeqq}{\end{equation*}}
\newcommand{\beaa}{\begin{eqnarray*}}
\newcommand{\eeaa}{\end{eqnarray*}}

 \newcommand{\Tr}{{\,\rm Tr}\:}

\newcommand{\td}[1]{{\tilde{#1}}}

\newcommand{\Pint}{{\int\kern -1.em -\kern-.25em}}

\renewcommand{\Re}{{\mathrm{Re}}}

\newcommand\Res{\mathop{{\rm Res}}}

\begin{document}

\sloppy

\pagestyle{empty}
\vspace{10pt}
\begin{center}
{\large \bf {Matrix models, Toeplitz determinants and recurrence times for powers of random unitary matrices}}
\end{center}
\vspace{3pt}
\begin{center}
\textbf{O. Marchal$^\dagger$}
\end{center}
\vspace{15pt}

$^\dagger$ \textit{Universit\'{e} de Lyon, CNRS UMR 5208, Universit\'{e} Jean Monnet, Institut Camille Jordan, France}
\footnote{olivier.marchal@univ-st-etienne.fr}

\vspace{30pt}

{\bf Abstract}:
The purpose of this article is to study the eigenvalues $u_1^{\, t}=e^{it\theta_1},\dots,u_N^{\,t}=e^{it\theta_N}$ of $U^t$ where $U$ is a large $N\times N$ random unitary matrix and $t>0$. In particular we are interested in the typical times $t$ for which all the eigenvalues are simultaneously close to $1$ in different ways thus corresponding to recurrence times in the issue of quantum measurements. Our strategy consists in rewriting the problem as a random matrix integral and use loop equations techniques to compute the first orders of the large $N$ asymptotic. We also connect the problem to the computation of a large Toeplitz determinant whose symbol is the characteristic function of several arc segments of the unit circle. In particular in the case of a single arc segment we recover Widom's formula. Eventually we explain why the first return time is expected to converge towards an exponential distribution when $N$ is large. Numerical simulations are provided along the paper to illustrate the results.

\vspace{15pt}
\pagestyle{plain}
\setcounter{page}{1}


\section{Summary of the article and main results}
The purpose of the article is to study the probability $P_N(t)$ that the $t^{\text{th}}$ power of a large random unitary matrix $U_N$ of size $N\times N$ returns close to the identity matrix when the size of the matrix tends to infinity. In the following section, we present three usual definitions characterizing the notion of being ``close'' to the identity matrix. The first and second definitions consist in saying that respectively the real part (real-part return time) or modulus (weak return time) of the average of the eigenvalues is close to $1$. The third alternative definition states that all eigenvalues belong to a small arc-interval around $1$ (strong return time). We also present in this section the connection with quantum physics where these questions naturally occur. In section \ref{Section2} we recall known results about the trace of powers of random unitary matrices and use them to estimate the probability of a real-part or weak return time. The rest of the paper is dedicated exclusively to the more interesting case of strong return times. Section \ref{Appendix A} provides the connection with the asymptotic of large Toeplitz determinants that is useful for numerical simulations. The central aspects of the paper are then presented in section \ref{Section4} where we use matrix models techniques to compute the probability of a strong return time. In particular our main results are:
\begin{itemize} \item In theorem \ref{Theo1} we prove that the probability to have a strong return time satisfies the technical conditions required in the main result of \cite{BorotGuionnetKoz} (briefly presented in proposition \ref{PropBorot}) so that the general form of the asymptotic expansion at large $N$ is known explicitly.
\item Since small times are shown to give a one-cut case (genus $0$ spectral curve), we carry out the explicit computations of the first orders of $\frac{1}{N^2}\ln P_N(t)$ up to $O\left(\frac{1}{N^6}\right)$ for small times. These results are presented in theorem \ref{Theo2} and provide a new derivation (as well as the next orders) of Widom's results \cite{Widom} regarding the asymptotic of certain Toeplitz determinants.
\item For integer times (i.e. integer powers of random unitary matrices) an additional symmetry allows the explicit computations of $\frac{1}{N^2}\ln P_N(t)$ up to order $O\left(\frac{1}{N}\right)$. Results are summarized in theorem \ref{Theo3}
\end{itemize}
In section \ref{MBIA} we propose an Average Block Interaction Approximation to address the issue of non-integer times that turns to be very efficient numerically at leading order. Confrontation of our various results with numerical simulations are presented in section \ref{Numerics}. Eventually in section \ref{Section7} we propose two conjectures (conjectures \ref{Yeah} and \ref{Yeahh}) regarding the distribution of the first strong return time. More specifically we conjecture that the first strong return time obeys an exponential distribution whose parameter $\lambda$ is also conjectured. We provide heuristic arguments as well as numerical simulations to support our conjectures.  

\section{Powers of random unitary matrices and definition of various return times}

\subsection{Recurrence time and random unitary matrices}

In quantum physics one of the long standing problems is the issue of finding a proper interpretation of measurement that is compatible both with the probabilistic interpretation of quantum mechanics and the possibility to observe a unique outcome for each experiment. This issue has been discussed at length since the early days of quantum physics and we refer to \cite{Balian1} for a recent review about the problems at stake. Although this paper is not directly aimed towards this problem, it deals with a question raised at the end of paragraph $6.1$ (page 59) of \cite{Balian1}. Indeed, in this paragraph, the authors explained that in the attempt to explain quantum measurements as a properly treated interaction between a quantum system $S$ and a measurement apparatus $A$ in a thermal bath, the discreteness of the spectrum of the operator $\hat{m}$ describing the pointer variable of the (finite but large) apparatus $A$ gives rise to \textit{recurrence times} when waiting long enough. Indeed if after a short time, the interaction between $S$ and $A$ destroys all correlations between the two systems (a phenomenon called \textit{truncation} happening during a typical time $\tau_{\text{trunc}}$), it happens that because the spectrum is discrete (and finite) there exists a Poincar\'{e} recurrence time $\tau_{\text{rec}}$ for which the pointer variable of $A$ comes back close to its original state. This means that even if the correlations between $S$ and $A$ are rapidly dephased (after a time of order $\tau_{\text{trunc}}$), the memory of the correlations is not lost for good and emerges back periodically (with period $\tau_{\text{rec}}$). In \cite{Balian1} a simple model with independent and random eigenvalues for $\hat{m}$ is developed and leads to the conclusion that the recurrence time is \textit{inaccessibly large} compared to the truncation time because the ratio is of order $\frac{\tau_{\text{rec}}}{\tau_{\text{trunc}}}=\pi\sqrt{2}e^{Qf^2}$ (formula $6.20$ of \cite{Balian1}) where $Q$ stands for the number of independent randomly distributed eigenvalues of the system and $f$ stands is the parameter controlling the size of the window defining that the system is close enough to its initial state. However the authors also indicate that their (independent) model is too simple and cannot be physically relevant since the eigenvalues cannot be completely random and uncorrelated in a general setting. In fact they suggest that a much proper setting would correspond to consider the operator $\hat{m}$ as a random matrix whose eigenvalues are known to be correlated. Eventually, they claimed that using matrix models the recurrence time is expected to be smaller than the one found in their simple model but still sufficiently large so that is remains \textit{inaccessibly large} in practice. In this spirit, this article can be seen as a confirmation of their argument in the case of a random unitary operator and direct comparison with their results is provided in section \ref{Discuss}.\\   
More specifically, we want to study the distribution of the eigenvalues of a unitary matrix $U_N$ of size $N$ sampled uniformly according to the Haar measure of the unitary group $\mathcal{U}(N)$. Let us denote $\mathcal{C}$ the unit circle and $u_i=e^{i\theta_i} \in \mathcal{C}$ the $i^{\text{th}}$ eigenvalue of the matrix $U_N$ with $\theta_i$ chosen in $[-\pi,\pi]$. It is well-known that the Haar measure on the unitary group is equivalent to the following distribution for the eigenvalues:
\beq \label{premiere}\td{Z}_N=\int_{[-\pi,\pi]^N} d\theta_1\dots d\theta_N \left(\prod_{i<j}^N |e^{i\theta_i}-e^{i\theta_j}|^2\right)\eeq
Equivalently, using the exponential variables $u_i=e^{i \theta_i}$ the last partition function is equivalent to:
\beq \label{EPF} Z_N=\int_{\mathcal{C}^N} du_1\dots d u_N \Delta(u_1,\dots,u_N)^2 e^{-N\underset{k=1}{\overset{N}{\sum}} \ln u_k}=(-1)^{\frac{N(N-1)}{2}}i^N\td{Z}_N\eeq
where $\Delta(u)=\underset{i<j}{\prod} \left(u_i-u_j\right)$ is the usual Vandermonde determinant. Indeed we have:
\bea \td{Z}_N&=&\frac{1}{i^N}\int_{\mathcal{C}^N} \frac{du_1}{u_1}\dots \frac{du_N}{u_N} \left(\prod_{i<j}^N |u_i-u_j|^2\right)\cr
&=&\frac{1}{i^N}\int_{\mathcal{C}^N} \frac{du_1}{u_1}\dots \frac{du_N}{u_N} \left(\prod_{i<j}^N(u_i-u_j)\left(\frac{1}{u_i}-\frac{1}{u_j}\right)\right)\cr
&=&\frac{1}{i^N}(-1)^{\frac{N(N-1)}{2}}\int_{\mathcal{C}^N} \frac{du_1}{u_1}\dots \frac{du_N}{u_N} \left(\prod_{i<j}^N \frac{(u_i-u_j)^2}{u_iu_j}\right)\cr
&=&(-1)^{\frac{N(N+1)}{2}}i^N\int_{\mathcal{C}^N} du_1\dots d u_N \Delta(u_1,\dots,u_N)^2 e^{-N\underset{k=1}{\overset{N}{\sum}} \ln u_k}\cr
&=&(-1)^{\frac{N(N+1)}{2}}i^N Z_N
\eea
The partition function \eqref{EPF} can be seen as the diagonalized form of a matrix integral over a set of $H_N(\Gamma)$ of normal matrices where the eigenvalues are restricted to the contour $\Gamma=\mathcal{C}$:
\beq H_N(\Gamma)= \{ M = \Lambda X \Lambda^\dagger \in \mathcal{M}_N(\mathbb{C})  \,|\, \Lambda \Lambda^\dagger = I_N \text{ and }  X=\text{diag}(x_i )  x_i\in \Gamma  \}\eeq
The measure on $H_N(\Gamma)$ is given by $\Delta(X)^2\,d\Lambda \,dX$ with $d\Lambda$ the Haar measure on $\mathcal{U}(N)$ and $dX=\underset{i=1}{\overset{N}{\prod}}dx_i$ is the curvilinear measure defined by:
\beqq dx=f'(t)dt \text{ for any parametrization } x=f(t) \,,\, t\in [0,1] \text{ of the curvilinear coordinate}\eeqq
In particular the measure $dX$ is invariant under the choice of the parametrization $f(t)$. The main interest of using \eqref{EPF} is that the loop equations and topological recursion method developed for hermitian matrix ensembles (i.e. $\Gamma=\mathbb{R}$) is mostly applicable for normal matrices. The only difficulty lies in the determination of the spectral curve for which additional ``hard edges'' terms will appear. Note also that the potential in \eqref{EPF} is logarithmic but not polynomial as usually assumed in hermitian matrix models.

\subsection{Time evolution and definition of various return times}

Let $\left(u_i\right)_{1\leq i\leq N}=\left(e^{i\theta_i}\right)_{1\leq i\leq N}$ be the eigenvalues of the unitary matrix $U_N$ drawn from the measure given in \eqref{EPF} and define $\theta_i(t)=t\theta_i$ corresponding to the eigenvalues of $U_N^{\, t}$. For convenience, we will always consider angles $\theta_i(t)$ belonging to the interval $\left[-\pi,\pi\right]$. At time $t=0$ it is obvious that all the eigenvalues are located at $1$ (i.e. $\theta_i=0$) and when $t$ increases, the eigenvalues $\theta_i(t)=t\theta_i$ start diluting around the unit circle. Note that since we are only interested in the eigenvalues we can extend the definition $\theta_i(t)=t\theta_i$ for any real positive times $t$ (and not only integers) even if the definition of $U_N^{\, t}$ could be problematic. When waiting long enough, it may happen that all the eigenvalues come back close to $1$ at the same time corresponding to a unitary matrix $U_N^{\, t}$ close to the identity matrix $I_N$. More precisely in this paper we are interested in the following quantities:
\begin{itemize}
\item Take $\epsilon>0$, we say that the eigenvalues have \textbf{strongly returned} around $1$ at time $t$ if 
$$ \forall \, 1\leq i\leq N\, : \,u_i(t)=e^{it\theta_i}\in \left[e^{-i\pi \epsilon},e^{i\pi \epsilon}\right]\subset \mathcal{C} $$
where the last interval notation is to be understood as the arc-interval on the unit circle $\{e^{it}, -\pi \epsilon\leq t\leq \pi \epsilon\}$.
\item Denote $U_N(t)=U_N^{\, t}$, we say that the eigenvalues have \textbf{weakly returned} around $1$ at time $t$ if: 
$$S_N(t)=\left|\frac{1}{N}\Tr U_N(t)\right|=\left|\frac{1}{N}\sum_{i=1}^N e^{it \theta_i}\right| \geq 1-\delta$$
where $\delta>0$ is a given fixed small constant. 
In the same way, we say that the eigenvalues have \textbf{real-part returned} at time $t$ if: 
$$R_N(t)=\frac{1}{N}\sum_{i=1}^N \Re \left( \Tr U_N(t) \right)=\frac{1}{N}\sum_{i=1}^N \cos \left(t\theta_i\right) \geq 1-\delta$$
This definition is the one used in \cite{Balian1} to define the recurrence time.
\item  \textbf{First return time}: For a given sampling of $U_N$, we define the first strong (resp. weak, real-part) return time to be the smallest time for which we have a new strong (resp. weak, real-part) return time: 
\beqq \tau_{N,\text{strong}}=\underset{t>0}{\text{Inf}}\,(t\text{ strong return time and } \exists \, 0<s<t \,\mid\, s\text{ is not a strong return time}) \eeqq
The last condition is necessary to avoid trivially that $\tau_\text{strong}=0$. It can also be replaced by requiring $t>1$ since we expect return times to be rare when $N$ is large. Note in particular that the recurrence time $\tau_{\text{rec}}$ introduced in \cite{Balian1} corresponds to $\tau_{N,\text{real-part}}$.
\end{itemize}
We observe that we have the following inequalities:
\beq \tau_{\text{weak}}(\delta) \leq \tau_{\text{real-part}}(\delta) \,\text{ and } \tau_{\text{real-part}}(\delta=1-\cos \pi\epsilon)\leq \tau_\text{strong}(\epsilon)\eeq
In other words, it is more common to observe a weak return time than a real-part return than a strong return time after a trivial identification of the parameters used in the definitions. Eventually, we also introduce the probabilities of each return event:
\bea P_{N,\text{strong}}(t)&=& \mathbb{P}(t \text{ is a strong return time})\cr
P_{N,\text{weak}}(t)&=& \mathbb{P}(t \text{ is a weak return time})\cr
P_{N,\text{real}}(t)&=& \mathbb{P}(t \text{ is a real-part return time})\cr
\eea
Note here that the probabilities are computed relatively to the Haar measure defined earlier for the matrix $U_N=U_N(t=1)$. Since we are only interested in the eigenvalues of the matrix, the problem can also be illustrated as follow: take $N$ particles on the unit circle and make them rotate, starting at $t=0$ at the point $1$, with a constant velocity $v_i=\theta_i \in[-\pi,\pi]$ taken randomly according to the measure induced by \eqref{premiere}. Then our problem consists in studying the times when all the particles come back close to the point $1$ in the different ways presented above. For example the strong return time problem can be illustrated with the following picture:

\medskip

\begin{center}
\includegraphics[width=10cm]{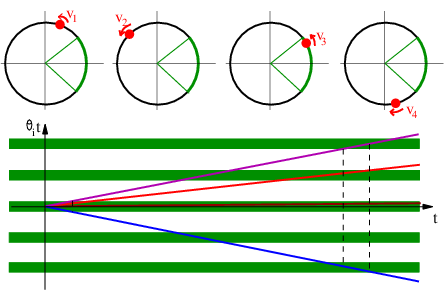}

\textit{Fig. $1$: Illustration of the strong return problem for $N=4$. Dashed lines give a strong return time interval.}
\end{center}

\medskip

In this moving particles perspective, a more natural setting would correspond to draw the velocities uniformly and independently on $[-\pi,\pi]$ rather than using the measure induced by \eqref{premiere}. However since the two measures are different the results cannot directly be extended from one setting to the other. However we will see that for some quantities both problems can be thought as equivalent.

\section{\label{Section2}Weak and real-part return times}

Eigenvalues of unitary matrices have been studied for quite a long time both at the macroscopic (distribution of the eigenvalues) and microscopic (local correlation between the eigenvalues) levels. Here we recall a few of the many results available in the literature. First since the works of Diaconis and Shashahani \cite{Diaconis} we know that the following theorems hold:
\begin{proposition} \label{asconvergence}Let $U_N$ be a sequence of $N \times N$ unitary matrices. Then the eigenvalues $\lambda_N=(\lambda_{1,1},\dots,\lambda_{1,N})$ tend almost surely to independent uniform variables on the circle. 
\end{proposition}
Moreover we get from \cite{DJ}:
\begin{proposition} \label{traceconvergence} Let $Z$ be a standard complex normal distributed random variable, and let $k(N)$ be an increasing sequence of integers. Then we have:
\beq \frac{1}{\sqrt{\text{min}(k(N),N)}}\Tr \left( U^{k(N)}_N \right) \underset{\text{ in distribution}}{\overset{N \to \infty}{\to}} Z\eeq
\end{proposition}
In particular there are two different regimes depending on whether the exponent increases more rapidly than the size of the matrix or not:
\bea &&\text{for } l\leq N: \frac{1}{\sqrt{l}}\Tr U^l_N\underset{\text{ in distribution}}{\overset{N \to \infty}{\to}} Z\cr
&&\text{for } l>N: \frac{1}{\sqrt{N}}\Tr U^l_N\underset{\text{ in distribution}}{\overset{N \to \infty}{\to}} Z
\eea
Eventually, results regarding the eigenvalues for a finite $N$ are known (\cite{Rains,Meckes}): 
\begin{proposition}\label{Indep} For any $m>N$, the eigenvalues $u_{1,m},\dots,u_{N,m}$ of $U_N^m$ are independent and uniformly distributed along the unit circle 
\end{proposition}
\begin{proposition}\label{NumberEigen} Let $m\leq N$ and $\theta\in[0,\pi]$ and denote $\mathcal{N}_{m,N}(\theta)$ the number of eigenvalues of $U_N^m$ inside the arc-interval $\left[1,e^{i\theta}\right]$. Then we have $\forall \, x>0$:
\beq P\left(\left|\mathcal{N}_{m,N}-\frac{N\theta}{2\pi}\right|>x\right)\leq 2 \,\textnormal{exp}\left(-\textnormal{Min}\left(\frac{x}{2},\frac{x^2}{4m\left(1+\ln\frac{m}{N}\right)}\right)\right)\eeq
\end{proposition}

Results \ref{traceconvergence} are interesting when studying the weak return times that involve the trace of the powers of the matrix $U_N$. We will specify the corresponding results in the next section. However, these theorems do not provide direct information about the strong return times (which are rarer then weak return times) and do not specify the convergence rates or exact bounds at finite but large $N$. In particular, for a given time $t$, the size of the matrix will eventually becomes larger than $t$ so that results from \eqref{Indep} will become useless.

\subsection{Threshold results for weak and real-part return times at large $N$}

We can use the two previous theorems in order to obtain some estimates of real-part and weak return times. Indeed, it is trivial that if the $\theta_i$'s are independent uniform variables on $[-\pi,\pi]$ then the variables $t \theta_i$'s are also independent uniform variables on $[-t\pi,t\pi]$. Moreover an elementary computations shows that:
\beq \int_{-\pi t}^{\pi t} \cos(\theta) \frac{d\theta}{2\pi t}=\frac{\sin \pi t}{\pi t}=\int_{-\pi t}^{\pi t} e^{i\theta} \frac{d\theta}{2\pi t}\eeq
For convenience we define the cardinal sine function to be:
\beq \label{sinecardinal}\sin_c(x)\overset{\text{definition}}{=}\frac{\sin x}{x}\eeq
Since it is known from \ref{asconvergence} that our eigenvalues almost surely tend to independent uniform variables we get that $\forall t>0$:
\bea R_N(t)&\overset{a.s.}{\underset{N\to \infty}{\to}}& \frac{\sin \pi t}{\pi t}\cr
 S_N(t)  &\overset{a.s.}{\underset{N\to \infty}{\to}}& \frac{\sin \pi t}{\pi t}
\eea
Since the r.h.s. decreases as $\frac{1}{t}$ we get the following:
\bea \forall \,t\leq t_c&:& P_{N,\text{weak}}(t) \text{ and } P_{N,\text{real}}(t)= \underset{N \to \infty}{\to} 1\cr
\forall \, t>t_c&:& P_{N,\text{weak}}(t) \text{ and } P_{N,\text{real}}(t)= \underset{N \to \infty}{\to} 0
\eea
where $t_c$ is the only positive time for which $\text{sin}_c(t_c)=1-\delta$ when $\delta$ is small. This threshold result and in particular for $t>t_c$ means that for a large but finite $N$ the weak or real-part return probability at a given time $t$ should be extremely small and rapidly decreasing when $N$ increases. In fact it seems natural to use theorem \eqref{traceconvergence} to obtain an estimate of the return probability for a large but given $N$.

\subsection{Estimates for the weak and real-part probability return for large $N$}

Let us use \ref{traceconvergence} in order to get an estimate of $R_N(t)$ and $S_N(t)$ when $t$ is a strictly positive integer. Theorem \ref{traceconvergence} implies that the leading order of $P_{N,\text{weak}}(t)$ as $N\to \infty$ should be given by:
\beaa &&\text{for a given } t>0 : P_{N,\text{weak}}(t)\underset{N\to \infty}{\sim} \mathbb{P}\left(|Z| \geq \frac{(1-\delta)N}{\sqrt{t}}\right)\cr
&&\text{for a sequence of times } t_N>N : P_{N,\text{weak}}(t_N)\underset{N\to \infty}{\sim} \mathbb{P}\left(|Z| \geq (1-\delta)\sqrt{N} \right)\cr
\eeaa
where $Z$ is a complex standard normal distribution. The r.h.s. can be computed explicitly since the complex density of $Z$ is given by $f(z)=\frac{1}{\pi}e^{-|z|^2}$. Note that after integrating over the angle, $|Z|$ has a Rayleigh distribution with parameter $\sigma^2=\frac{1}{2}$ whose density is given by $g(r)dr=2re^{-r^2}dr$. A straightforward computation shows that:
\beq \mathbb{P}\left( |Z| \geq \frac{(1-\delta)N}{\sqrt{t}}\right)=\int_{\frac{(1-\delta)N}{\sqrt{t}}}^{+\infty}re^{-r^2}dr=e^{-\frac{(1-\delta)^2N^2}{t}}\eeq
In other words, we find that for a given large $N$:
\bea \label{Pweak} \text{for a given } t>0 &:& \, P_{N,\text{weak}}(t)\underset{N\to \infty}{\sim}e^{-\frac{(1-\delta)^2N^2}{t}}  \cr
\text{for a sequence of times } t_N\geq N &:&\, P_{N,\text{weak}}(t_N)\underset{N\to \infty}{\sim}e^{-(1-\delta)^2N}
\eea
We can use similar techniques for the real-part return times. Note that the real part of $Z$ is by definition a real normal distribution $\mathcal{N}(0,\sigma^2=\frac{1}{2})$. We find:
\bea P_{N,\text{real}}(t)&=&\mathbb{P}\left(\frac{1}{N}\sum_{i=1}^N \Re \Tr U(t) \geq 1-\delta\right)=\mathbb{P}\left(\Re \Tr U_N^{\, t} \geq (1-\delta)N\right)\cr
&\underset{N\to \infty}{\sim}& \mathbb{P}\left(\Re\, Z\geq \frac{(1-\delta)N}{\sqrt{t}} \right)=\int_{\frac{(1-\delta)N}{\sqrt{t}}}^{+\infty} \frac{1}{\sqrt{\pi}}e^{-x^2}dx=\frac{1}{2}\text{erfc}\left(\frac{(1-\delta)N}{\sqrt{t}}\right)\cr
\eea
where $\text{erfc}(x)$ is the complementary error function whose asymptotic expansion is given by:
\beq \text{erfc}(x)=\frac{e^{-x^2}}{x\sqrt{\pi}}\sum_{n=0}^\infty \frac{(-1)^n (2n-1)!!}{(2x^2)^n}\underset{x\to \infty}{\sim}\frac{e^{-x^2}}{x\sqrt{\pi}}\eeq
Eventually we find:
\bea \label{Preal}\text{for a given } t>0&:&\, P_{N,\text{real}}(t)\underset{N\to \infty}{\sim}\frac{\sqrt{t}}{2N(1-\delta)\sqrt{\pi}}e^{-\frac{(1-\delta)^2N^2}{t}} \cr
\text{for a sequence of times } t_N\geq N &:&\, P_{N,\text{real}}(t_N)\underset{N\to \infty}{\sim}\frac{1}{2\sqrt{N}(1-\delta)\sqrt{\pi}}e^{-(1-\delta)^2N} 
\eea

The general form of \eqref{Pweak} and \eqref{Preal} for a given $0<t<N$ is consistent with the expected form of the series expansion of the random matrix integrals like \eqref{EPF}. Indeed, for random matrix integrals, it was proved in \cite{Hiai} that the leading order of the partition functions should be of order:
\beq \ln Z_N=N^2F^{[-2]}+ O(\ln N) \eeq
This result is also compatible with the full expansion provided by Borot, Guionnet and Kozlowski in the main theorem of \cite{BorotGuionnetKoz} that we will present below.

\subsection{The case of strong return times}

It is tempting to apply the previous method to the strong return times and to consider the eigenvalues $\theta_i$'s as i.i.d. uniform variables on $[-\pi, \pi]$ when $N$ is large. Under this assumption a simple computation shows that we would get:
\beq  \label{SimpleModel} \td{P}_{N,\text{strong}}(t)= \left|\frac{I(t)}{2\pi}\right|^N \eeq
where $I(t)$ is a union of intervals characterizing the different possibilities that $t\theta\in[-\pi\epsilon,\pi\epsilon] \text{ mod } 2\pi$. In particular the $N$ dependence in this case is of the form $\td{P}_{N,\text{strong}}(t)=e^{N\ln \left|\frac{I(t)}{2\pi}\right|}$ which is no longer consistent with the expected $e^{-N^2 F_0(t)}$. As we will see in the next sections, this simple model does not give the right answer for $P_{N,\text{strong}}(t)$ that we find to be of the same form as $P_{N,\text{weak}}(t)$ and $P_{N,\text{real}}(t)$. \textbf{The main difference} with the other two cases \textbf{is that for the real and weak return times, the presence of the trace} (i.e. the sum over all eigenvalues) \textbf{averages quantities} sufficiently so that the simple independent and uniform variable model is a very good approximation. On the contrary strong return times do not involve averaged quantities and exhibit a peculiar behavior. It is the purpose of this article to study the large $N$ asymptotic of these strong return times using first Toeplitz determinants and more importantly matrix models techniques. 

\section{\label{Appendix A} Strong return times: Exact computations at finite $N$ with Toeplitz determinants}

\subsection{Toeplitz determinants}

In this section, we use the general theory of Toeplitz determinants (see \cite{Rev} for a simple review) to express the \text{exact} value of $P_{N,\text{strong}}(t)$. We start with the definition:
\beq \label{Toep}P_{N,\text{strong}}(t)=\frac{1}{\td{Z}_N}\int_{\td{I}(t)^N} \prod_{i<j} |e^{i\theta_i}-e^{i\theta_j}|^2 d\theta_1\dots d\theta_N\eeq
where (Cf \eqref{Normalization} for the following computation) the normalization is given by:
\beq \td{Z}_N=\int_{[-\pi,\pi]^N} \prod_{i<j} |e^{i\theta_i}-e^{i\theta_j}|^2 d\theta_1\dots d\theta_N=(2\pi)^NN! \eeq
The domain of integration $\td{I}(t)$ is a union of intervals corresponding to the fact that $e^{it\theta_i}$ must belong to the segment of unit circle $\left[e^{- i\pi \epsilon},e^{i \pi \epsilon}\right]$. A straightforward computation shows that for any integer $k\geq 1$:
\bea \label{IntervalsInTime}
\forall t\in[2k+\epsilon,2(k+1)-\epsilon]&:& \td{I}(t)=\underset{j=-k}{\overset{k}{\bigcup}} \left[\frac{2\pi j-\pi \epsilon}{t},\frac{2\pi j+\pi \epsilon}{t} \right]\cr 
 \forall t\in[2k-\epsilon,2k+\epsilon]&:& \td{I}(t)=\left(\underset{j=-k+1}{\overset{k-1}{\bigcup}} \left[\frac{2\pi j-\pi \epsilon}{t},\frac{2\pi j+\pi \epsilon}{t} \right] \right)\cup \left[\frac{2\pi k-\pi \epsilon}{t},\pi \right]\cr
&&\cup \left[-\pi,-\frac{2\pi k-\pi \epsilon}{t} \right]
\eea
In the first case, we get $2k+1$ complete intervals of length $\frac{2\pi\epsilon}{t}$ whereas in the first cases we get $2k-1$ complete intervals and two pieces that reach the endpoints $\pi$ and $-\pi$. We illustrate the situation with the following picture:

\begin{center}
\includegraphics[width=10cm]{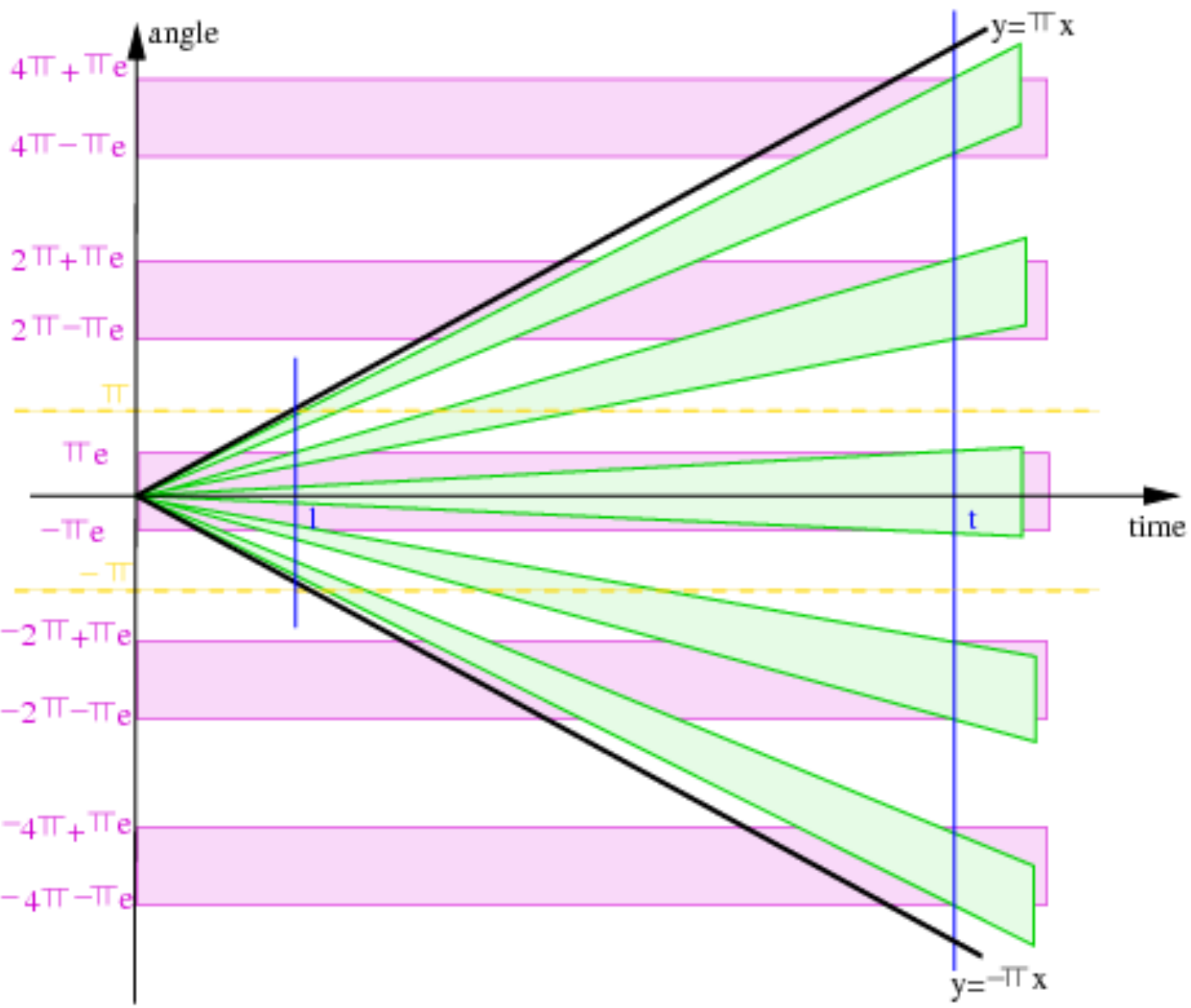}

\textit{Fig. $2$: Evolution of the angles $t\theta_i$}
\end{center} 

Integrals of type \eqref{Toep} are known as Toeplitz integrals. Toeplitz integrals can be rewritten as the determinant of a Toeplitz matrix (i.e. a matrix whose entries $M_{i,j}$ only depend on $j-i$) that can be obtained from the discrete Fourier transform of their symbol. We remind quickly the reader of the standard formalism. Define:
\beq \label{Toep2}I_f=\int_{[-\pi,\pi]^N} \prod_{i=1}^N f(e^{i\theta_i}) \prod_{i<j} |e^{i\theta_i}-e^{i\theta_j}|^2 d\theta_1\dots d\theta_N\eeq
where $f$ is an integrable function on the unit circle called the \textit{symbol} of the integral $I_f$. Then, the general theory of Toeplitz matrices implies that the integral $I_f$ can be expressed as the determinant of a Toeplitz matrix of size $N\times N$ whose entries are given by the Fourier coefficients of $f$. Namely:
\beq I_f=(2\pi)^N N! \det\left(T_{i,j}(f)=t_{i-j}\right)_{1\leq i,j\leq N} \,\,\text{ with }\,\, t_k=\frac{1}{2\pi}\int_0^{2\pi}f(e^{i\theta})e^{-ik\theta}d\theta\eeq
The matrix $T_N(f)$ is a Toeplitz matrix and one can reconstruct $f$ on the unit circle from its Fourier coefficients by the usual formula:
\beq f(e^{i\theta})=\sum_{k=-N}^N t_k e^{ik\theta}\eeq
making the correspondence one to one between a Toeplitz matrix and its symbol $f$ on the unit circle. The general theory can first be applied for the normalizing constant $\td{Z}_N$ for which the symbol is $f(x)=1$ and gives $(2\pi)^N N!$. We rederive this result using permutation techniques in appendix. Moreover, the numerator in \eqref{Toep} is also a Toeplitz integral associated with the symbol $f_t(e^{i\theta})=\mathbf{1}_{\td{I}(t)}(\theta)$ that is to say a piecewise constant function on the unit circle. Note that since $\td{I}(t)$ can be decomposed into unions of disjoint intervals, one could rewrite $f_t$ as a product: $f_t(e^{i\theta})=\prod_{j} \mathbf{1}_{\td{I}_{j,t}}(\theta)$. Unfortunately, the products of symbols is not known to behave nicely in the theory of Toeplitz determinants so we did not find a direct way to use this observation. However, since we know exactly the symbol $f$ of our integral, we can rewrite our strong return probability as a Toeplitz determinant by computing the Fourier coefficients directly. The computation is relatively straightforward and we propose in appendix another equivalent way to recover the results by using permutations of $S_N$. We find:

\bea \label{formula1bis}  P_N\left(t\in [2R+\epsilon,2(R+1)-\epsilon]\right)&=&\det \left[ \frac{\sin\frac{(j-i)(2R+1)\pi}{t} \sin\frac{(j-i)\pi\epsilon}{t}}{\pi(j-i)\sin\frac{(j-i)\pi}{t}} \right]_{1\leq i,j\leq N}\cr
P_N \left(t\in [2R-\epsilon,2R+\epsilon]\right)&=&\det \left[ \delta_{j-i=0}-\frac{ \sin \frac{2(j-i)\pi R}{t} \sin \frac{(1-\epsilon)(j-i)\pi}{t}}{\pi(j-i)\sin \frac{(j-i)\pi}{t}} \right]_{1\leq i,j\leq N}\cr
\eea



The problem is now to get the large $N$ expansion of such determinants. It is also worth noticing that numerically such determinants are much easier to handle than the computation the initial integrals. Note also that in our case the Toeplitz matrices involved are symmetric. 

\subsection{Simplification at integer times}

At integer times, the last formulas substantially simplify and we get:
\beq \label{OddInteger}  P_N\left(t=2k+1\right)=\epsilon^N\det \left[ \sin_c\left(\frac{(j-i)\pi \epsilon}{2k+1}\right) \, \delta_{(j-i)\, \equiv \,0\,[2k+1]} \right]_{1\leq i,j\leq N} 
\eeq
\beq \label{EvenInteger}  P_N\left(t=2k\right)=\det \left[ \delta_{j-i}-(1-\epsilon)\sin_c\left(\frac{(j-i)\pi (1-\epsilon)}{2k}\right) \, \delta_{(j-i)\, \equiv \,0\,[2k]} \right]_{1\leq i,j\leq N} 
\eeq

Note that in the even integer case, the first delta function is necessary to force the diagonal entries of the matrix to be $\epsilon$ (and not $\epsilon-1$ as the $\sin_c$ would imply). As far as we know, there exists no general formula in the literature for such determinants although most of the entries are vanishing. Getting the asymptotic of such determinants is a difficult problem but as we will see in \ref{IntegerTimes}, matrix models techniques and the use of the specific rotation symmetry at these times give the possibility to actually compute the first terms of the series expansion when $N\to \infty$.

\subsection{Simple case when $t>N$}

Let us consider an integer time $t>N$ (or a sequence of times $t_N>N$). In this case the previous determinants simplify because the condition $\delta_{(j-i)\, \equiv \,0\,[2k]}$ is only satisfied for $i=j$ giving a diagonal determinant. It trivially gives:
\beq \label{exactintegertimes}\forall\, t>N \in \mathbb{N}\,:\, P_N(t)=\epsilon^{N}=e^{-N\left|\ln \epsilon\right|}  \eeq
This formula is exact and recover the result of the simple independent and uniform model \eqref{SimpleModel}. We also note that this result has the same $N$ dependence as the results \eqref{Pweak} and \eqref{Preal}. However in this case, the probability does no longer depend on $t$ and remains constant. Result \eqref{exactintegertimes} is exact for integer times but can be adapted for non-integer times to get the leading order as $N\to \infty$ but no longer an exact formula. Let us consider a time $t>N$ (not necessarily in $\mathbb{N}$). The exact return probability is given by the Toeplitz determinant \eqref{formula1bis}. In the first line of formula \eqref{formula1bis}, we have: $R=\lfloor \frac{t-\epsilon}{2}\rfloor$ while in the second we have $R=\lfloor\frac{t+\epsilon}{2}\rfloor$. Observe also that in the determinants, the diagonal contributions are special. For large $t$, a simple expansion of the determinant shows that the diagonal terms represent the leading order since they are the only ones not to involve factors of $\frac{1}{t}$. Hence we find that:
\beq \label{Wild} \forall\, t>N \,:\, P_N(t)\underset{N\to \infty}{\sim} \epsilon^N\left(\frac{2}{t}\lfloor \frac{t}{2}\rfloor\right)^N=e^{N\epsilon\ln\left(\frac{2}{t}\lfloor \frac{t}{2}\rfloor\right)}\eeq
Note that we always have:
\beq 1-\frac{2}{t}\leq \frac{2}{t}\lfloor \frac{t}{2}\rfloor\leq 1 \eeq
so that for integer times $t>N$ we recover that the leading order is $\epsilon^N$ in agreement with the exact formula \eqref{exactintegertimes}. The function $t\mapsto \frac{2}{t}\lfloor \frac{t}{2}\rfloor$ gives damped oscillations of period $2$ and looks like:
 
\begin{center}
\includegraphics[width=10cm]{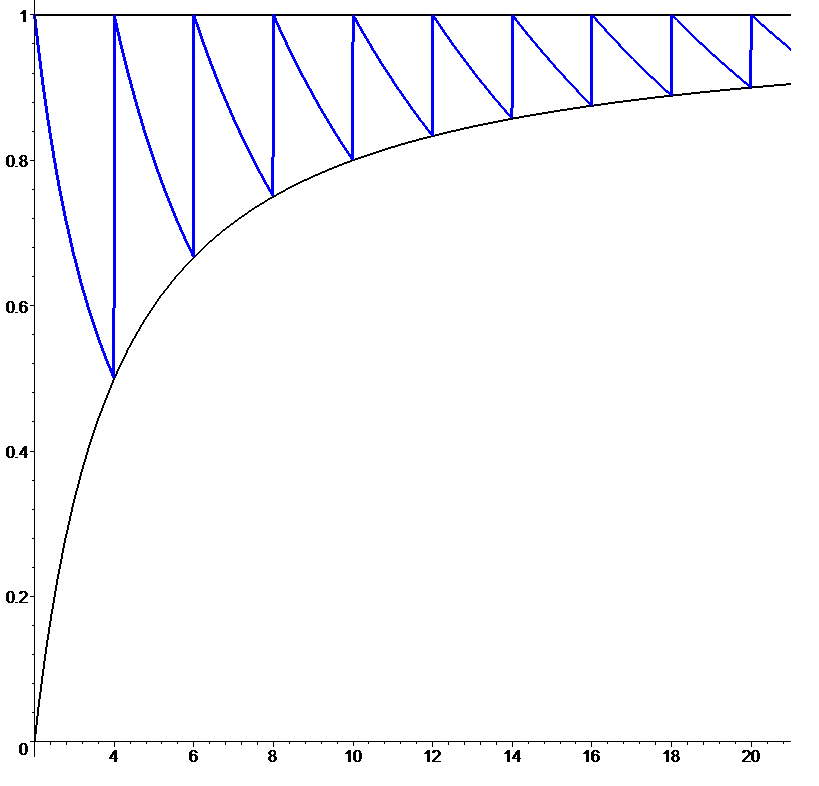}

\textit{Fig. $3$: Plot of the curves $t\mapsto \frac{2}{t}\lfloor \frac{t}{2}\rfloor$ (blue) and $y=1-\frac{2}{t}$ and $y=1$ (black)} 
\end{center}

The oscillations around the value $\epsilon^N$ are easily recovered in numerical simulations for $t\approx N$ but are damped for very large time $t>>N$ like $\frac{2}{t}$.

\subsection{General theorem for the asymptotic of a Toeplitz determinant}

We now need to understand the situation when $t\leq N$. Toeplitz determinants have been studied intensively \cite{Szego,Widom,Widom2,BS,BW,FishHart} but as far as we know our general case is not part of the general results of the theory. Indeed, the first standard result about Toeplitz determinants is the following:
\begin{proposition} (Szeg\"{o})\label{theoToep} If $f$ is strictly positive and continuous on the unit circle then we have:
\beq \frac{1}{N}\ln \det T_N\overset{N\to \infty}{\to} \frac{1}{2\pi}\int_0^{2\pi}\ln(f(\lambda)) d\lambda\eeq
\end{proposition}
Unfortunately this general result cannot apply to our case because our symbol $f$ is not continuous nor strictly positive on the unit circle (it is the indicator function of a union of intervals). This is in fact not surprising because we expect the leading order in $N$ to be of the form $N^2$ while the last theorem would only give a $N$ dependence. Since the works of Szeg\"{o}, many refinements of theorem \ref{theoToep} have been carried out. For example, Fisher-Hartwig singularities \cite{BW,FishHart} can be introduced giving formulas for the asymptotic of a Toeplitz determinant whose symbol is a strictly positive function $f$ having jump discontinuity or isolated zeros. Unfortunately, these types of singularities are still not sufficient for our purpose since we have intervals over which our symbol function $f$ is vanishing. Indeed, we stress here that the \textbf{main difficulty does not lie in the discontinuity of the characteristic functions but rather in the fact that the functions identically vanishes on certain segments}. To our knowledge, the closest known case regarding our problem has been carried out by Widom (See main theorem of \cite{Widom}) for a symbol $f$ supported on the arc interval $\left[\alpha,2\pi-\alpha\right]$. In this case the quantity $\frac{1}{N^2} \ln \det T_N(f)$ is convergent in contrast with \ref{theoToep} for which $\frac{1}{N} \ln \det T_N(f)$ is convergent. This change of behavior in the asymptotic is typical of a symbol vanishing on a segment (and as we will see is logical from a matrix model perspective). In the next section, we will recover this result (and extend it to any interval $\left[\theta_0,\theta_1\right]$) and provide a way to obtain all the subleading terms recursively by the use of the symplectic invariants developed by Eynard and Orantin in \cite{EO}. Unfortunately the general case of a union of intervals is much more difficult even with the use of the symplectic invariants because the genus of the spectral curve increases rapidly. To our knowledge the problem of finding exact values or large $N$ asymptotic of Toeplitz determinants whose symbol vanishes on multiple arc-intervals has never been answered in the literature. In the next section, we connect the strong return time probability problem with the series expansion of a suitable matrix model providing an algebraic characterization of the solution. In practice, this solution is mainly formal since when the spectral curve is of genus $g$ we need to determine ``optimal filling fractions" to get the complete spectral curve, a problem which is known to be transcendent and highly difficult to perform theoretically and even numerically. However, in the case of integer times, an additional rotation symmetry of the intervals allows a complete determination of the spectral curve so that we can get the asymptotic expansion of the corresponding Toeplitz determinant. Moreover, for general (non necessarily integer valued) $t<N$, we give an accurate way to approximate the leading order of the asymptotic of the Toeplitz determinant.   

\section{\label{Section4}Probability of a strong return time using matrix models techniques}

In the last section, we have formulated $P_{N,\text{strong}}(t)$ as a Toeplitz determinant with a piecewise symbol $f$ vanishing on a union of intervals. Unfortunately, getting a large $N$ asymptotic of such determinants is complicated and we propose to study the problem with random matrix models techniques.
Let us observe that:
\beq \label{MMIntegral2} P_{N,\text{strong}}(t)=\frac{1}{Z_N}\int_{I(t)^N} du_1\dots d u_N \Delta(u_1,\dots,u_N)^2 e^{-N\underset{k=1}{\overset{N}{\sum}} \ln u_k}\eeq
(where $Z_N$ is given by \eqref{EPF}) is exactly an hermitian matrix integral in its diagonalized form and therefore can be tackled through matrix models techniques. Here the set $I(t)$ is given by:
\bea \label{Intervalsss}\forall t\in[2k+\epsilon,2(k+1)-\epsilon]\,:\, I(t)&=&\underset{j=-k}{\overset{k}{\bigcup}} [e^{i\frac{2\pi j-\pi \epsilon}{t}},e^{i\frac{2\pi j+\pi \epsilon}{t}} ]\cr 
\forall t\in[2k-\epsilon,2k+\epsilon]\,:\, I(t)&=&\left(\underset{j=-k+1}{\overset{k-1}{\bigcup}} [e^{i\frac{2\pi j-\pi \epsilon}{t}},e^{i\frac{2\pi j+\pi \epsilon}{t}} ] \right)\cup [e^{i\frac{2\pi k-\pi \epsilon}{t}},e^{i\pi} ]\cr
&&\cup \, [e^{-i\pi},e^{-i\frac{2\pi k-\pi \epsilon}{t}} ]
\eea
In particular from matrix model considerations we expect $P_{N,\text{strong}}(t)$ to be of order $\exp(N^2 F^{[-2]}(t))$ when $N$ is large. The last matrix integral exhibits two main differences from a traditional hermitian matrix model. First, the integration set is compact and thus in matrix models terminology our integral has several ``hard edges'' corresponding to the endpoints of the intervals of $I(t)$. Secondly, the potential is only logarithmic but not polynomial (this would be impossible for hermitian matrix integrals because the integral would diverge). We will see later that these issues must be taken into account in the analysis but do not prevent the usual theory to work properly at least at leading order. We also recall that the normalizing constant $Z_N$ is known explicitly: $Z_N=(2\pi)^N N!(-1)^{\frac{N(N+1)}{2}}i^N$ and is expected to be subleading compared to the $e^{-N^2 F^{[-2]}}$ term.

\subsection{General form of the expansion at large $N$}
Computations of integrals of the form:
\beq \td{Z}_{N}=\int_{\td{I}^N} \prod_{i<j} |e^{i\theta_i}-e^{i\theta_j}|^2 d\theta_1\dots d\theta_N\eeq
given in \eqref{Toep} where $\td{I}$ is a union of $g+1$ intervals, falls into the category of mean field models with Coulomb gas interactions (i.e. a repulsive interaction equivalent to $(\theta_i-\theta_j)^2$ when $\theta_j\to \theta_i$) whose large $N$ asymptotic has been studied in \cite{BorotGuionnetKoz}. In our context, we have $\beta=2$ and only hard edges at the extremities of our intervals. The main theorem of \cite{BorotGuionnetKoz} (theorem $1.1$ page $6$) states that:

\begin{proposition} \label{PropBorot}Main theorem of Borot-Guionnet-Kozlowski in \cite{BorotGuionnetKoz} with $\beta=2$ and hard edges. For a measure supported on $g+1$ segments and under specific conditions (described and verified in appendix \ref{AppendixNew}) we have the following large $N$ expansion:
\bea \label{LaGrosse}\td{Z}_N&=&N^{N+\frac{1}{4}(g+1)} exp\left(\sum_{k=-1}^\infty N^{-2k}F_{\boldsymbol{\epsilon}^\star}^{[2k]}\right)\cr
&&\left\{\sum_{m\geq 0}\,\sum_{\begin{subarray}{l}l_1,\dots,l_m\geq 1\\ k_1,\dots,k_m\geq -1 \\ \underset{i=1}{\overset{m}{\sum}} l_i+2k_i>0\end{subarray}} \frac{N^{-\underset{i=1}{\overset{m}{\sum}} l_i+2k_i}}{m!}
\left(\underset{i=1}{\overset{m}{\bigotimes}} \frac{ F^{[2k_i],(l_i)}_{\boldsymbol{\epsilon}^\star}}{l_i!}\right) \cdot \nabla_\nu^{\otimes\left(\underset{i=1}{\overset{m}{\sum}} l_i\right)} \right\}\Theta_{-N\boldsymbol{\epsilon}^\star}\left(\mathbf{0}\big|F^{[-2],(2)}_{\boldsymbol{\epsilon}^\star}\right)\cr
\eea
where $\Theta$ is the Siegel theta function:
\beq \Theta_{\boldsymbol{\gamma}}(\boldsymbol{\nu},\mathbf{T})=\sum_{\mathbf{m}\in \mathbb{Z}^g} exp\left(-\frac{1}{2}(\mathbf{m}+\boldsymbol{\gamma})\cdot \mathbf{T}\cdot(\mathbf{m}+\boldsymbol{\gamma})+\boldsymbol{\nu}\cdot (\mathbf{m}+\boldsymbol{\gamma})\right)\eeq
and $F^{[2k],(l)}_{\boldsymbol{\epsilon}}$ are defined as the $l^{\text{th}}$ derivative of the coefficient $F_{\boldsymbol{\epsilon}}^{[2k]}$ relatively to the filling fractions (See paragraph $8.1$ of \cite{BorotGuionnetKoz}).
\end{proposition}
Details of the quantities in the last proposition can be found in \cite{BorotGuionnetKoz} but since we will not need the full expansion we only stress on the following points:
\begin{itemize} \item The exponential part $\underset{k=-1}{\overset{\infty}{\sum}} N^{-2k}F_{\boldsymbol{\epsilon}^\star}^{[2k]}$ is usually known in the literature as the perturbative (or sometimes called formal) series expansion of $\ln \td{Z}_N$. For $\beta=2$, only even powers of $N$ appears in the expansion and the numbers $F_{\boldsymbol{\epsilon}^\star}^{[2k]}$ are called symplectic invariants. They only depend on the spectral curve and can be computed using the so-called topological recursion described in \cite{EO}. We stress here that there is a difference in the notation with \cite{EO} where powers are indexed $F^{(k)}$ with $k\geq 0$. To avoid confusion we will keep the bracket notation $F^{[k]}$ for quantities referring to \cite{BorotGuionnetKoz} formalism. For $k\geq 0$, notation $F^{(k)}$ in Eynard-Orantin formalism \cite{EO} corresponds to $-F^{[2k-2]}$ in Borot-Guionnet-Kozlowski formalism. 
\item For $g=0$ (i.e. only one cut that is to say genus $0$ spectral curves), the Theta function is identically zero and we recover a $\frac{1}{N}$ expansion of $\ln \td{Z}_N$.
\item Since the Siegel Theta function is evaluated at $-N\boldsymbol{\epsilon}^\star$, we get a pseudo-periodic behavior at each order in $\frac{1}{N}$ including $N^0$.
\item The $g+1$ dimensional vector $\boldsymbol{\epsilon}^\star$ is the vector of optimal filling fractions $\epsilon_i$ to spread over the various intervals. In general it is a solution of difficult equations and thus is usually impossible to determine analytically. 
\end{itemize}

In our case, we are only interested in the first terms of the expansion and we get:

\begin{theorem} \label{Theo1} The probability to have a strong return time \textbf{at times} $\mathbf{t}\boldsymbol{\in} \boldsymbol{\mathbb{N}}^\ast$ (additional symmetry) \textbf{or} $\boldsymbol{\epsilon}\boldsymbol{<}\mathbf{t}\boldsymbol{<}\mathbf{2}\boldsymbol{-}\boldsymbol{\epsilon}$ (genus $0$ curve) satisfies conditions of \cite{BorotGuionnetKoz} for $\beta=2$ and hard edges so in the $g+1$ segments regime we have:
\beq \label{asympto} \ln \td{Z}_N=N^2F^{[-2]}_{\boldsymbol{\epsilon}^\star}+N\ln N+\frac{1}{4}(g+1)\ln N +F^{[0]}_{\boldsymbol{\epsilon}^\star}+\ln\left(\Theta_{-N\boldsymbol{\epsilon}^\star}\left(\mathbf{0}\big|F^{[-2],(2)}_{\boldsymbol{\epsilon}^\star}\right)\right)+ O\left(\frac{1}{N}\right)\eeq
where $\textbf{T}=F^{[-2],(2)}_{\boldsymbol{\epsilon}^\star}$ is the matrix of the second derivatives of $F^{[-2]}_{\boldsymbol{\epsilon}}$ relatively to the filling fractions, i.e.
the $g\times g$ matrix whose entries are given by:
\beq \textbf{T}_{i,j}=\partial_{t_1=0}\,\partial_{t_2=0}\,F^{[-2]}_{\boldsymbol{\epsilon}^\star +t_1\boldsymbol{\eta}_i +t_2\boldsymbol{\eta}_j} \text{ with } \boldsymbol{\eta}_k=\left(\underbrace{0,\dots,0}_{k-1},1,\underbrace{0\dots,0}_{g-k},-1\right)^t \in \mathbb{R}^{g+1}\eeq
For general times $t\in \mathbb{R}_+^\ast$, only the off-criticality condition of the equilibrium measure remains unproven though we conjecture that it should hold except at times $2k\pm\epsilon$ with $k\in \mathbb{N}^\ast$ where we have an isolated point $e^{i\pi}$ (for $2k-\epsilon$) or the splitting of an interval ($2k+\epsilon$).
\end{theorem}

The definitions and proofs of the technical conditions proposed by Borot, Guionnet and Kozlowski are presented in appendix \ref{AppendixNew}. Independently of their result, it is worth mentioning that the leading order $N^2F^{[-2]}_{\boldsymbol{\epsilon}^\star}$ follows directly from potential theory (See \cite{Johansson} and \cite{Hiai}). In particular, this leading term does not require the off-criticality condition and is defined even for singular equilibrium measure. This is helpful to deal with singular times $t=2k\pm \epsilon$ for which conditions of Borot, Guionnet and Kozlowski may not apply. This is also in agreement with our numerical simulations (see section \ref{Numerics} and our Average Block Interaction Approximation (see section \ref{MBIA}) where the leading order is a continuous function of $t$ including at those times.
\newline
Our strategy is now to compute the spectral curve of the problem in order to obtain the quantities $F_{\boldsymbol{\epsilon}^\star}^{[k]}$ at least for $k=-2$ and $k=-1$. In particular for integer times, an additional symmetry gives the possibility to compute $\boldsymbol{\epsilon}^\star$ exactly even when we have multiple cuts. Since for non-integer times computing $\boldsymbol{\epsilon}^\star$ is analytically impossible, we propose instead an approximation to get back to genus $0$ quantities.

\subsection{\label{generalities}Loop equations and spectral curve}
We plan to apply the standard techniques of the topological recursion to obtain the asymptotic expansion of \eqref{MMIntegral2}. We refer the reader to \cite{EO} for a review on this method. In random matrix models it is standard to introduce the following correlation functions:
\bea \label{deff}Z_{N,t}&=&\int_{I(t)^N}du_1\dots d u_N \Delta(u_1,\dots,u_N)^2 e^{-N\underset{k=1}{\overset{N}{\sum}} \ln u_k}\cr
W_{1,t}(x)&=&\left<\sum_{i=1}^N \frac{1}{x-u_i}\right>^t\cr
&=&\frac{1}{Z_{N,t}}\int_{I(t)^N}du_1\dots d u_N \left(\sum_{i=1}^N \frac{1}{x-u_i}\right) \Delta(u_1,\dots,u_N)^2 e^{-N\underset{k=1}{\overset{N}{\sum}} \ln u_k}\cr
W_{p,t}(x_1,\dots,x_p)&=&=\left<\sum_{i_1,\dots,i_p=1}^N \frac{1}{x_1-u_{i_1}}\dots\frac{1}{x_p-u_{i_p}}\right>_{c}^t
\eea
where the average $\left<\,\,\,\,\right>^t$ is taken relatively to the measure $Z_{N,t}$. The subscript $_c$ means the connected or cumulant part. In general, the correlation functions are useful when taking the $x\to \infty$ expansion that provide a generating series of the moments of the eigenvalues. These definitions recover the standard case of the one-matrix models integrals with hard edges at the endpoints of $I(t)$. The first loop equation is therefore:
\beq \label{FirstLoop} W_{1,t}^2(x)+W_{2,t}(x,x)-\frac{N}{x}W_{1,t}(x)-\frac{N}{x}\left<\sum_{i=1}^N \frac{1}{u_i}\right>^t=\sum_{a_k \text{hard edges of } I(t)} \frac{\alpha_k}{x-a_k}\eeq
where $\alpha_k$ are undetermined constants depending on $N$ and $t$ (they can be written formally as a matrix integral of size $N-1$, Cf. \eqref{Alphak}). The proof of \eqref{FirstLoop} is standard in random matrix theory and is presented in appendix \ref{Appendix B} for completeness. In random matrix theory, it is known that the so called $\frac{1}{N}$ expansion (also called perturbative expansion) of the correlation functions can be determined only with the spectral curve. The spectral curve is obtained from the projection of \eqref{FirstLoop} at leading order. Since we only need the leading order of the expansion, results of \cite{Johansson} and  \cite{Hiai} can be used even if the equilibrium measure is critical. We get:
\bea \label{dev}\ln Z_{N,t}&=&F^{[-2]}(t)N^2+O(N)\cr
W_{1,t}(x)&=&W_{1,t}^{[1]}(x)N+O(1)\cr
W_{p,t}(x_1,\dots,x_p)&=&W_{p,t}^{[2-p]}(x_1,\dots,x_p) N^{2-p}+O(N^{1-p})
\eea

As usual in matrix models integrals, it is convenient to translate the function $W_{1,t}^{[1]}(x)$ to define:
\beq y_t(x)\overset{\text{def}}{=} W_{1,t}^{[1]}(x)-\frac{1}{2x}\eeq
Projecting the first loop equation \eqref{FirstLoop} on $N^2$ gives the so-called \textbf{spectral curve} of the model:
\beq \label{SpectralCurve}y_t^2(x)=\frac{1}{4x^2}+\frac{c}{x}+\sum_{a_k \text{hard edges of } I(t)} \frac{\alpha_k^{(0)}}{x-a_k} \eeq
where $c$ and $\alpha_k^{(0)}$ are defined as:
\beq \label{EquationC} c=\lim_{N\to \infty} \frac{1}{N}\left<\sum_{i=1}^N \frac{1}{u_i}\right>^t \,\, ,\,\, \alpha_k^{(0)}=\lim_{N\to \infty} \frac{1}{N^2} \alpha_k\eeq
Here, $c$, $a_k$ and $\alpha_k$ as well as $\alpha_k^{(0)}$ are implicitly assumed to depend on $t$ but for compactness we will omit to rewrite down the explicit $t$ dependence.

From random matrix theory we know that the knowledge of this spectral curve should completely determine all functions $F_{N,t}^{[g]}, W_{1,t}^{[g]}(x)$ and $W_{p,t}^{[g]}(x_1,\dots,x_p)$ through the implementation of the \textbf{topological recursion} of this spectral curve as described in \cite{EO}. Unfortunately, the knowledge of the spectral curve \eqref{SpectralCurve} is incomplete so far since we have not determined constants $c$ and $\alpha_k^{(0)}$ in the formula. The general theory tells us that these constants can be fixed uniquely by specifying the following conditions:
\begin{itemize}
\item Asymptotic at infinity:
\beq \label{Asymp} y_t(x)\underset{x\to \infty}{=}\frac{1}{2x}+\frac{c}{x^2}+O\left(\frac{1}{x^3}\right)\,\Rightarrow y^2(x)=\frac{1}{4x^2}+\frac{c}{x^3}+O\left(\frac{1}{x^4}\right)\eeq
\item If we decompose the domain of integration in the form of union of arc intervals of the form $I(t)=\underset{k=1}{\overset{q}{\bigcup}} [a_k,b_k]=\underset{k=1}{\overset{q}{\bigcup}} I_k(t)$, the filling fractions $\epsilon_1,\dots,\epsilon_q$ (that also depend on $t$) defined by:
\beq \label{fillingfraction}\int_{I_k(t)} y_t(x)dx=2i\pi \epsilon_k\eeq
are fixed dynamically by minimizing the so-called chemical potential:
\beq  \label{equilibrium} \forall \, 1\leq k\leq q-1\,:\,\, \int_{b_k}^{a_{k+1}} y_t(x)dx=0\eeq
From these definitions it is clear that the filling fraction $\epsilon_k$ represents the proportion of eigenvalues in the arc interval $I_k(t)$. The previous minimization physically represents the equilibrium achieved by transferring some eigenvalues from one interval to another (tunneling). Note here that all integrals are restricted on the unit circle and therefore can be also expressed on intervals for the corresponding angles.
\end{itemize} 
In general, the first condition on the asymptotic is quite easy to deal with and provides simple relations between the unknown coefficients of the spectral curve. By contrast, the conditions on the filling fractions are known to be very difficult to handle on a computational perspective. The first condition on the asymptotic comes from the fact that by definition:
\beq W_{1,t}(x)=\left<\sum_{i=1}^N\frac{1}{x}+\frac{u_i}{x^2} +O\left(\frac{1}{x^3}\right)\right>^t\eeq
Now, since $u_i$ belongs to the unit circle we get that $\frac{1}{u_i}=\bar{u_i}$ and since the domain of integration $I(t)$ is symmetric under complex conjugate, we get that $\left<\underset{i=1}{\overset{N}{\sum}}u_i\right>^t=\left<\underset{i=1}{\overset{N}{\sum}} \frac{1}{u_i}\right>^t \in \mathbb{R}$. Projecting on the order of $N$ gives the constant $c$ introduced earlier and the asymptotic \eqref{Asymp} for $y(x)$. A straightforward computation combining \eqref{Asymp} and \eqref{SpectralCurve} for $y^2(x)$, shows that the asymptotic is satisfied if and only if the following constraints are verified:
\bea \label{AsympConstrains} 0&=&c+\sum_{a_k \text{hard edges of } I(t)} \alpha_k^{(0)} \,\,\text{  (Order } \frac{1}{x} \text{ of } y_t^2(x)\text{ )}\cr
    0&=&\sum_{a_k \text{hard edges of } I(t)}a_k\alpha_k^{(0)} \,\,\text{  (Order } \frac{1}{x^2} \text{ of } y_t^2(x)\text{ )}\cr
c&=&\sum_{a_k \text{hard edges of } I(t)}a_k^2\alpha_k^{(0)} \,\,\text{  (Order } \frac{1}{x^3} \text{ of } y_t^2(x)\text{ )}
\eea

\medskip

From \eqref{SpectralCurve}, we know that the spectral curve should be of the form:
\beq y_t(x)=\frac{1}{2x}\sqrt{\frac{P_d(x)}{\underset{k=1}{\overset{d}{\prod}}(x-a_k)(x-\bar{a_k})}}\eeq
where $d$ is the number of intervals in $I(t)$. $P_d(x)$ is a monic polynomial of degree $2d$. Let us denote $P_d(x)=\underset{i=1}{\overset{2d}{\prod}}(x-s_i)$ where $s_i$ are the complex zeros of $P_d$. Part of the coefficients of $P_d(x)$ are determined with equations \eqref{AsympConstrains} but in general degrees of freedom remain. Three situations may happen: 
\begin{itemize} \item $P_d(x)$ has an odd zeros outside the cuts defined by the intervals $[a_k,b_k]$. This creates an additional branchpoint and therefore a new cut where eigenvalues will concentrate. This cannot happen since we restricted the eigenvalues inside the original cuts. Hence if $P_d(x)$ has some odd zeros, they must lie inside the the union of intervals $[a_k,b_k]$.
\item $P_d(x)$ has two simple (or more generally odd) zeros noted $s_0$ and $s_1$ inside one cut $[a_k,b_k]$. In that case the only possible choice of cuts is $[a_k,s_0]$ and $[s_1,b_{k}]$ where $s_0$ and $s_1$ are soft edges (possibly singular if the zeros are not simple). This means that there is a whole interval gap $[s_0,s_1]$ inside the original cut $[a_k,b_{k}]$. In the case where $s_0=a_k$ and $s_1=b_{k}$ the cut disappears (or becomes a soft edges cut if $a_k$ and $b_k$ or multiple odd zeros). Note that in the case of simple zeros, the equilibrium measure remains non-critical but the number of cuts is just strictly greater than $d$.
\item $P_d(x)$ admits a double zero $s_0$ inside the cut $[a_k,b_k]$. In that case the equilibrium measure is vanishing at this point and thus becomes critical. This phenomenon usually appears when two intervals are merging or one interval is splitting into two.
\end{itemize}
In conclusion, we see that the multiplicity and location of the zeros of $P_d(x)$ are crucial to determine the number of cuts and the criticality of the equilibrium measure. In theory imposing \eqref{equilibrium} or solving the equilibrium measure problem presented in \eqref{energyfunct} should determine completely $y_t(x)$ or its corresponding equilibrium density. However in practice solving analytically one of these problem is out of reach with current tools. We conjecture that for generic times $t\neq 2\mathbb{N}\pm \epsilon$, $y_t(x)$ should only have double zeros outside the domain $I(t)$ (one in each gap between intervals $[a_k,b_k]$) and thus the corresponding equilibrium density is non-critical. As we will see in the following sections, there are two situations, namely when $I(t)$ consists only in one interval (i.e. $0<t<2-\epsilon$) or when $t\in \mathbb{N}^\ast$, where additional symmetry considerations are sufficient to determine the complete expression of $y_t(x)$. In particular we can verify our conjecture in those cases and the equilibrium density is non-critical. 

\subsection{The one cut case: Exact results for $t\leq 2-\epsilon$}
Let us start with the following observation:
\begin{theorem}
\beq \label{trivial} \forall\, 0\leq t \leq \epsilon\,, \, t \text{ is a strong return time and } Z_{N,t}=1 \eeq
\end{theorem}
Indeed all angles $\theta_i\in[-\pi,\pi]$ will be contracted to $t\theta_i\in[-t\pi,t\pi]\subset[-\pi \epsilon,\pi \epsilon]$. \newline
Let us now take $t\in [\epsilon,2-\epsilon]$ for which we have $d=1$. In appendix \ref{AppendixD} we prove that it gives a spectral curve of genus $0$ for which computations can be carried out completely. In particular, conditions of proposition \ref{PropBorot} are met. In appendix \ref{AppendixD} we study the more general case (we will need it below) given by :
\beq Z_{N,\epsilon_0,\theta_0,\theta_1}=\int_{[e^{i\theta_0},e^{i\theta_1}]^{N\epsilon_0}}du_1\dots du_{N\epsilon_0} \Delta(u_1,\dots,u_{N\epsilon_0})^2 e^{-N\underset{i=1}{\overset{N\epsilon_0}{\sum}} \ln u_i}\eeq
for which we find the leading orders of $\ln Z_{N,\epsilon_0,\theta_0,\theta_1}$ to be:
\bea \label{Widdd}-F^{[-2]}(N,\epsilon_0,\theta_0,\theta_1)&=&-(2\epsilon_0-1)^2\ln \left(\sin \frac{\left|\theta_1-\theta_0\right|}{4}\right)\cr
-F^{[0]}(N,\epsilon_0,\theta_0,\theta_1)&=&-\frac{1}{24}\ln 2-\frac{1}{12}\ln \epsilon_0+  \frac{1}{24}\ln\left(\tan\frac{\left|\theta_1-\theta_0\right|}{4}\right)-\frac{1}{8}\ln\left(\sin \frac{\left|\theta_1-\theta_0\right|}{2}\right)\cr
-F^{[2]}(N,\epsilon_0,\theta_0,\theta_1)&=&\frac{3\cos\left(\frac{\theta_1-\theta_0}{2}\right)-1}{128\,\epsilon_0^2\cos^2\left(\frac{\theta_1-\theta_0}{4}\right)}
\eea
Note here that it is important to select the angles within the same interval of length $2\pi$ (for example $[-\pi,\pi]$). As expected from rotation invariance the result only involves the length of the interval $\theta_1-\theta_0$ but not $\theta_0$ or $\theta_1$ directly and only real quantities are involved (since the partition function $Z_{N,\epsilon_0,\theta_0,\theta_1}$ is real, it is natural to check that its expansion are large $N$ only involves real quantities). The case $t\in [\epsilon,2-\epsilon]$ is obtained from \eqref{Widdd} by taking $\epsilon_0=1$ and $\theta_1=-\theta_0=\frac{\pi \epsilon}{t}$. Moreover since there is only one cut, no quasi-periodic terms are present in \eqref{LaGrosse} so we get:

\begin{theorem}\label{Theo2} For $\forall\, t\,\in [\epsilon,2-\epsilon]$, the probability to have a strong return time is given by:
\bea&& \label{F0t} \frac{1}{N^2}\ln P_{N,\text{strong}}(t)+\frac{1}{N^2}\ln((2\pi)^N N!)\underset{N\to \infty}{=}  \ln \left(\sin \frac{\pi\epsilon}{2t}\right)+\frac{\ln N}{N}+\frac{1}{4}\frac{\ln N}{N^2}\cr
&&+\frac{1}{24N^2}\ln \left(\frac{2\sin^3 \frac{\pi\epsilon}{t} }{\tan \frac{\pi\epsilon}{2t}}\right)+\frac{1}{64N^4}\frac{1-3\cos\left(\frac{\pi \epsilon}{t}\right)}{1+\cos(\frac{\pi \epsilon}{t})} +O\left(\frac{1}{N^6}\right) \eea
\end{theorem}

Details of the computations can be found in \ref{AppendixD}. As claimed earlier, this result is also in agreement with Widom's result \cite{Widom} on the asymptotic expansion of a Toeplitz determinant with a symbol $f$ supported on an arc interval $\left[\alpha,2\pi-\alpha\right]$. In other words, we have just provided here a \textbf{new derivation of Widom's result} with topological recursion techniques as well as the next order in the expansion. In particular we observe that $\sin \frac{\pi\epsilon}{2t}$ is smaller than $1$ so that the probability is indeed exponentially small when $N$ gets large. Moreover at $t=\epsilon$ we recover that $P_{N,\text{strong}}(t=\epsilon)\sim 1$ which is consistent with \eqref{trivial}. We can verify numerically that the last formula is correct by computing the exact value of $P_{N,\text{strong}}(t)$ for low values of $N$ using the Toeplitz determinant \eqref{formula1bis}. We find the following picture:

\begin{center}
\includegraphics[width=10cm]{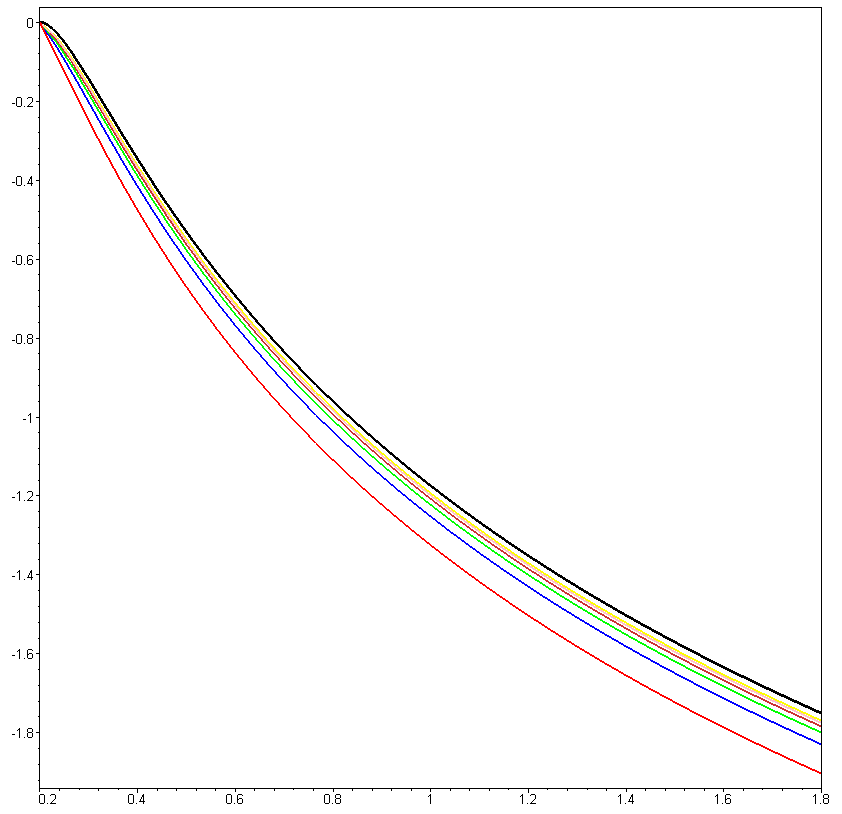}

\textit{Fig. $4$: Plot of $t \mapsto \frac{1}{N^2}\ln P_{N,\text{strong}}(t)$ for $\epsilon\leq t \leq 2-\epsilon$ with $\epsilon=\frac{1}{5}$ for $N\in\{1,2,\dots, 7\}$ computed exactly with the formula \eqref{EqComp}. The black curve is the expected value $\ln\left(\sin\frac{\pi\epsilon}{2t}\right)$ for large $N$.}
\end{center} 

We clearly observe the convergence to formula \eqref{F0t}. The simulations validate also the rate of convergence of order $O\left(\frac{1}{N^2}\right)$. Since the spectral curve is of genus $0$ and explicit (See \eqref{rototo}) one could apply the topological recursion on this curve to obtain all $F^{[2k]}(N,\epsilon_0,\theta_0,\theta_1)$ terms and therefore reconstruct up to any order the $\frac{1}{N}$ expansion of $\ln P_{N,\text{strong}}(t)$.

\subsection{\label{IntegerTimes}Exact Results for integer times}

At \textbf{integer times} the problem exhibits an additional discrete rotation symmetry giving extra information that is precious to solve the filling fractions problem. In particular it gives the opportunity to compute \textbf{exactly} the spectral curve and the optimal filling fractions. Consequently we can then compute the symplectic invariants associated to this spectral curve giving the $\frac{1}{N}$ series expansion of the corresponding Toeplitz determinant. Let us illustrate what happens in the case of an odd integer $t=2k+1$. For an odd integer time, we have $2k+1$ intervals of size $\frac{2\pi \epsilon}{2k+1}$ that are distributed along the unit circle with a discrete rotation invariance of angle $\theta=\frac{2\pi}{2k+1}$. For example for $k=4$, the situation is illustrated by:

\begin{center}
\includegraphics[width=5cm]{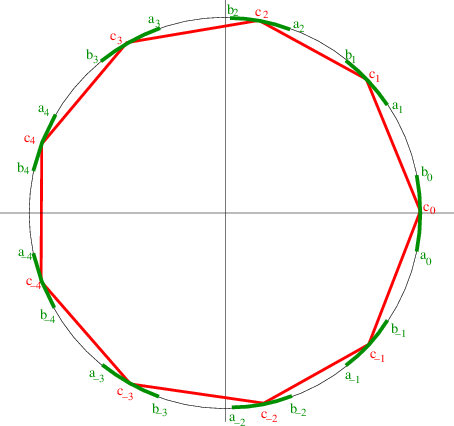}

\textit{Fig. $5$: Situation at time $t=2k+1$ with $k=4$. The $9$ distinct arc intervals are represented in green.} 
\end{center}

As explained earlier the general form of the spectral curve coming from loop equations should be of the form:
\beq \label{rototo2}y_{2k+1}^2(x)=\frac{P_{2k+1}(x)}{4x^2\prod_{j=-k}^k (x-a_j)(x-b_j)}=\frac{P_{2k+1}(x)}{4x^2\left(x^{2k+1}-e^{-i\pi \epsilon}\right)\left(x^{2k+1}-e^{i\pi \epsilon}\right)}\eeq
with $a_j=e^{\frac{2i\pi j -i\pi \epsilon}{2k+1}}$ and $b_j=e^{\frac{2i\pi j+i\pi \epsilon}{2k+1}}$ and $P_{2k+1}(x)$ is a monic polynomial of degree $2(2k+1)$. The rotation symmetry of the problem should apply also to our spectral curve. Since the denominator is already invariant under such rotations, the numerator should also share this symmetry. This implies that:
\beq \label{rototobis} P_{2k+1}(x)=\prod_{j=-k}^{k}\left((x-e^{\frac{2i\pi j}{2k+1}}a)(x-e^{\frac{2i\pi j}{2k+1}}b)\right)\eeq
Since the situation is also invariant by complex conjugation, it means that $b=\bar{a}$ so that:
\beq \label{rototo3bis} P_{2k+1}(x)=\prod_{j=-k}^{k}\left((x-e^{\frac{2i\pi j}{2k+1}}a)(x-e^{\frac{2i\pi j}{2k+1}}\bar{a})\right)\eeq
We can then use a similar symmetry arguments as the one developed in \eqref{Eqqq2}. Let us take $-k\leq j\leq k$, we can compute (defining $\td{u}_i=e^{-i\frac{2\pi j}{2k+1}}u_i$):
\bea W_1\left(-e^{\frac{2i\pi j}{2k+1}}\right)&=&-e^{-\frac{2i\pi j}{2k+1}}\left<\sum_{i=1}^N \frac{1}{1+\td{u}_i}\right>\overset{\text{symm.}}{=}-e^{-\frac{2i\pi j}{2k+1}}\left<\sum_{i=1}^N \frac{1}{1+\frac{1}{\td{u}_i}}\right>\cr
&=&-\frac{N}{2}e^{-\frac{2i\pi j}{2k+1}}
\eea
Hence 
\beq W_1\left(-e^{\frac{2i\pi j}{2k+1}}\right)=-\frac{N}{2}e^{-\frac{2i\pi j}{2k+1}} \,\, \Rightarrow y\left(-e^{\frac{2i\pi j}{2k+1}}\right)=0\eeq
Consequently we must have:
\beq \label{rototo3} P_{2k+1}(x)=\prod_{j=-k}^{k}\left((x+e^{\frac{2i\pi j}{2k+1}})^2)\right)\eeq
and the spectral curve reads:
\beq \label{SpecCurveOddInteger} y_{2k+1}^2(x)=\frac{\left(x^{2k+1}+1\right)^2}{4x^2\left(x^{2k+1}-e^{-i\pi \epsilon}\right)\left(x^{2k+1}-e^{i\pi \epsilon}\right)}\eeq
In particular this shows that the equilibrium density at $t_k=2k+1$ is supported on the the whole domain $I(t_k)$ and is non-critical (zeros of $y(x)$ fall into $\mathcal{C}\setminus I(t_k)$). This spectral curve looks very similar to the case of one interval with the exception that $x$ is now replaced by $x^{2k+1}$. A natural Zhukowsky parametrization would be:
\beq \left\{
\begin{array}{lcl}
  x^{2k+1}(z)&=&\cos \pi \epsilon+\frac{1}{2}\sin \pi \epsilon\left(z-\frac{1}{z}\right)\cr
y(z)&=&\frac{1+x^{2k+1}(z)}{x(z)\left(z+\frac{1}{z}\right)\sin \pi \epsilon}
\end{array}
\right.\eeq

One must be careful here since $x(z)$ is not well defined with the first equation. In fact around each interval, one should specify the choice of the branch used. In other words, the spectral curve \eqref{SpecCurveOddInteger} defines a Riemann surface that is obtained as the gluing of $2k+1$ identical genus $0$ Riemann surfaces. Define $X(z)=x^{2k+1}(z)$ and $Y(z)=\frac{y(z)}{(2k+1)x^{2k}(z)}$ satisfying $dx\wedge dy= dX\wedge dY$. Then \eqref{SpecCurveOddInteger} is equivalent to $2k+1$ copies of the following spectral curve:
\beq \label{EqSpecCurveOdd}\left\{
\begin{array}{lcl}
  X(z)&=&\cos \pi \epsilon+\frac{1}{2}\sin \pi \epsilon\left(z-\frac{1}{z}\right)\cr
  Y(z)&=&\frac{1+X(z)}{(2k+1)X(z)\left(z+\frac{1}{z}\right)\sin \pi \epsilon}
\end{array}
\right.\eeq

The spectral curve \eqref{EqSpecCurveOdd} defines a genus $0$ Riemann surface of the same nature as the one studied in appendix \ref{AppendixD} with identical filling fractions $\epsilon_0=\frac{1}{2k+1}$ (as symmetry implies) and $\theta_1=-\theta_0=\pi \epsilon$. Thus, we can directly use results given in appendix \ref{AppendixD} (keeping in mind that we have $2k+1$ identical copies) to compute $F^{[-2]}_{\boldsymbol{\epsilon}^\star}$ and $F^{[0]}_{\boldsymbol{\epsilon}^\star}$. Moreover in expansion \eqref{LaGrosse} we have contributions from the Siegel Theta function terms with $\boldsymbol{\epsilon}^\star=\left(\frac{1}{2k+1},\dots,\frac{1}{2k+1}\right)\in \mathbb{R}^{2k+1}$. Eventually we get:

\begin{theorem} \label{OODD}At integer times $t=2k+1$ with $k\in \mathbb{N}$, the probability to have a strong return time is given by:
\beaa &&\frac{1}{N^2}\ln P_{N,\text{ strong}}(2k+1)+\frac{1}{N^2}\ln((2\pi)^NN!)= \frac{1}{2k+1}\ln\left(\sin \frac{\pi \epsilon}{2}\right)+\frac{\ln N}{N}+\frac{2k+1}{4}\frac{\ln N}{N^2} \cr
&&-\frac{2k+1}{24N^2}\left(2\ln (2k+1)+\ln\left(4\tan\frac{\pi \epsilon}{2}\right)\right)+ \frac{1}{N^2}\ln\left(\Theta_{-N\boldsymbol{\epsilon}^\star}\left(\mathbf{0}\big|F^{[-2],(2)}_{\boldsymbol{\epsilon}^\star}\right)\right)+ o\left(\frac{1}{N^2}\right)\cr
\eeaa
with $\boldsymbol{\epsilon}^\star=\left(\frac{1}{2k+1},\dots,\frac{1}{2k+1}\right)\in \mathbb{R}^{2k+1}$
\end{theorem}

We stress that this result is \textbf{exact} and that one could obtain the next orders $F^{[g]}_{\boldsymbol{\epsilon}^\star}$ by applying the topological recursion to the curve \eqref{EqSpecCurveOdd} (see for example \eqref{F22}). Moreover, these results are consistent with the simulations presented in section \ref{Numerics} as well as the computations for general times $t \notin \mathbb{N}$ presented in section \ref{MBIA}. Nevertheless it is worth mentioning that our method does not provide a way to compute explicitly $F^{[-2],(2)}_{\boldsymbol{\epsilon}^\star}$ in theorem \ref{OODD}. Indeed, it would require to compute explicitly the energy functional (and its Hessian) or the spectral curve (and its first symplectic invariant) in a neighborhood of $\boldsymbol{\epsilon}^\star$. Since our method relies mostly on the symmetry at $\boldsymbol{\epsilon}^\star$ it cannot be extended easily in a neighborhood where the symmetry of the polynomial $P_{2k+1}$ will be lost. On the bright side, we do not expect the support of the equilibrium density to change, only the position of the zeros of $P_{2k+1}$ should change a little to accommodate the change in filling fractions.

\medskip

At even integer times $t=2k \in \mathbb{N}$, the situation is very similar and the same method can be applied. The spectral curve is given by:
\beq \label{blan} y_{2k}^2(x)=\frac{\left(x^{2k}+1\right)^2}{4x^2\left(x^{2k}-e^{-i\pi \epsilon}\right)\left(x^{2k}-e^{i\pi \epsilon}\right)}\eeq
Computations for this curve provides similar results as the odd case. Eventually we find that the last theorem can be extended to all integers:

\begin{theorem}\label{Theo3} At any integer time $t=k \in \mathbb{N}^\ast$, the probability to have a strong return time is given by:
\beaa &&\frac{1}{N^2}\ln P_{N,\text{ strong}}(k)+\frac{1}{N^2}\ln((2\pi)^NN!)= \frac{1}{k}\ln\left(\sin \frac{\pi \epsilon}{2}\right)+\frac{\ln N}{N}+\frac{k}{4}\frac{\ln N}{N^2} \cr
&&-\frac{k}{24N^2}\left(2\ln (k)+\ln\left(4\tan\frac{\pi \epsilon}{2}\right)\right)+ \frac{1}{N^2}\ln\left(\Theta_{-N\boldsymbol{\epsilon}^\star}\left(\mathbf{0}\big|F^{[-2],(2)}_{\boldsymbol{\epsilon}^\star}\right)\right)+ o\left(\frac{1}{N^2}\right)\cr
\eeaa
with $\boldsymbol{\epsilon}^\star=\left(\frac{1}{k},\dots,\frac{1}{k}\right)\in \mathbb{R}^{k}$.
\end{theorem}

We stress again that this result is \textbf{exact} and that one could apply the topological recursion to \eqref{SpecCurveOddInteger} and \ref{PropBorot} to get the next orders of the series expansion of $\ln P_{N,\text{ strong}}(k)$.

\section{\label{MBIA}Non-integer times: Average Block Interaction Approximation} 

When $t$ increases, the number of cuts also increases and therefore we have to deal with a spectral curve with a strictly positive genus. We have seen in last section that for integer times $t\in \mathbb{N}$ then an additional symmetry (rotation) gives us the opportunity to compute the exact spectral curve and thus the series expansion of $\ln P_{N,\text{strong}}(t)$. In theory, one could compute the corresponding $F^{[-2]}$ in a similar way than the one presented in the genus $0$ case. As mentioned earlier, the main difficulty for non-zero genus curves is that some coefficients of the spectral curve are determined by conditions \eqref{equilibrium} on the filling fractions. In practice, these conditions are impossible to solve except in some rare and exceptional cases with additional symmetries. A possible way to avoid this difficulty may consist in computing $F^{[-2]}$ for a spectral curve with unknown coefficients $\epsilon_1,\dots \epsilon_d$ for example using formulas developed in \cite{EO}. However this seems difficult since at some point we need to integrate $ydx$, that is to say the Abel map of the Riemann surface between two points. Such computations are usually very hard for a given Riemann surface and become unreasonable when the Riemann surface depends on some formal parameters $\epsilon_1,\dots,\epsilon_d$. Even if this could be performed and one could get an explicit formula for $F^{[-2]}(\epsilon_1,\dots,\epsilon_d)$, we would still have to minimize this function relatively to $(\epsilon_1,\dots,\epsilon_d)$ in order to obtain the correct dynamical filling fractions. Since the function $F^{[-2]}(\epsilon_1,\dots,\epsilon_d)$ is not expected to be simple, this last step is likely to be impossible analytically. In order to avoid such complications we propose to obtain some suitable approximations of $\frac{1}{N^2}\ln P_N(t)$ where only genus $0$ computations appear. Of course the price to pay is that we will only get an approximation (suitable for computation) but not the exact value.

\subsection{General Average Block Interaction Approximation}

In order to have a strong-return time at time $t$, we need to draw the eigenvalues $(\theta_1,\dots,\theta_N)$ from a union of $r(t)$ intervals (with $r(t)$ increasing with $t$). Let us denote generically $\left(\left[\theta_i^m(t),\theta_i^M(t)\right]\right)_{1\leq i\leq r(t)}$ the $r(t)$ arc-intervals given by \eqref{IntervalsInTime}. We label them so that: $-\pi< \theta_1^m(t)<\theta_1^M(t)<\theta_2^m(t)<\dots<\theta_r^m(t)<\theta_r^M(t)\leq \pi$. The probability to have a strong return time at time $t$ is thus given by:
\beq P_{N,\text{strong}}(t)=\frac{1}{Z_N}\int_{\left(\underset{K=1}{\overset{r(t)}{\bigcup}}[\theta^m_K(t),\theta^M_K(t)]\right)^N} d\theta_1\dots d \theta_N \prod_{i<j} \left|e^{i\theta_i}-e^{i\theta_j}\right|^2\eeq
The idea is then the following: we impose for each $1\leq i\leq r(t)$ to have $N\epsilon_i$ eigenvalues in the arc interval $[\theta^m_i(t),\theta^M_i(t)]$ (To be precise since we need the number of eigenvalues to be an integer we should take $\lfloor N\epsilon_i\rfloor$ eigenvalues in the interval $[\theta^m_i(t),\theta^M_i(t)]$ but to avoid unnecessary notational complication we will omit the $\lfloor\, \rfloor$ symbols) We denote by $M(\epsilon_1,\dots,\epsilon_r,N)$ the number of ways to select the eigenvalues according to this restriction. Hence we define:
\bea \label{yo}P_{N,\text{strong}}(t,\epsilon_1,\dots,\epsilon_r)&=&M(\epsilon_1,\dots,\epsilon_{r(t)},N) \td{P}_{N,\text{strong}}(t,\epsilon_1,\dots,\epsilon_{r(t)})\cr
M(\epsilon_1,\dots,\epsilon_{r(t)},N)&=&\frac{N!}{(N\epsilon_1)!(N\epsilon_2)!\dots (N\epsilon_{r(t)})!}\eea
Here $\td{P}_{N,\text{strong}}(t,\epsilon_1,\dots,\epsilon_{r(t)})$ is computed with the first $\theta_1,\dots, \theta_{N\epsilon_1}$ eigenvalues in $[\theta^m_1(t),\theta^M_1(t)]$, the next $\theta_{N\epsilon_1+1},\dots \theta_{N(\epsilon_2-\epsilon_1)}$ eigenvalues in $[\theta^m_2(t),\theta^M_2(t)]$ and so on. Now computing exactly $\td{P}_{N,\text{strong}}(t,\epsilon_1,\dots,\epsilon_{r(t)})$ is still as hard as the initial problem therefore we need to find a good approximation to compute it in practice. We denote by $\theta_{i,K}$ the $i^{\text{th}}$ eigenvalue in the $K^{\text{th}}$ interval. The idea is to decompose the interaction in the following form: first we have interactions between eigenvalues belonging to the same interval $\underset{i<j}{\prod} \left|e^{i\theta_{i,K}}-e^{i\theta_{j,K}}\right|^2$. Then we also have interactions between eigenvalues belonging to different intervals: $\underset{k<k'}{\prod}\underset{i=1}{\overset{N\epsilon_k}{\prod}}\underset{j=1}{\overset{N\epsilon_{k'}}{\prod}}\left|e^{i\theta_{i,k}}-e^{i\theta_{j,k'}}\right|^2$. Since the intervals are small compared to the distance between them (they are of length $\frac{2\pi \epsilon}{t}$ and distant from $\frac{2\pi}{t}$), we can approximate in the last term every eigenvalue by a constant which we take as the center of its corresponding interval:

\begin{definition} We call the Average Block Interaction Approximation (ABIA) the following approximation:
\beq \prod_{k<k'}\prod_{i=1}^{N\epsilon_k}\prod_{j=1}^{N\epsilon_{k'}}\left|e^{i\theta_{i,k}}-e^{i\theta_{j,k'}}\right|^2 \approx \prod_{k<k'}|c_k(t)-c_{k'}(t)|^{2\epsilon_k\epsilon_{k'}N^2} \eeq
where $c_k(t)=e^{i\frac{\theta^m_k(t)+\theta^M_k(t)}{2}}$ is the center of the arc interval $[e^{i\theta^m_k}(t),e^{i\theta^M_k(t)}]$.
\end{definition} 
\textbf{This assumption is equivalent to average the interactions between the eigenvalues of different intervals} (hence the name average block interaction approximation). The initial problem is now significantly simplified. Indeed, we have $N\epsilon_j$ eigenvalues in each interval submitted to the usual interaction between themselves (but not the other ones) and an additional potential term coming from the averaged interaction with the other intervals. More precisely the probability is given by:
\bea P_{N,\text{strong}}(t,\epsilon_1,\dots,\epsilon_{r(t)})&\overset{\text{ABIA}}{\approx}&\frac{M(\epsilon_1,\dots,\epsilon_{r(t)},N)}{Z_N}\left(\prod_{k<k'=1}^{r(t)}|c_k(t)-c_{k'}(t)|^{2\epsilon_k\epsilon_{k'}N^2}\right)\cr
&&\left(\prod_{k=1}^{r(t)}\int_{[\theta^m_k(t),\theta^M_k(t)]^{N\epsilon_k}}d\theta_1\dots d\theta_{N\epsilon_k} \prod_{1\leq i<j\leq N\epsilon_k}\left|e^{i\theta_i}-e^{i\theta_j}\right|^2 \right)\cr
&=&\frac{M(\epsilon_1,\dots,\epsilon_{r(t)},N)}{Z_N}\left(\prod_{k<k'=1}^r|c_k(t)-c_{k'}(t)|^{2\epsilon_k\epsilon_{k'}N^2}\right)\cr
&&\left(\prod_{k=1}^{r(t)}\int_{[a_k(t),b_k(t)]^{N\epsilon_k}}du_1\dots du_{N\epsilon_k} \Delta(u_1,\dots,u_{N\epsilon_k})^2 e^{-N\epsilon_k\underset{i=1}{\overset{N\epsilon_k}{\sum}} \ln u_i}\right)\cr
\eea
where we have defined $a_(t)=e^{i\theta^m_k(t)}\in \mathcal{C}$ and $b_k(t)=e^{i\theta^M_k(t)} \in \mathcal{C}$. The last quantity is a one interval (genus $0$ curve) computation with only $N\epsilon_k$ eigenvalues and is computed in appendix \ref{AppendixD}. At leading order in $N\to \infty$ the quantities $Z_N$ and $M(\epsilon_1,\dots,\epsilon_{r(t)},N)$ do not contribute. Indeed a Stirling expansion in \eqref{yo} with the fact that $Z_N=(2\pi)^NN!$ easily shows that the contributions are subleading compared to a $e^{-N^2 F^{[-2]}}$ term. Hence we are left at leading order with:
\beq \label{Approxx}\frac{1}{N^2}\ln P_{N,\text{strong}}(t,\epsilon_1,\dots,\epsilon_r)\approx 2\epsilon_k\epsilon_{k'} \sum_{k<k'=1}^{r(t)} \ln |c_k(t)-c_{k'}(t)|-\sum_{k=1}^rF^{[-2]}(a_k(t),b_k(t),\epsilon_k,N) +O\left(\frac{1}{N}\right)\eeq
We remind the reader here that to obtain $P_{N,\text{strong}}(t)$ one needs to minimize the previous result relatively to the filling fractions. As we will see below, \eqref{Approxx} is a quadratic form in the filling fractions $(\epsilon_1,\dots,\epsilon_{r(t)})$ and hence can easily be minimized by standard techniques. 

\subsection{Case when $t\in[2k+\epsilon,2(k+1)-\epsilon]$}

In this case, we have a union of arc segments given by $I(t)=\underset{j=-k}{\overset{k}{\bigcup}} [e^{\frac{i(2\pi j-\pi \epsilon)}{t}},e^{\frac{i(2\pi j+\pi \epsilon)}{t}} ]$. There are $2k+1$ intervals: $a_j=e^{\frac{i(2\pi j-\pi \epsilon)}{t}}$ and $b_j=e^{\frac{i(2\pi j+\pi \epsilon)}{t}}$ with $j\in [-k,k]$. In particular we observe here that for every interval we have $\frac{\theta_1-\theta_0}{2}=\frac{\pi \epsilon}{t}$ which does not depend on the interval. Therefore from appendix \ref{AppendixD} we get that at leading order each interval contributes with a factor: 
\beq F^{[-2]}_j=\epsilon_j^2\ln \left(\sin \frac{\pi \epsilon}{2t}\right) +O\left(\frac{1}{N}\right) \eeq
Then we take $c_j=e^{\frac{2i\pi j}{t}}$ as the central point of each interval. Therefore we have a contribution of the form:
\bea \left(\prod_{j<j'=-k}^k\left|e^{\frac{2i\pi j}{t}}-e^{\frac{2i\pi j'}{t}}\right|^{2\epsilon_j\epsilon_{j'}N^2}\right)&=&2^{N^2\underset{j<j'=-k}{\overset{k}{\sum}} \epsilon_j\epsilon_{j'}}\prod_{j<j'=-k}^k\left(1-\cos \frac{2\pi(j-j')}{t}\right)^{N^2\epsilon_j\epsilon_{j'}}\cr
&=&4^{N^2\underset{j<j'=-k}{\overset{k}{\sum}} \epsilon_j\epsilon_{j'}}\prod_{j<j'=-k}^k\left(\sin^2 \frac{\pi(j-j')}{t}\right)^{N^2\epsilon_j\epsilon_{j'}}
\eea
In the end we find (we denote $\vec{\epsilon}=(\epsilon_{-k},\dots,\epsilon_k)^t$):
\bea \frac{1}{N^2} \ln P_{N}(t,\vec{\epsilon})&\approx&\ln \left(\sin \frac{\pi \epsilon}{2t}\right) \sum_{j=-k}^k \epsilon_j^2+ 2\ln 2\underset{j<j'=-k}{\overset{k}{\sum}} \epsilon_j\epsilon_{j'} \cr
&&+ \underset{j<j'=-k}{\overset{k}{\sum}}\epsilon_j\epsilon_{j'}  \ln \left(\sin^2 \frac{\pi(j-j')}{t}\right) +O\left(\frac{1}{N}\right)
\eea
We stress here that the filling fractions vector $\vec{\epsilon}=(\epsilon_{-k},\dots,\epsilon_k)$ should not be confused with the initial scalar parameter $\epsilon$ corresponding to the size of the window allowed around $\theta=0$ to declare a strong return time. As explained earlier we now need to optimize the previous function relatively to the filling fractions $\vec{\epsilon}$. Moreover from symmetry consideration (the problem is invariant by complex conjugation) we have $\epsilon_{-k}=\epsilon_k$ so the dimension of the problem can be lowered from $2k+1$ down to $k+1$. Let us introduce:
\beq g(\vec{\epsilon})=\ln \left(\sin \frac{\pi \epsilon}{2t}\right) \sum_{j=-k}^k \epsilon_j^2+ \underset{j < j'=-k}{\overset{k}{\sum}} \epsilon_j\epsilon_{j'}\left(2\ln 2+\ln \sin^2 \frac{\pi(j-j')}{t}\right)\eeq
It is clearly a \textbf{quadratic form} in the filling fractions. However by definition of the filling fractions we have an affine constraint $\underset{j=-k}{\overset{k}{\sum}}\epsilon_j=1$ that from symmetry can be recast as $1-\epsilon_0-2\underset{j=1}{\overset{k}{\sum}} \epsilon_j=0$. Therefore we introduce a symmetric matrix $A$ of size $(k+2)\times(k+2)$:
\bea \label{A}
A_{1,1}&=&2\ln\left(\sin \frac{\pi \epsilon}{2t}\right)\cr
A_{1,k+2}&=&A_{k+2,1}=-1\cr
A_{j,k+2}&=&A_{k+2,j}=-2 \,\, ,\,\, \forall\, 2\leq j\leq k+1 \cr
A_{1,j}&=&A_{j,1}=4\ln 2+2\ln\left(\sin^2 \frac{\pi (j-1)}{t}\right) \,\, ,\,\, \forall\, 2\leq j\leq k+1 \cr
A_{j,j}&=& 4\ln\left(\sin \frac{\pi \epsilon}{2t} \right)+4 \ln 2+2\ln\left(\sin^2 \frac{2\pi (j-1)}{t}\right)\,\,,\,\, \forall\, 2\leq j\leq k+1\cr
A_{k+2,k+2}&=&0\cr
A_{i,j}&=&8\ln 2+2\ln\left(\sin^2 \frac{\pi(j-i)}{t}\right)+2\ln\left(\sin^2 \frac{\pi(j+i-2)}{t}\right)\,\,,\,\, \forall\, 2\leq i\neq j\leq k+1\cr
\eea
We also introduce the vector $\vec{b}$ of size $k+2$ given by:
\beq \label{b} \vec{b}=(0,\dots,0,-1)^t\eeq
and finally we introduce the vector of filling fractions and Lagrange multiplier $\vec{x}$ of size $k+2$:
\beq \vec{x}=\left(\epsilon_0,\dots,\epsilon_k,\lambda\right)^t \eeq
So that we get:
\beq  h(\vec{x})=g(\vec{\epsilon})+\lambda\left(1-\sum_{j=-k}^k \epsilon_j\right)=\frac{1}{2}\vec{x}^{\,\,t}A\vec{x} -\vec{x}^{\,\,t}\vec{b}\eeq
Our problem is now equivalent to find the extreme values of $h(\vec{x})$. The general theory of quadratic forms gives:
\beq \vec{x}_{\text{extr}}=A^{-1} \vec{b}\,\,\text{ and }\,\,h(\vec{x}_{\text{extr}})=-\frac{1}{2}\vec{b}^{\,\,t}A^{-1}\vec{b}=-\frac{1}{2}\left(A^{-1}\right)_{k+2,k+2}\eeq
Note that we can also compute $\vec{x}_{\text{extr}}$ by $\left(\vec{x}_{\text{extr}}\right)_i=(-1)^{i+k+1}\frac{\det\left(\td{A}_{i,k+2}\right)}{2\det A}$ where $\td{A}_{i,k+2}$ is the submatrix of $A$ with the $i^\text{th}$ line and $(k+2)^{\text{th}}$ column removed. (The last expression is interesting because determinants are more stable and faster to compute than the inverse of a matrix) In the end we find:
\bea \label{Approxtnormal} \forall t\in[2k+\epsilon,2(k+1)-\epsilon]\,\,: \frac{1}{N^2}\ln P_{N,\text{strong}}(t) &\overset{\text{ABIA}}{\approx}& -\frac{1}{2}\vec{b}^{\,\,t}A^{-1}\vec{b}=-\frac{1}{2}\left(A^{-1}\right)_{k+2,k+2}\cr
&=&-\frac{1}{2}\frac{\det \td{A}_{k+2,k+2}}{\det A} \eea
with $A$ and $\vec{b}$ defined in \eqref{A} and \eqref{b} and $\td{A}_{k+2,k+2}$ being the submatrix of $A$ where we have excluded the last line and column.

\medskip

We would like now to see if the former result is compatible with exact results obtained at integer times in theorem \ref{OODD}. At $t=2k+1\in \mathbb{N}$, symmetry considerations imply that all filling fractions are identical. In that specific case, our ABIA approximation gives:
\bea \frac{1}{N^2} \ln P_{N,\text{Strong}}(2k+1) &\overset{\text{ABIA}}{\approx}& \frac{1}{2k+1}\ln \left(\sin \frac{\pi \epsilon}{2(2k+1)}\right)+\frac{2k}{2k+1}\ln 2\cr
&&+\frac{1}{(2k+1)^2}\sum_{j=1}^{2k}j\ln\left(\sin^2\frac{\pi j}{2k+1}\right)\eea
From formula \eqref{TrigoSum} we can compute the last sum and we obtain:
\beq \label{tequal2k1}\frac{1}{N^2} \ln P_{N,\text{Strong}}(2k+1) \overset{\text{ABIA}}{\approx} \frac{1}{2k+1}\ln \left((2k+1)\sin \frac{\pi \epsilon}{2(2k+1)}\right) \eeq
We note here that the ABIA fails to reproduce the exact formula of theorem \ref{OODD} but is very close to it. Indeed, as soon as the approximation $\sin \frac{\pi \epsilon}{2(2k+1)}\sim \frac{\pi \epsilon}{2(2k+1)}$ can be made we recover theorem \ref{OODD}. In particular this is the case for $\epsilon\to 0$ or $k\to \infty$.  
\medskip
\medskip
\begin{remark} A more drastic approximation would correspond to remove all interactions between different intervals instead of averaging them. In this case we would get:
\beq \label{VeryStrongApprox}\forall \, j\in[-k,k] \,:\,\, \epsilon_j=\frac{1}{2k+1}  \text{ and } \frac{1}{N^2}\ln P_{N,\text{strong}}(t)\approx \frac{1}{2k+1}\ln \left(\sin \frac{\pi \epsilon}{2t}\right) \eeq
However this strong approximation provides a completely wrong answer since at large $k$ (and hence large $t$) its behavior is very different from theorem \ref{OODD}. Consequently removing interactions between different intervals fails to reproduce the refined structure of the situation whereas averaging them is a much better approximation.
\end{remark}

\subsection{Case when $t\in[2k-\epsilon,2k+\epsilon]$}

The case when $t\in[2k-\epsilon,2k+\epsilon]$ is trickier than the previous one since we have to deal with an incomplete interval around $e^{i\pi}=-1$ which gives a contribution $\ln\left(\cos\frac{2\pi k-\pi\epsilon}{2t}\right)$. Indeed, we remind the reader that we must choose a suitable determination of the angles that is compatible with the cut in order to obtain $F^{[-2]}$ and thus it would be wrong to take $\theta_0=-\frac{2\pi k-\pi\epsilon}{2t}$ and $\theta_1=\frac{2\pi k-\pi\epsilon}{2t}$. In the present situation we should take $\theta_0=\frac{2\pi k-\pi\epsilon}{2t}$ and $\theta_1=2\pi -\frac{2\pi k-\pi\epsilon}{2t}$ for which we find $\sin \left(\frac{\pi}{2}-\frac{2\pi k-\pi\epsilon}{2t}\right)=\cos \frac{2\pi k-\pi\epsilon}{2t}$.

Then, the approximation of the interaction between intervals is given by: 
\bea \frac{1}{N^2}\ln \prod_{j<j'=-(k-1)}^k |c_j-c_{j'}|^2&=&2\ln 2\sum_{j<j'=-(k-1)}^{k-1} \epsilon_j \epsilon_{j'} +\sum_{j<j'=-(k-1)}^{k-1} \epsilon_j \epsilon_{j'}\ln \left(\sin^2 \frac{\pi(j-j')}{t}\right)\cr
&&+2\ln 2\sum_{j=-(k-1)}^{k-1} \epsilon_j \epsilon_k +\sum_{j=-(k-1)}^{k-1}\epsilon_j \epsilon_k \ln \left(\cos^2\frac{\pi j}{t}\right)\cr
\eea
In the end we find the following approximation:
\bea \label{BetterApprox}\frac{1}{N^2} \ln P_{N}(t,\vec{\epsilon})&\overset{\text{ABIA}}{\approx}&\ln \left(\sin \frac{\pi \epsilon}{2t}\right) \sum_{j=-(k-1)}^{k-1} \epsilon_j^2+ \epsilon_k^2\ln \left(\cos \left( \frac{\pi k}{t}-\frac{\pi \epsilon}{2t}\right)\right) \cr
&&+ \underset{j<j'=-(k-1)}{\overset{k-1}{\sum}}\epsilon_j\epsilon_{j'} \left(2\ln 2+ \ln \left(\sin^2 \frac{\pi(j-j')}{t}\right) \right)\cr
&&+ \sum_{j=-(k-1)}^{k-1}\epsilon_j \epsilon_k \left(2\ln2+ \ln \left(\cos^2\frac{\pi j}{t}\right) \right)+O\left(\frac{1}{N}\right)\cr
\eea
This is again a quadratic form in $\vec{\epsilon}$. Moreover, since the problem is invariant under complex conjugation, the filling fractions must satisfy $\epsilon_{-j}=\epsilon_j$ for $j\in[1,k-1]$. This observation only leaves $k+1$ unknown filling fractions. Moreover, the Lagrange multiplier looks like $\lambda\left(1-\epsilon_0-\epsilon_k-2\underset{j=1}{\overset{k-1}{\sum}} \epsilon_j\right)$. Therefore we introduce the following $(k+2)\times (k+2)$ symmetric matrix $A$:
\bea \label{Abis}
A_{k+2,k+2}&=&0\cr
A_{1,k+2}&=&A_{k+2,1}=-1\cr
A_{k+1,k+2}&=&A_{k+2,k+1}=-1\cr
A_{i,k+2}&=&A_{k+2,i}=-2 \,\, ,\,\, \forall\, 2\leq i\leq k\cr 
A_{1,1}&=&2\ln\left(\sin \frac{\pi \epsilon}{2t}\right)\cr
A_{1,j}&=&A_{j,1}=4\ln 2+2\ln\left(\sin^2 \frac{\pi (j-1)}{t}\right) \,\, ,\,\, \forall\, 2\leq j\leq k \cr
A_{1,k+1}&=&A_{k+1,1}=2\ln 2\cr
A_{j,j}&=& 4\ln\left(\sin \frac{\pi \epsilon}{2t} \right)+4 \ln 2+2\ln\left(\sin^2 \frac{2\pi (j-1)}{t}\right)\,\,,\,\, \forall\, 2\leq j\leq k\cr
A_{k+1,k+1}&=&2\ln \left(\cos \left(\frac{\pi k}{t}-\frac{\pi \epsilon}{2t}\right)\right)\cr
A_{i,j}&=&8\ln 2+2\ln\left(\sin^2 \frac{\pi(j-i)}{t}\right)+2\ln\left(\sin^2 \frac{\pi(j+i-2)}{t}\right)\,\,,\,\, \forall\, 2\leq i\neq j\leq k\cr
A_{j,k+1}&=&4\ln 2+2\ln \left(\cos^2\frac{\pi (j-1)}{t}\right) \,\,,\,\, \forall\, 2\leq j\leq k
\eea
The vector $\vec{b}$ of size $k+2$ is now given by:
\beq \label{bbis} \vec{b}=\left(0,\dots,0,-1\right) \eeq
and finally we introduce the reduced (by symmetry) vector of filling fractions and Lagrange multiplier $\vec{x}$ of size $k+2$:
\beq \vec{x}=\left( \epsilon_0,\dots,\epsilon_k,\lambda\right)^t \eeq
so that we get again the following function to extremize:
\beq  h(\vec{x})=\frac{1}{2}\vec{x}^{\,\,t}A\vec{x} -\vec{x}^{\,\,t}\vec{b}\eeq
We find:
\beq \label{SpecialCaseApprox}\forall t\in[2k-\epsilon,2k+\epsilon]\,\,: \frac{1}{N^2}\ln P_{N,\text{strong}}(t) \overset{\text{ABIA}}{\approx} -\frac{1}{2}\left(A^{-1}\right)_{k+2,k+2}=-\frac{1}{2}\frac{\det \td{A}_{k+2,k+2}}{\det A} \eeq
where $A$ and $\vec{b}$ are defined in \eqref{Abis} and \eqref{bbis} and $\td{A}_{k+2,k+2}$ is the submatrix of $A$ where we have excluded the last line and column. At time $t=2k$, it is easy to observe by symmetry that all filling fractions are again identical (and take the value $\frac{1}{2k}$). Therefore we can compute explicitly an approximation of the probability for these times. We get from \eqref{BetterApprox}:
\bea \frac{1}{N^2}\ln P_{N,\text{strong}}(t=2k)&\overset{\text{ABIA}}{\approx}& \frac{1}{2k}\ln \left(\sin \frac{\pi \epsilon}{4k}\right) \cr
&&+ \frac{2(2k-1)(k-1)}{(2k)^2}\ln 2+ \frac{1}{(2k)^2}\underset{j<j'=-(k-1)}{\overset{k-1}{\sum}}\ln \left(\sin^2 \frac{\pi(j-j')}{2k}\right) \cr
&&+ \frac{2(2k-1)}{(2k)^2}\ln 2+\frac{1}{(2k)^2}\left(\ln k -(2k-2)\ln 2\right)\cr
&=& \frac{1}{2k}\ln \left(\sin \frac{\pi \epsilon}{4k}\right) +\frac{2(2k-1)k-(2k-2)}{(2k)^2}\ln 2+ \frac{1}{(2k)^2}\ln k\cr
&&+ \frac{1}{(2k)^2}\underset{j<j'=-(k-1)}{\overset{k-1}{\sum}}\ln \left(\sin^2 \frac{\pi(j-j')}{2k}\right) \cr
&=& \frac{1}{2k}\ln \left(\sin \frac{\pi \epsilon}{4k}\right) +\frac{2(2k-1)k-(2k-2)}{(2k)^2}\ln 2+ \frac{1}{(2k)^2}\ln k\cr
&&+ \frac{1}{(2k)^2}\sum_{j=1}^{2k-1}(j-1)\ln\left(\sin^2\frac{\pi j}{2k}\right)
\eea
Eventually using \eqref{TrigoSum} we find:
\beq \label{tequal2k} \frac{1}{N^2}\ln P_{N,\text{strong}}(2k)\overset{\text{ABIA}}{\approx}\frac{1}{2k}\ln \left(2k\sin\frac{\pi \epsilon}{4k}\right) \eeq
As in the case of odd integer times, we do not recover completely the exact value of theorem \ref{Theo3} but in the limit where $\sin\frac{\pi \epsilon}{4k}\sim \frac{\pi \epsilon}{4k}$ (i.e. $\epsilon\to 0$ or $k\to \infty$) we recover it properly.


\subsection{General remarks about our approximation}

Our average block interaction approximation satisfies various interesting aspects:
\begin{itemize}
\item The filling fractions and the probability computed with our approximation are both continuous functions of $t$ including at $t=2k-\epsilon$ where the dimension of the matrix increases. This is coherent with the fact that we do not expect any specific singular behavior at $t=2k-\epsilon$.
\item At $t\in \mathbb{N}$ we recover identically distributed filling fractions and our approximation almost gives the right formula. Indeed, our approximation gives $\frac{1}{t}\ln \left(t\sin \frac{\pi \epsilon}{2t}\right)$ while the exact computation gives $\frac{1}{t}\ln \left(\sin \frac{\pi\epsilon}{2}\right)$. As one can see, the difference is small and vanishes at first order in $\epsilon$ or $t\to \infty$.
\item At $t=2-\epsilon$ the probability is continuous and we know that the left limit is exact because it is a genus $0$ computation. Therefore it means that our approximation is consistent at this point.
\item Numerical simulations (Cf. \ref{Numerics}) show very good agreement between the approximate curve and the exact values of the probability computed at finite large $N$ by Toeplitz determinants.
\item When $t=N$ we recover that the leading order of the approximation is given by \eqref{exactintegertimes}
\end{itemize}

We also mention that our approximation could be improved in the following way: in order to carry out the approximation we had to take $\lfloor N\epsilon_i \rfloor$ eigenvalues in each interval. However in the rest of the analysis, we considered $\epsilon_i$ to be a real parameter in $[0,1]$ whereas it can of the form $\frac{p}{N}\in \mathbb{Q}$. This aspect is not problematic in the matrix model perspective developed in appendix \ref{AppendixD} since it is valid for any $\epsilon_i\in \mathbb{R}$ but it may be important in the optimization process. However optimizing a quadratic form on a discrete space $\{\frac{p}{N}, p\in\left[0,N\right]\}^N$ seems much more complicated than on a continuous space. In particular this discrete/continuous issue will be important when $t$ (which basically gives the number of cuts) is of order $N$. In that case, we can only have a few eigenvalues (and possibly none) in each interval and the analysis developed should be adapted.

\section{\label{Numerics} Numerical simulations for strong return times}

We want to compare numerical evaluations of the probability for finite values of $N$ with the theoretical solutions or approximations developed in the previous sections. We take $\epsilon=\frac{1}{5}$ and we are able to compute exact values of the probability up to $N=9$ for $t<8$. We have several following possibilities:
\begin{itemize}
\item Exact $N$ computations given by the determinantal formulas \eqref{formula1bis} that allow numerical computations of exact finite $N$ values of the probability at any time $t$. Since the matrices involved are Toeplitz matrices, the determinant computation is much faster than usual. We computed the exact value up to $N=35$.
\item Specific exact cases for finite $N$ at integer times given by the reduction of the determinantal formulas at these points in \eqref{EvenInteger} and \eqref{OddInteger}.
\item The exact limiting curve for $t<2-\epsilon$ given by \eqref{F0t}
\item For $t>2-\epsilon$ we can use our average block interaction approximation. The results are given in \eqref{Approxtnormal} and \eqref{SpecialCaseApprox} and can be carried out numerically for small values of $N$.  
\end{itemize}

Numerical simulations are presented here and show that our \textbf{average block interaction approximation matches perfectly the large $N$ exact computations}. The simulations give the following results for $\epsilon=\frac{1}{5}$ (the choice of $\frac{1}{5}$ is purely conventional but the picture is the same with other values of $\epsilon$): 

\begin{center}
\includegraphics[width=15cm]{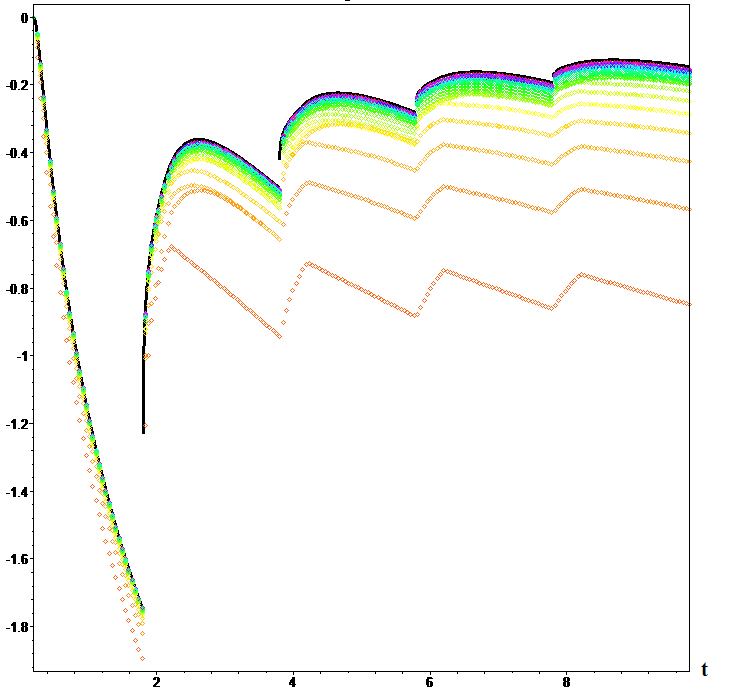}

\textit{Fig. $6$: Plot of $t \mapsto \frac{1}{N^2}\ln P_{N,\text{strong},\epsilon=\frac{1}{5}}(t)$ computed exactly with the formula \eqref{EqComp} for values of $N$ from $2$ to $35$ (colored points). The black curve is our average block interaction approximation given by \eqref{Approxtnormal} and \eqref{SpecialCaseApprox}.} 
\end{center}

For a given $N$ (i.e. one of the dot-curve presented above), we clearly observe that at time $t=N$ the behavior of the curve changes and is given by \eqref{Wild}. In particular it exhibits a quasi-periodic behavior of period $2$ and the probability at integer times remains constant as claimed in \eqref{exactintegertimes}.

\section{\label{Section7}First strong return time}

In the previous section we have computed the probability that the eigenvalues come back into the interval $\left[e^{-i\pi \epsilon}, e^{i\pi \epsilon}\right]$ at time $t$. A natural question is now to find information about the distribution of the first strong return time $\tau_{N,\text{strong}}$ when $N$ becomes large. When $N$ is large, having all the eigenvalues back into $\left[e^{-i\pi \epsilon}, e^{i\pi \epsilon}\right]$ is a very rare event since each eigenvalue only spends a proportion of time $\epsilon$ in the interval and that they are not synchronized a priori. Therefore, we expect the first-return time to be much larger than the size of the matrix $N$ (a typical time is expected to be of order $\epsilon^{-N}$). Moreover it is known from \ref{Indep} that for times $t>N$ the eigenvalues can be considered independent and uniformly distributed on the unit circle. Thus we expect the first-return time $\tau_{N,\text{strong}}$ to be equivalent to the first-return time for the following problem:

\begin{definition} Denote $\left\|a\right\|$ the distance to the nearest integer of a real number $a$. Take $X_1,\dots,X_N$ i.i.d. uniform variables on $[-\frac{1}{2},\frac{1}{2}]$ (it would correspond to take $X_i=\frac{\theta_i}{2\pi}$ in the unitary matrix interpretation). Define
\beq S_t=\mathbf{1}_{ \left\|t X_1\right\|\leq \frac{\epsilon}{2},\dots,\left\|t X_N\right\|\leq \frac{\epsilon}{2} } \eeq
 and the first return time $\td{\tau}_{N,\epsilon}$:
\beq \td{\tau}_{N,\epsilon}=\underset{t>0}{\text{Inf}}\,\{t \text{ such that } S_t=1 \text{ and } \exists\, s<t \text{ such that } S_s=0 \}\eeq
(It is the exact analogue of our first strong return time in this i.i.d. setting)
\end{definition}

Then we conjecture that:
\begin{conjecture} \label{Yeah} In the large $N$ limit we should have:
\beq \frac{N\tau_{N,\text{strong}}}{4\epsilon^{-(N-1)}}\underset{N\to \infty}{\overset{Law}{\to}} \mathcal{E}(1)\text{   and   } 
\frac{N\td{\tau}_{N,\epsilon}}{4\epsilon^{-(N-1)}}\underset{N\to \infty}{\overset{Law}{\to}} \mathcal{E}(1)\eeq
where $\mathcal{E}(1)$ stands for an exponential distribution with parameter $\lambda=1$. 
\end{conjecture}

The heuristic idea behind the previous proposition is the following: First observe that for a given sampling of the initial angles $\theta_i$ the only possible first return times are given by:
\beq t_{i,k}=\frac{2\pi k}{|\theta_i|}-\frac{\epsilon}{2} \text{Sign}(\theta_i) \,\,,\,\, \forall\, 1\leq i \leq N \text{ and } k\in \mathbb{N}^\ast\eeq
Indeed a simple continuity argument of the trajectories implies that one of the eigenvalues must be reentering the interval at $\tau_{N,\text{strong}}$ (otherwise there would be a immediate earlier return time). Then let us consider the variables:
\beq \label{yoyo} S_{i,k}=\mathbf{1}_{ \left\|t_{i,k} \frac{\theta_1}{2\pi}\right\|\leq \frac{\epsilon}{2},\dots,\left\|t_{i,k} \frac{\theta_N}{2\pi}\right\|\leq \frac{\epsilon}{2} } \eeq
Since the $\theta_i$'s are independent and uniform variables, it is well known that for a given $t>1$ the variables $\left\|t \frac{\theta_i}{2\pi}\right\|$ are independent and uniformly distributed on $\left[0,\frac{1}{2}\right]$. Hence the variables $S_{i,k}$ are Bernoulli variables with parameter $p_{N,\epsilon}=\epsilon^{N-1}$ (note that by construction of the times $t_{i,k}$ we know that the $i^{\text{th}}$ eigenvalue is in the success zone and thus only $N-1$ eigenvalues remain to be tested). Unfortunately, the variables $S_{i,k}$ are not independent because knowing the position of an eigenvalue at time $t$ provides information on its location at other times. However we note that $S_{i,k}=1$ is a very rare event when $N$ is large. Indeed, conjecture \ref{Yeah} states that the average strong return time is expected to be of order $\frac{4}{N\epsilon^{N-1}}$. Additionally, it is known that rotating around the circle is a mixing system (in the sense that a little change of the initial velocity makes the position very different after some time, thus making the position of a particle unpredictable after a sufficiently long time if the velocity is known approximately). Hence it seems reasonable that the dependence between the positions of an eigenvalue at time $t_0$ and at time $t_1$ should rapidly decrease when the times $t_0$ and $t_1$ become distant (typically $\Delta t=t_1-t_0$ should be greater than the so-called mixing time of the system). Consequently it seems reasonable that a memoryless exponential distribution should appear if the typical mixing time is lower than the typical strong-return time. Anyway, keeping the discussion rigorous, we have to compute the first success time for dependent Bernoulli variables corresponding to rare events. Recent results using Stein's method about the convergence of dependent Bernoulli variables for rare events towards Poisson variables (and exponential variables regarding the first success time) can be found in \cite{Arratia,Stein,RareEvent} (for example theorem $4.1$ of \cite{RareEvent}). In particular these theorems explain why is seems natural to recover an exponential distribution in the limit. The last difficulty is to get an estimate of the average first success which we claim to be $\frac{4}{N}\epsilon^{-(N-1)}$. Considering temporarily that the variables $S_{i,k}$ are independent trivially gives that the mean first success should be:
\beq \text{Mean First Success}=\frac{1}{\left(\Delta_{N,\epsilon} t\right) p_{N,\epsilon}}\eeq
where $\Delta_{N,\epsilon}\, t$ is the average time between two consecutive possible first return times $t_{i,k}$:
\beq \Delta_{N,\epsilon}\, t=\mathbb{E}\left(\underset{p\in \mathbb{N}^\ast}{\text{Min}} \left(s_{p+1}-s_p\right)\right)\eeq
where the time sequence $\left(s_p\right)_{p\in\mathbb{N}^\ast}$ is a reordering in increasing order of the possible return times $\left(t_{i,k}\right)_{i,k}$ (with $1\leq i\leq N$ and $k\in\mathbb{N}^\ast$). When $N$ increases the times $\left(s_p\right)_{n\in \mathbb{N}^\ast}$ become denser hence the typical difference between two of them decreases. A direct computation shows that we have $\Delta_{N,\epsilon}\, t=\frac{4}{N}$ hence giving conjecture \ref{Yeah}. The proposition is supported by the following numerical simulations:

\medskip

\begin{center}
\includegraphics[width=15cm]{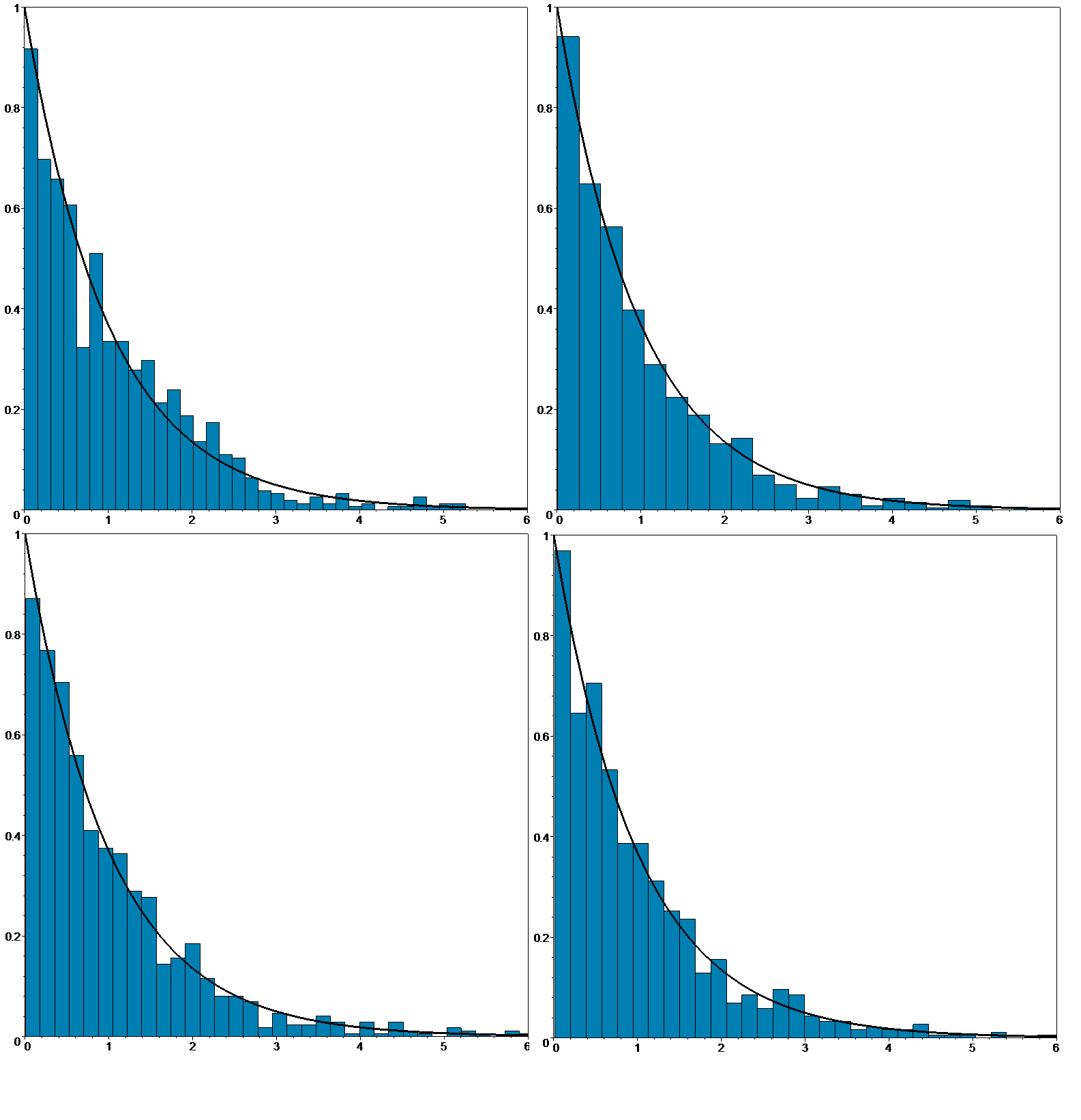}

\textit{Fig. $7$: Histograms of $\frac{N\tau_{N,\text{strong}}}{4\epsilon^{-(N-1)}}$ for $N=6$ and $\epsilon\in\{0.15,0.2,0.25,0.3\}$ from top-left to right-bottom obtained for $n=10^3$ independent samples. The black curve is the normalized exponential distribution $\mathcal{E}(1)$. Empirical estimation of parameter $\lambda$ ranges from $1.021$ ($\epsilon=0.3$) to $1.002$ ($\epsilon=0.20$).} 
\end{center}

\medskip

Similar histograms can be obtained for the continuous i.i.d. case confirming that the two problems are intimately related:

\medskip

\begin{center}
\includegraphics[width=15cm]{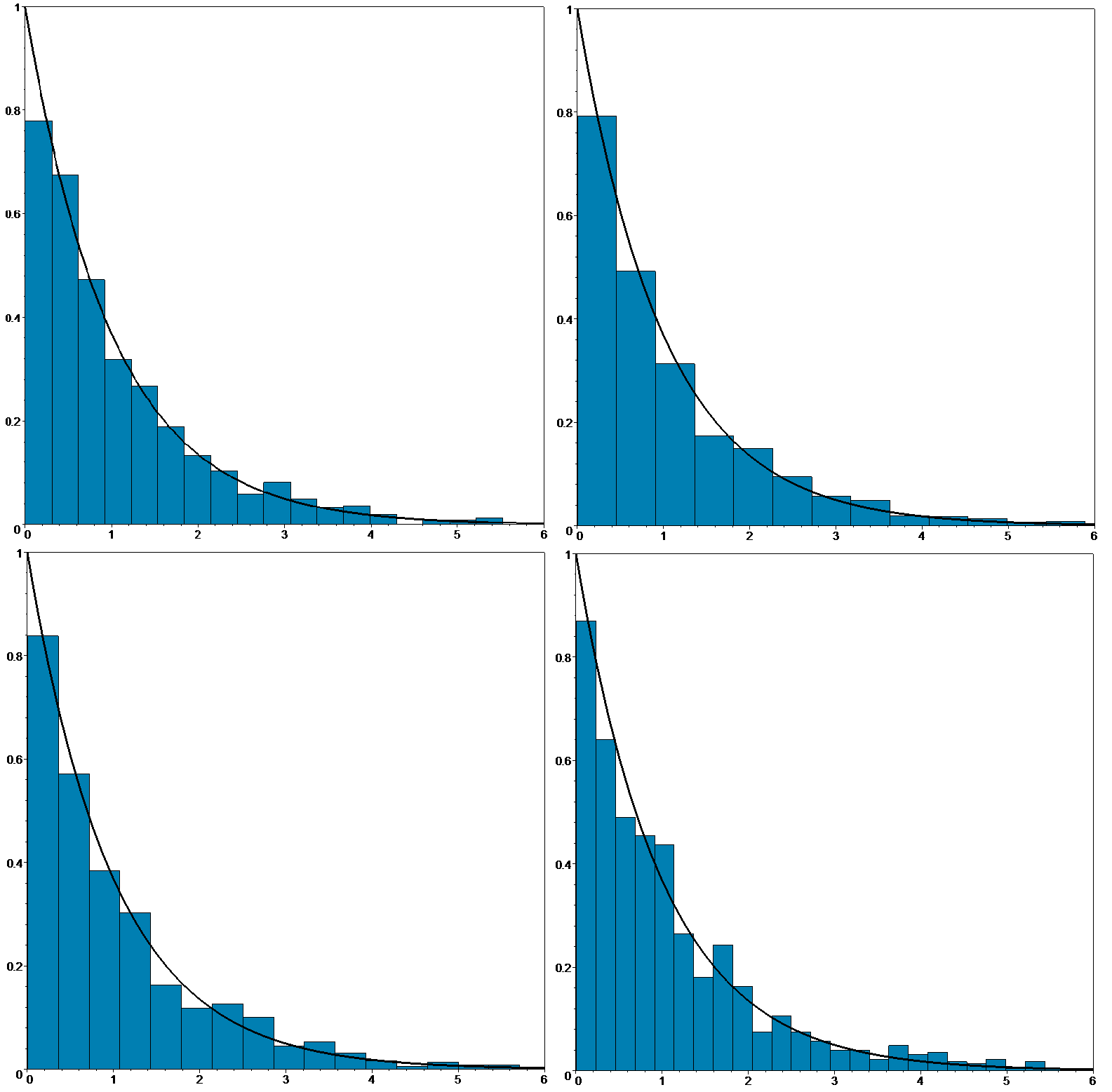}

\textit{Fig. $8$: Histograms of $\frac{N\td{\tau}_{N,\text{strong}}}{4\epsilon^{-(N-1)}}$ for $N=6$ and $\epsilon\in\{0.15,0.2,0.25,0.3\}$ from top-left to right-bottom obtained for $n=10^3$ independent samples. The black curve is the normalized exponential distribution $\mathcal{E}(1)$. Empirical estimation of parameter $\lambda$ ranges from $1.1$ ($\epsilon=0.3$) to $1.049$ ($\epsilon=0.15$).} 
\end{center}

\medskip
 
However we mention here that the results involving Stein's method found in \cite{Arratia,Stein,RareEvent} (and especially theorem $4.1$ of \cite{RareEvent}) cannot be directly used in our situation. Indeed, these results are very efficient when the range of dependence between the Bernoulli variables is bounded (or is rapidly decreasing with the distance between the variables) which is not the case here. They are also very efficient in the case where we have an explicit strong bound on the level of dependence on the Bernoulli variables which would require a precise analysis here.

\subsection{The discrete version of the first return time}

We mention here a discrete version of the first return time problem by only considering integer powers of the matrix $U_N$. Let us consider $n_0(N,\epsilon)$ the smaller integer $n\geq 1$ for which $U_N^n$ has all its eigenvalues in $\left[e^{-i\pi \epsilon}, e^{i\pi \epsilon}\right]$. Since we also expect $n_0(N,\epsilon)$ to be much larger than $N$, we can recast this problem into the following one: Take $X_1,\dots,X_N$ i.i.d. uniform variables on $\left[-\frac{1}{2},\frac{1}{2}\right]$ and define $\td{n}_0(N,\epsilon)$ the smallest integer $n \geq 1$ such that:
\beq \forall \,1\leq i\leq N \,:\, \left\|n X_i\right\|\leq \frac{\epsilon}{2}\eeq
Then similar arguments as the one developed in the continuous time setting give:

\begin{conjecture}\label{Yeahh} In the large $N$ limit we have:
\beq \frac{n_0(N,\epsilon)}{\epsilon^{-N}}\underset{N\to \infty}{\overset{Law}{\to}} \mathcal{E}(1)\text{   and   } \frac{\td{n}_0(N,\epsilon)}{\epsilon^{-N}}\underset{N\to \infty}{\overset{Law}{\to}} \mathcal{E}(1)\eeq
\end{conjecture} 

Note that the normalizing factor is now $\epsilon^{-N}$ and no longer $\frac{N}{4\epsilon^{-(N-1)}}$. This is because the difference between two consecutive testing times is now trivially $1$ (we test all integer times). Moreover for a given integer time $n$ we are no longer sure that one of the eigenvalues is located inside the success zone therefore giving $p_{N,\epsilon}=\epsilon^N$ as the success probability for variables $S_n$ defined similarly to \eqref{yoyo}. The discrete version of the first-return time is supported by the following simulations:

\medskip

\begin{center}
\includegraphics[width=15cm]{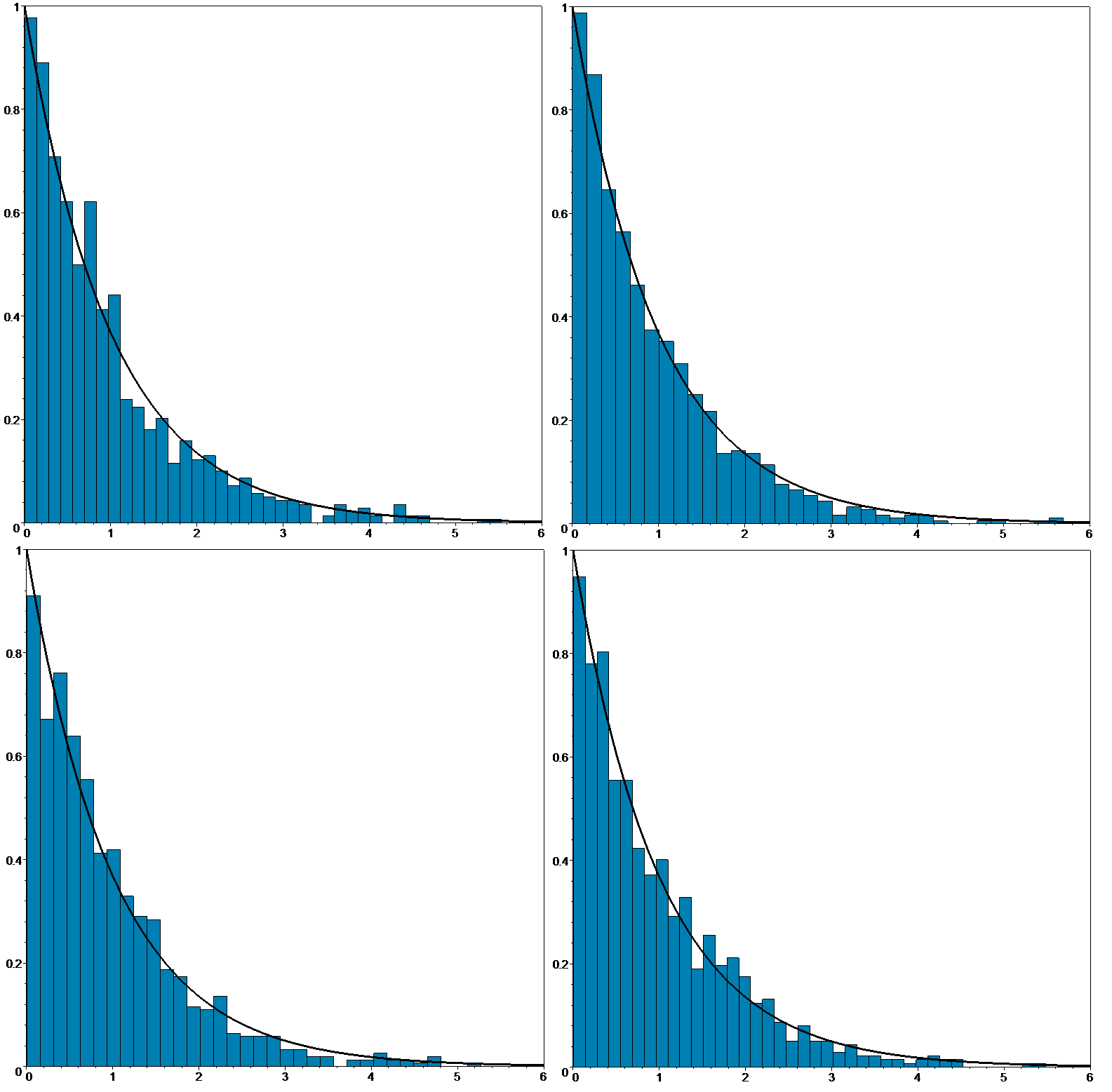}

\textit{Fig. $9$: Histograms of $\frac{n_0(N,\epsilon)}{\epsilon^{-N}}$ for $N=6$ and $\epsilon\in\{0.15,0.2,0.25,0.3\}$ from top-left to right-bottom obtained for $n=10^3$ independent samples. The black curve is the normalized exponential distribution $\mathcal{E}(1)$. Empirical estimation of parameter $\lambda$ ranges from $0.96$ ($\epsilon=0.3$) to $0.91$ ($\epsilon=0.20$).} 
\end{center}

\medskip

Similar histograms can be obtained for the discrete i.i.d. case:

\medskip

\begin{center}
\includegraphics[width=15cm]{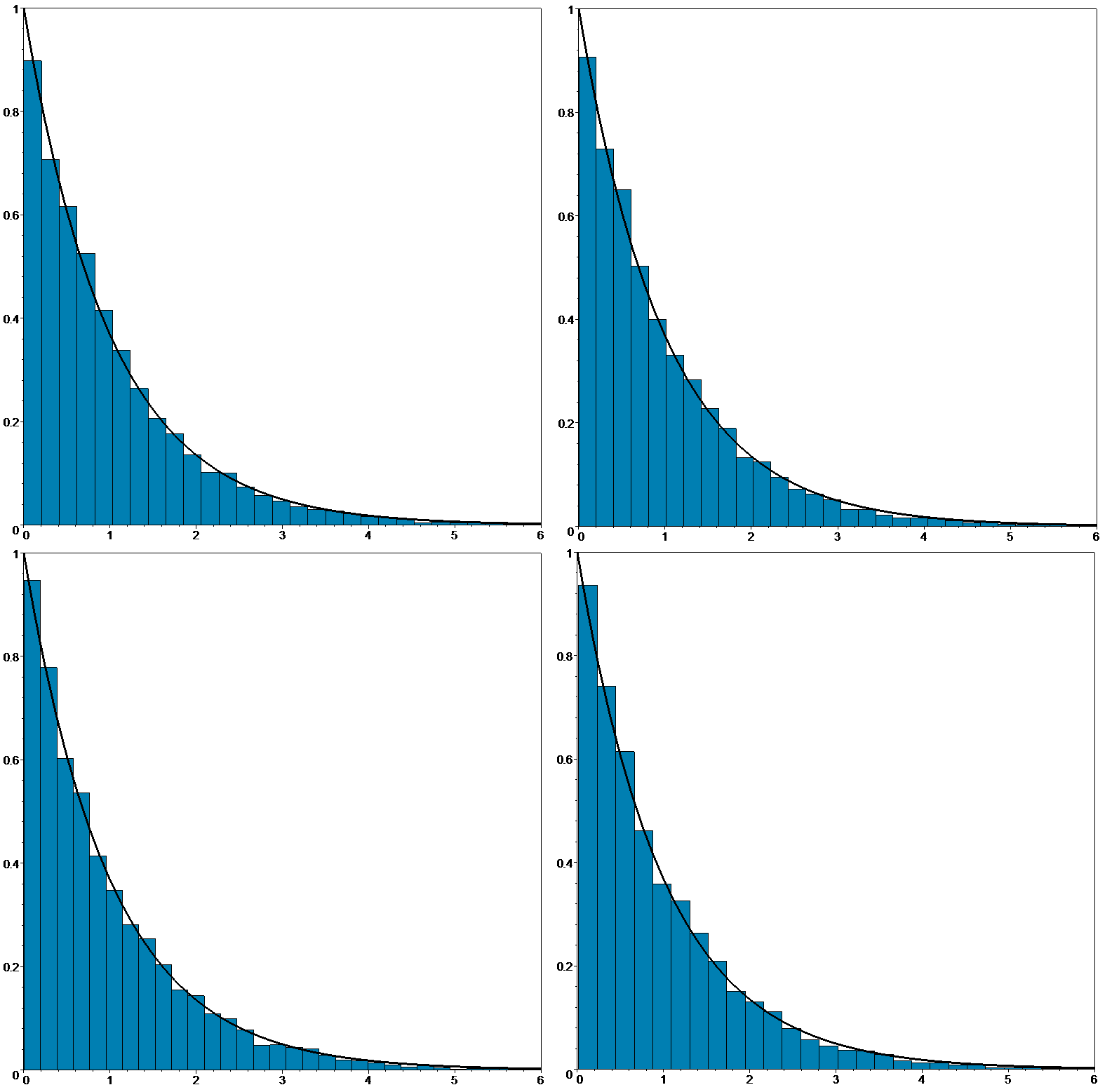}

\textit{Fig. $10$: Histograms of $\frac{\td{n}_0(N,\epsilon)}{\epsilon^{-N}}$ for $N=8$ and $\epsilon\in\{0.15,0.2,0.25,0.3\}$ from top-left to right-bottom obtained for $n=10^4$ independent samples. The black curve is the normalized exponential distribution $\mathcal{E}(1)$. Empirical estimation of parameter $\lambda$ ranges from $0.956$ ($\epsilon=0.3$) to $0.973$ ($\epsilon=0.15$).} 
\end{center}

\medskip

Note here that the simplified i.i.d. version of the discrete first-return time is a pure number theoretic problem. It can be recast into simultaneous Diophantine approximation of $N$ random independent real numbers and may have some interests for people in this field. In this spirit it seems possible to tackle similar number theory inspired problems (like the lonely runner conjecture, see \cite{Goddyn}) by extending them to unitary matrix models in first approximation (of course loosing the independence property since the eigenvalues are correlated) and apply known results there (like Toeplitz determinant or matrix models tools). Numerically we also observe that in the integer time setting always underestimate the value of $\lambda$ while the continuous time setting always overestimate the value of $\lambda$. This probably means that the subleading term of the large $N$ expansion is of different signs in the two settings.

\subsection{Connection with the recurrence time quantum measurement \label{Discuss}}

Let us try to interpret the results of the last section in relation with the recurrence time developed in \cite{Balian1}. Conjecture \ref{Yeah} suggests that the average first strong return time is of order $\tau_{\text{strong}}\approx \frac{4\epsilon^{-(N-1)}}{N}$ when $N$ is large. The first difficulty to compare this estimate with results from \cite{Balian1} is that $\tau_{\text{rec}}$ is related to the first real-part return time $\tau_{\text{real-part}}$ and not the first strong return time $\tau_{\text{strong}}$. In this paper we focused on the strong return time because it is easier to handle from the matrix models perspective as well as arguments presented in the previous section. However it is easy to see that strong return times will be less frequent than real-part return times whose definition is weaker. Hence $\tau_{\text{strong}}\approx\frac{4\epsilon^{-(N-1)}}{N}$ can be seen as an upper bound on the recurrence time. Let us now translate into our context the notation of \cite{Balian1}. In our case, we have $Q=N$ eigenvalues and the threshold required to have a strong (or real-part) return time is $f=1-\delta=\cos \pi \epsilon$. Hence results from from \cite{Balian1} states that we should have:
\beq \frac{\tau_{\text{rec}}}{\tau_{\text{trunc}}}=\pi \sqrt{2}e^{N\cos\pi \epsilon}\eeq
Our conjecture for $\tau_{\text{strong}}$ supports such result. Indeed even we did not compute the truncation time in this paper, authors in \cite{Balian1} claim that it should be of order $\tau_{\text{trunc}}=\frac{1}{\Delta \omega}$ where $\Delta \omega$ stands for the average difference between two consecutive eigenvalues around the initial time. A trivial lower bound in our situation is to take $\tau_{\text{trunc}}$ of order $1$, though as mentioned in \cite{Balian1} a more natural estimate is given by the uniform case: $\tau_{\text{trunc}}=\frac{1}{\Delta \omega}=\frac{N}{2\pi}$. In this last case we find that:
\beq \frac{\tau_{\text{rec}}}{\tau_{\text{trunc}}}=\frac{8\pi \epsilon^{-(N-1)}}{N^2}=\frac{8\pi}{N^2}e^{(N-1)|\ln \epsilon|}\eeq
Thus we recover the exponential dependence in $N$ of the ratio and only the dependence in the window parameter is modified from $\cos (\pi \epsilon)$ to $|\ln \epsilon|$ (both being decreasing functions of $\epsilon$ in $(0,1]$). In particular this makes the recurrence time rapidly inaccessible when $N$ increases. We believe that the typical time $\tau_{\text{real-part}}$ should not be very different from $\tau_{\text{strong}}$ because of the following argument: Let us consider a first strong return time $\tau_a$ in an arc-interval $\left[e^{-ia},e^{ia}\right]$. At $\tau_a$ one of the eigenvalues (say the $N^{\text{th}}$) is re-entering the arc-interval and therefore its real-part is $\cos a$. Since we expect $\tau_a>>N$, we know that the eigenvalues can be considered independent and uniformly distributed on the unit circle. Thus it seems reasonable at time $\tau_a$ to consider that the $N-1$ remaining eigenvalues $\theta_i^a$ are i.i.d. on the arc interval $\left[e^{-ia},e^{ia}\right]$. We now consider the random variable:
\beq S^a=\text{Re}\left(\frac{1}{N}\sum_{i=1}^N e^{i\theta_i}\right)=\frac{\cos a}{N}+ \frac{1}{N}\sum_{i=1}^{N-1} \cos \theta_i^a \eeq
where $\theta_i^a$ are i.i.d. uniform random variables on $\left[-a,a\right]$. In the large $N$ limit, we can approximate the last sum with a normal distribution using the central limit theorem. A simple computation shows that (we denote $\sin_c x=\frac{\sin x}{x}$):
\beq P(S>1-\delta)=1-F_0\left(\frac{1-\delta-\frac{\cos a}{N}-\frac{N-1}{N}\sin_c a}{\frac{1}{2}+\frac{1}{2}\sin_c(2\epsilon)-\sin_c^2\epsilon} \sqrt{N-1}\right)\eeq
where $F_0$ is the cumulative distribution function of the standard normal distribution. Hence when $N$ is large (we discard the $\frac{\cos a}{N}$ term and take $\frac{N-1}{N}\approx 1$) we observe the following threshold: when $1-\delta-\sin_c a<0$ we get that in the large $N$ limit the eigenvalues have also real-part returned in the zone $1-\delta$ at time $\tau_a$ while for $1-\delta-\sin_c a>0$ they have not real-part returned. Thus with this argument we can get an estimate of the first real-part return time with the knowledge of the first strong return time but evaluated with a suitable parameter $a$. We find that:

\medskip

When $N$ is large, a typical first real-part return time (corresponding to a recurrence time in \cite{Balian1}) should be of order $\frac{4a^{-(N-1)}}{N}$ where $a$ is the unique solution of $1-\delta-\sin_c a=0$ (i.e. for small values $a\sim\sqrt{6\delta}$) 
\medskip

In this approach we see that the inaccessibility claim in \cite{Balian1} is still correct since the last estimate is still exponentially large in $N$.

\section{Conclusion and outlooks}

In this article we presented general results regarding the eigenvalues of powers of large random unitary matrices using different classical tools like Toeplitz determinants and matrix models. In particular we focused our analysis on determining various return times into the arc-interval $\left[e^{-i\pi \epsilon}, e^{i\pi \epsilon}\right]$ for a fixed $\epsilon$. We supported the work with numerical simulations and recover some known results with different methods (like Widom's formula \eqref{F0t}). We also motivated our work with the recurrence time problem occurring in quantum measurement developed in \cite{Balian1}. An interesting extension of our work could be to vary the size of the return zone $\epsilon$ according to the power $n$ (or $t$ in the continuous setting) of the matrix or the size of the matrix $N$. We believe that the methods presented here could be used to deal with these problems. We would be very happy to work on these matters with anyone interested in these problems.

\section{Acknowledgements}
The author would like to thank Universit\'e Lyon $1$ and particularly Universit\'e Jean Monnet and Institut Camille Jordan for the opportunity to make this research possible. O. Marchal would also like to thank his family and friends for moral support during the preparation of this article as well as B. Eynard for fruitful discussions and R. Balian for references and explanations about the underlying quantum measurement physics problem. The author also thanks G. Borot for useful discussions about the existence of a unique minimizer and an unknown referee for suggesting the use of proposition \ref{PropBorot}.

\begin{appendices}

\section{Recovering the Toeplitz determinants with permutations}

In this section, we derive the standard results of Toeplitz integrals and determinants for our case. We use a permutation approach that is different from the standard Fourier approach known for the general case. 

\subsection{Computing the normalization factors}

We have the following: 
\bea \label{Normalization}\td{Z}_N&=&\int_{[-\pi,\pi]^N} \prod_{i<j} |e^{i\theta_i}-e^{i\theta_j}|^2 d\theta_1\dots d\theta_N\cr
&=&\int_{[-\pi,\pi]^N}d\theta_1\dots d\theta_N \det\left(\left(e^{i(j-1)\theta_{i}}\right)_{i,j=1..N}\right) \det\left(\left(e^{-i(j-1)\theta_{i}}\right)_{i,j=1..N}\right)\cr
&=&\int_{[-\pi,\pi]^N}d\theta_1\dots d\theta_N \sum_{\sigma,\tau \in S_N}(-1)^{|\sigma|+|\tau|} \prod_{k=1}^N e^{i(\sigma(k)-1)\theta_k}\prod_{r=1}^Ne^{-i(\tau(r)-1)\theta_r}\cr
&=&\sum_{\sigma,\tau \in S_N}(-1)^{|\sigma|+|\tau|}\int_{[-\pi,\pi]^N}d\theta_1\dots d\theta_N  \prod_{k=1}^N e^{i(\sigma(k)-\tau(k))\theta_k}\cr
&=&\sum_{\sigma,\tau \in S_N}(-1)^{|\sigma|+|\tau|}\prod_{k=1}^N\left(\int_{[-\pi,\pi]} d\theta_ke^{i(\sigma(k)-\tau(k))\theta_k}\right)\cr
&=&(2\pi)^NN!
\eea
Indeed, the integrals are only non zero when $\sigma=\tau$ for which the result is trivial. In particular we get by change of variables $u_i=e^{i\theta_i}$ that:
\beq Z_N=\int_{\mathcal{C}^N} du_1\dots d u_N \Delta(u_1,\dots,u_N)^2 e^{-N\underset{k=1}{\overset{N}{\sum}} \ln u_k}=(-1)^{\frac{N(N-1)}{2}} (2i\pi)^N N!\eeq

\subsection{Case for a single interval}

We can apply the same method for the following probability: $P_N(b)=\text{Prob}( \text{All eigenvalues are in the arc segment } [-b,b])$ where $b$ is a fixed real angle in $[0,\pi]$. We find:
\bea \label{EqComp} P_N(b)&=&\frac{1}{Z_N}\int_{[-b,b]^N} \prod_{i<j} |e^{i\theta_i}-e^{i\theta_j}|^2 d\theta_1\dots d\theta_N\cr
&=&\frac{1}{Z_N}\int_{[-b,b]^N}d\theta_1\dots d\theta_N \det\left(\left(e^{i(j-1)\theta_{i}}\right)_{i,j=1..N}\right) \det\left(\left(e^{-i(j-1)\theta_{i}}\right)_{i,j=1..N}\right)\cr
&=&\frac{1}{Z_N}\int_{[-b,b]^N}d\theta_1\dots d\theta_N \sum_{\sigma,\tau \in S_N}(-1)^{|\sigma|+|\tau|} \prod_{k=1}^N e^{i(\sigma(k)-1)\theta_k}\prod_{r=1}^Ne^{-i(\tau(r)-1)\theta_r}\cr
&=&\frac{1}{Z_N}\sum_{\sigma,\tau \in S_N}(-1)^{|\sigma|+|\tau|}\int_{[-b,b]^N}d\theta_1\dots d\theta_N  \prod_{k=1}^N e^{i(\sigma(k)-\tau(k))\theta_k}\cr
&=&\frac{1}{Z_N}\sum_{\sigma,\tau \in S_N}(-1)^{|\sigma|+|\tau|}\prod_{k=1}^N\left(\int_{[-b,b]} d\theta_ke^{i(\sigma(k)-\tau(k))\theta_k}\right)\cr
&=&\frac{1}{Z_N}\sum_{\sigma,\tau \in S_N}(-1)^{|\sigma|+|\tau|}\prod_{k=1}^N\left(\frac{1}{i(\sigma(k)-\tau(k))}\left[e^{i(\sigma(k)-\tau(k))b}-e^{-i(\sigma(k)-\tau(k))b}\right] \right)\cr
&=&\frac{N!}{Z_N}\sum_{\sigma \in S_N}(-1)^{|\sigma|}\prod_{k=1}^N\left(\frac{1}{i(\sigma(k)-k)}\left[e^{i(\sigma(k)-k)b}-e^{-i(\sigma(k)-k)b}\right] \right)\cr
&=&\left(\frac{b}{\pi}\right)^N\frac{1}{N!}\sum_{\sigma,\tau \in S_N}(-1)^{|\sigma|+|\tau|}\prod_{k=1}^N\sin_c\left( (\sigma(k)-\tau(k))b\right) \cr
&=&\left(\frac{b}{\pi}\right)^N\sum_{\sigma \in S_N}(-1)^{|\sigma|}\prod_{k=1}^N\sin_c\left( (\sigma(k)-k)b\right)\cr
&=&\det C[b,N]
\eea
where the previous expression of $Z_N$ was used. Moreover the $N\times N$ symmetric matrix $C[b,N]$ is given by:
\beq \label{MatrixDeter}\left(C[b,N]\right)_{i,j}=\frac{b}{\pi}\sin_c\left( (j-i)b\right) \,\,,\,\,\forall \,(i,j)\,\in [1,N]^2\eeq

In particular, the last formula is useful when dealing with a unique interval that is to say for $t\in[\epsilon,2-\epsilon]$. Indeed, in that case we only need to take $b=\frac{\pi\epsilon}{t}$ in order to recover $P_{N,\text{strong}}(t)$.

\subsection{Several intervals case and exact computation at time $t$}

When the domain of integration is on the unit circle of the form $\theta\in \bigcup_{r=-R}^R [a_r,b_r]$ with $a_{-r}=-b_r$, $b_{-r}=-a_r$ for $r>0$ and $a_0=-b_0$ we can apply a similar computation. We find:
\bea \label{EqComp2} P_N&=&\frac{1}{Z_N}\int_{I^N} \prod_{i<j} |e^{i\theta_i}-e^{i\theta_j}|^2 d\theta_1\dots d\theta_N\cr
&=&\frac{1}{(2\pi)^N}\sum_{\sigma \in S_N}(-1)^{|\sigma|}\prod_{k=1}^N\left(\int_{I} d\theta_k e^{i(\sigma(k)-k)\theta_k}\right)\cr
&=&\frac{1}{(2\pi)^N}\sum_{\sigma \in S_N}(-1)^{|\sigma|}\prod_{k=1}^N\Big(\frac{1}{i(\sigma(k)-k)}\left[e^{i(\sigma(k)-k)b_0}-e^{-i(\sigma(k)-k)b_0}\right]\cr
&&+\sum_{r=1}^R \frac{1}{i(\sigma(k)-k)}\left[e^{i(\sigma(k)-k)b_r}-e^{i(\sigma(k)-k)a_r}+e^{-i(\sigma(k)-k)a_r}-e^{-i(\sigma(k)-k)b_r}\right] \Big)\cr
&=&\frac{1}{\pi^N}\sum_{\sigma \in S_N}(-1)^{|\sigma|}\prod_{k=1}^N\Big(\frac{1}{\sigma(k)-k}\Big[\sin \left((\sigma(k)-k)b_0\right)\cr
&&+\sum_{r=1}^R \sin \left((\sigma(k)-k)b_r\right)-\sin \left((\sigma(k)-k)a_r\right) \Big]\Big)\cr
&=&\frac{1}{\pi^N}\sum_{\sigma \in S_N}(-1)^{|\sigma|}\prod_{k=1}^N\Big[b_0\sin_c \left((\sigma(k)-k)b_0\right)\cr
&&+\sum_{r=1}^R b_r\sin_c \left((\sigma(k)-k)b_r\right)-a_r\sin_c \left((\sigma(k)-k)a_r\right) \Big]\cr
&=&\det C[b_0,a_1,\dots,b_r,N]
\eea
where the matrix $C[b_0,a_1,\dots,b_r,N]$ is given by $\forall \,(i,j)\,\in [1,N]^2$:
\beq \label{MatrixDeterGeneral}\left(C[b_0,a_1,\dots,b_r,N]\right)_{i,j}=\frac{b_0}{\pi}\sin_c\left( (j-i)b_0\right)+\sum_{r=1}^R \frac{b_r}{\pi}\sin_c \left((j-i)b_r\right)-\frac{a_r}{\pi}\sin_c \left((j-i)a_r\right) \eeq

In particular the previous result can be applied to compute the strong returning time probability. There are two cases depending on whether $t\in [2k-\epsilon,2k+\epsilon]$ or not. Indeed, in this case, one of the interval is incomplete and one has to be careful with the determination of the angles used. We have:
\bea
\forall t\in[2R+\epsilon,2(R+1)-\epsilon]&:& I(t)=\underset{j=-R}{\overset{R}{\bigcup}} [e^{i\frac{2\pi j-\pi \epsilon}{t}},e^{i\frac{2\pi j+\pi \epsilon}{t}} ]\cr 
\forall t\in[2R-\epsilon,2R+\epsilon]&:& I(t)=\left(\underset{j=-R+1}{\overset{R-1}{\bigcup}} [e^{i\frac{2\pi j-\pi \epsilon}{t}},e^{i\frac{2\pi j+\pi \epsilon}{t}} ] \right)\cup [e^{i\frac{2\pi R-\pi \epsilon}{t}},e^{i\pi} ]\cr
&&\cup [e^{-i\pi},e^{-i\frac{2\pi R-\pi \epsilon}{t}} ]
\eea

\medskip

In the first case we find $a_r=\frac{2\pi r-\pi \epsilon}{t}$ and $b_r=\frac{2\pi r+\pi \epsilon}{t}$ so that:
\bea \sum_{r=-R}^R e^{i(\sigma(k)-k)b_r}-e^{i(\sigma(k)-k)a_r}&=&\frac{\cos \frac{(\sigma(k)-k)\pi(2R+1+\epsilon)}{t} -\cos \frac{(\sigma(k)-k)\pi(2R+1-\epsilon)}{t}}{i\sin \frac{(\sigma(k)-k)\pi}{t}}\cr
&=& \frac{2i \sin \frac{(\sigma(k)-k)\pi(2R+1)}{t} \sin \frac{(\sigma(k)-k)\pi\epsilon}{t}}{\sin \frac{(\sigma(k)-k)\pi}{t}}
\eea
so that we get:
\beq  P_N\left(t\in [2R+\epsilon,2(R+1)-\epsilon]\right)=\frac{1}{\pi^N}\sum_{\sigma \in S_N}(-1)^{|\sigma|}\prod_{k=1}^N\left( \frac{ \sin \frac{(\sigma(k)-k)\pi(2R+1)}{t} \sin \frac{(\sigma(k)-k)\pi\epsilon}{t}}{(\sigma(k)-k)\sin \frac{(\sigma(k)-k)\pi}{t}}\right)
\eeq
which can be rewritten with a determinant:
\beq \encadremath{ P_N\left(t\in [2R+\epsilon,2(R+1)-\epsilon]\right)=\det \left[ \frac{\sin\frac{(j-i)(2R+1)\pi}{t} \sin\frac{(j-i)\pi\epsilon}{t}}{\pi(j-i)\sin\frac{(j-i)\pi}{t}} \right]_{1\leq i,j\leq N} }
\eeq
Note here that in the last formula the case when $j=i$ is special since the sine product have to be replaced $\frac{\pi \epsilon(2R+1)}{t}$ which corresponds to its Taylor expansion and is coherent with the integral formulation.

\medskip

The second case is a little different since the two extremal intervals are limited by $-\pi$ and $\pi$. Observe that in the sum this two terms simplify so that we find:
\bea &&\sum_{r=-R+1}^{R-1} e^{i(\sigma(k)-k)b_r}-e^{i(\sigma(k)-k)a_r}+e^{-i\frac{2\pi R+\pi \epsilon}{t}}-e^{i\frac{2\pi R-\pi \epsilon}{t}}\cr
&&=\frac{\cos \frac{(\sigma(k)-k)\pi(2R-1+\epsilon)}{t} -\cos \frac{(\sigma(k)-k)\pi(2R+1-\epsilon)}{t}}{i\sin \frac{(\sigma(k)-k)\pi}{t}}\cr
&&=-2i\frac{ \sin \frac{2(\sigma(k)-k)\pi R}{t} \sin \frac{(2-\epsilon)(\sigma(k)-k)\pi}{t}}{\sin \frac{(\sigma(k)-k)\pi}{t}} 
\eea
so we get:
\beq P_N \left(t\in [2R-\epsilon,2R+\epsilon]\right)=\frac{1}{(2\pi)^N}\sum_{\sigma \in S_N}(-1)^{|\sigma|}\prod_{k=1}^N\left( 2\pi\delta_{\sigma(k)=k}-\frac{ 2\sin \frac{2(\sigma(k)-k)\pi R}{t} \sin \frac{(1-\epsilon)(\sigma(k)-k)\pi}{t}}{(\sigma(k)-k)\sin \frac{(\sigma(k)-k)\pi}{t}}\right) 
\eeq
Here one has to be careful with the case $\sigma(k)=k$ for which the last formula does not make direct sense. In the integral formulation, the integral is equal to $|I|=2\pi\frac{2R\epsilon}{t}+2\pi(1-\frac{2R}{t})$ whereas the Taylor expansion of the sine functions does not contain the factor $2\pi$. This is why we have inserted the factor $2\pi\delta_{\sigma(k)-k}$ in order to recover the right factor when taking the usual Taylor expansion. The determinant representation is:
\beq \encadremath{ P_N \left(t\in [2R-\epsilon,2R+\epsilon]\right)=\det \left[ \delta_{j-i=0}-\frac{ \sin \frac{2(j-i)\pi R}{t} \sin \frac{(1-\epsilon)(j-i)\pi}{t}}{\pi(j-i)\sin \frac{(j-i)\pi}{t}} \right]_{1\leq i,j\leq N}  } 
\eeq
In both cases, we have been able to rewrite an exact formula to compute the strong return probability as the determinant of a symmetric $N\times N$ matrix \eqref{formula1bis}.

\section{\label{Appendix B} First loop equation}

The first loop equation presented in \eqref{FirstLoop} is obtained from the following observation. Define:
\beq A_t(x)=\frac{1}{Z_N}\int_{I(t)^N}du_1\dots d u_N \sum_{j=1}^N\frac{d}{du_j}\left(\frac{1}{x-u_j} \Delta(u_1,\dots,u_N)^2 e^{-N\underset{i=1}{\overset{N}{\sum}} \ln u_i}\right)\eeq
Then we can compute this quantity in two different ways. First, since it is a total derivative, it can be directly integrated and since we have hard edges, we get a contribution of the form:
\beq A_t(x)=\sum_{a_k \text{ hard edge of } I(t)} \frac{\alpha_k}{x-a_k}\eeq
where we can write explicitly:
\bea \label{Alphak} \alpha_k&=&\pm\frac{1}{Z_N}\int_{I(t)^{N-1}} \left(\prod_{i\neq k} du_i\right)  \Delta_{N-1}(u_1,\dots,u_{k-1},u_{k+1},\dots,u_N)^2\left(\prod_{i\neq k}(u_i-a_k)^2\right) \cr
&&e^{-N\ln a_k} e^{-N\underset{i\neq k}{\overset{N}{\sum}} \ln u_i}\eea
Here the sign is $+1$ if the hard edge is an upper bound of an interval or is $-1$ if the hard edge is a lower bound of an interval. The second way to evaluate $A_t$ is to make the derivative acts on every term and identify the terms with their corresponding correlation functions. We find:
\bea  A_t(x)&=&\left<\sum_{k=1}\frac{1}{(x-u_k)^2}\right>^t+2\left<\sum_{i,j\neq i} \frac{1}{x-u_i}\frac{1}{x-u_j}\right>^t-N\left<\sum_{j=1}^N\frac{1}{(x-u_j)u_j}\right>^t\cr
&=&-W_{1,t}'(x)+W_{2,t}(x,x)+W_{1,t}(x)^2+W_{1,t}'(x)-\frac{N}{x}W_{1,t}(x)-\frac{N}{x}\left<\sum_{j=1}^N \frac{1}{u_j}\right>^t\cr
&=&W_{2,t}(x,x)+W_{1,t}(x)^2-\frac{N}{x}W_{1,t}(x)-\frac{N}{x}\left<\sum_{j=1}^N \frac{1}{u_j}\right>^t
\eea
thus proving \eqref{FirstLoop}.

\section{\label{AppendixD}The one-cut case with two arbitrary points on the unit circle and an arbitrary filling fraction}

\subsection{Spectral curve}
In this section, we are interested in computing:
\beq I(a,b,\epsilon_0)=\int_{[a,b]^{N\epsilon_0}}du_1\dots du_{N\epsilon_0} \Delta(u_1,\dots,u_{N\epsilon_0})^2 e^{-N\epsilon_0\underset{i=1}{\overset{N\epsilon_0}{\sum}} \ln u_i}\eeq
We also define the correlation functions as well as their $\frac{1}{N}$ expansion similarly to \eqref{deff} and \eqref{dev}. (but with all sums going from $1$ to $\epsilon_0 N$ instead of $N$).
We restrict to the case where the endpoints are chosen on the unit circle and write $a=e^{i\theta_0}$ and $b=e^{i\theta_1}$ with $\theta_0$ and $\theta_1$ chosen in $[-\pi,\pi]$. Let us study:
\beq A_{t,a,b,\epsilon_0}(x)=\frac{1}{I(a,b,\epsilon_0)}\int_{[a,b]^{N\epsilon_0}}du_1\dots du_{N\epsilon_0} \sum_{j=1}^{N\epsilon_0}\frac{d}{d u_j}\left(\frac{1}{x-u_j}\Delta(u_1,\dots,u_{N\epsilon_0})^2 e^{-N\epsilon_0\underset{i=1}{\overset{N\epsilon_0}{\sum}} \ln u_i}\right)\eeq
Since it is a total derivative, it can be evaluated at the endpoints of the integral and we get:
\beq A_{t,a,b,\epsilon_0}(x)=\frac{\mu}{x-a}+\frac{\nu}{x-b}\eeq
Alternatively, we can apply the derivative to each term and we get: 
\bea A_{t,a,b,\epsilon_0}(x)&=&\left<\sum_{i=1}^{N\epsilon_0}\frac{1}{(x-u_i)^2}\right>+2\left<\sum_{i,j\neq i=1}^{N\epsilon_0} \frac{1}{x-u_i}\frac{1}{x-u_j}\right>-N\epsilon_0\left<\sum_{i=1}^{N\epsilon_0}\frac{1}{(x-u_i)u_i}\right>\cr
&=&-W_{1,a,b,\epsilon_0}'(x)+W_{2,a,b,\epsilon_0}(x,x)+W_{1,a,b,\epsilon_0}(x)^2+W_{1,a,b,\epsilon_0}'(x)\cr
&&-\frac{N\epsilon_0}{x}W_{1,a,b,\epsilon_0}(x)-\frac{N\epsilon_0}{x}\left<\sum_{j=1}^N \frac{1}{u_j}\right>\cr
&=&W_{2,a,b,\epsilon_0}(x,x)+W_{1,a,b,\epsilon_0}(x)^2-\frac{N\epsilon_0}{x}W_{1,a,b,\epsilon_0}(x)-\frac{Nc}{x}
\eea
We thus introduce the following shift:
\beq y(x)=W_{1,a,b,\epsilon_0}^{[1]}(x)-\frac{\epsilon_0}{2x}\eeq 
so that the spectral curve is given by:
\beq \label{JeFais}y^2(x)=A_{t,a,b,\epsilon_0}(x)+\frac{c}{x}+\frac{\epsilon_0^2}{4x^2}=\frac{\epsilon_0^2}{4x^2}+\frac{c}{x}+\frac{\mu}{x-a}+\frac{\nu}{x-b}\eeq
By definition of $W_{1,a,b,\epsilon_0}(x)$ the asymptotic at infinity must be of the form:
\beq W_{1,a,b,\epsilon_0}^{[1]}\sim \frac{\epsilon_0}{x}+O\left(\frac{1}{x^2}\right) \,\Rightarrow y(x)\sim \frac{\epsilon_0}{2x}+O\left(\frac{1}{x^2}\right) \eeq
So far we get:
\beq \label{Eqqq2}y^2(x)=\frac{\epsilon_0^2(x-\alpha)(x-\gamma)}{4x^2(x-a)(x-b)}=\frac{\epsilon_0^2}{4x^2}+\frac{c}{x}+\frac{\mu}{x-a}+\frac{\nu}{x-b} \eeq
We want to prove that $\alpha=\gamma$, i.e. that the spectral curve does not admit simple zeros. Indeed, simple zeros would imply that the equilibrium density is supported on two intervals included in $[\theta_0,\theta_1]$ on not the whole $[\theta_0,\theta_1]$ segment. Let us compute (we define $\td{u}_i=e^{-i\frac{\theta_0+\theta_1}{2}}u_i$):
\bea W_{1,a,b,\epsilon_0}\left(-e^{i\frac{\theta_0+\theta_1}{2}}\right)&=&-e^{-i\frac{\theta_0+\theta_1}{2}}\left<\sum_{i=1}^{N\epsilon_0} \frac{1}{1+e^{-i\frac{\theta_0+\theta_1}{2}}u_i}\right>\cr
&=&-e^{-i\frac{\theta_0+\theta_1}{2}}\left<\sum_{i=1}^{N\epsilon_0} \frac{1}{1+\td{u}_i}\right>_{\td{u}_i\in \left[e^{-i\frac{\theta_1-\theta_0}{2}},e^{i\frac{\theta_1-\theta_0}{2}}\right]}
\eea
If we translate the origin of angles at $e^{i\frac{\theta_0+\theta_1}{2}}$ we get by symmetry under complex conjugation (with $\frac{1}{u}=\bar{u}$ on the unit circle) that:
\beq  \left<\sum_{i=1}^{N\epsilon_0} \frac{1}{1+\td{u}_i}\right>_{\td{u}_i\in \left[e^{-i\frac{\theta_1-\theta_0}{2}},e^{i\frac{\theta_1-\theta_0}{2}}\right]}=\left<\sum_{i=1}^{N\epsilon_0} \frac{1}{1+\frac{1}{\td{u}_i}}\right>_{\td{u}_i\in \left[e^{-i\frac{\theta_1-\theta_0}{2}},e^{i\frac{\theta_1-\theta_0}{2}}\right]}\eeq
Hence:
\bea 2W_{1,a,b,\epsilon_0}\left(-e^{i\frac{\theta_0+\theta_1}{2}}\right)&=&-e^{-i\frac{\theta_0+\theta_1}{2}}\left<\sum_{i=1}^{N\epsilon_0} \frac{1}{1+\td{u}_i}+\frac{\td{u}_i}{\td{u}_i+1}\right>_{\td{u}_i\in \left[e^{-i\frac{\theta_1-\theta_0}{2}},e^{i\frac{\theta_1-\theta_0}{2}}\right]}\cr
&=&-N\epsilon_0 e^{-i\frac{\theta_0+\theta_1}{2}}\eea
In particular we must have:
\beq W_{1,a,b,\epsilon_0}^{[1]}\left(-e^{i\frac{\theta_0+\theta_1}{2}}\right)=-\frac{\epsilon_0}{2} e^{-i\frac{\theta_0+\theta_1}{2}} \,\, \Rightarrow y\left(-e^{i\frac{\theta_0+\theta_1}{2}}\right)=0\eeq
Inserting this result into \eqref{Eqqq2} gives:
\beq \label{Eqq2}y^2(x)=\frac{\epsilon_0^2(x+e^{i\frac{\theta_0+\theta_1}{2}})^2}{4x^2(x-a)(x-b)}=\frac{\epsilon_0^2}{4x^2}+\frac{c}{x}+\frac{\mu}{x-a}+\frac{\nu}{x-b} \eeq
or equivalently:
\beq \label{rototo}\encadremath{ y(x)=\frac{\epsilon_0(x-\alpha)}{2x\sqrt{(x-a)(x-b)}} \text{ where } \alpha=-e^{i\frac{(\theta_0+\theta_1)}{2}} }\eeq
This shows that the equilibrium measure is supported on the whole interval $[\theta_0,\theta_1]$ and is non-critical (the double zero of $y(x)$ is outside $[\theta_0,\theta_1]$). We can then use a Zhukowsky parametrization:
\bea\label{Zhi} x(z)&=&\frac{a+b}{2}+\frac{(b-a)}{4i}\left(z-\frac{1}{z}\right)=e^{i\frac{(\theta_0+\theta_1)}{2}}\left[\cos\frac{\theta_1-\theta_0}{2}+\frac{1}{2}\sin \frac{\theta_1-\theta_0}{2}\left(z-\frac{1}{z}\right)\right]  \cr
y(z)&=&\frac{\epsilon_0\left(x(z)e^{-i\frac{(\theta_0+\theta_1)}{2}}+ 1\right)}{x(z)\sin \frac{\theta_1-\theta_0}{2} (z+\frac{1}{z})}=\frac{\epsilon_0\left( 1+\cos\frac{\theta_1-\theta_0}{2}+\frac{1}{2}\sin \frac{\theta_1-\theta_0}{2}\left(z-\frac{1}{z}\right)\right)}{x(z)\sin \frac{\theta_1-\theta_0}{2} (z+\frac{1}{z})}
\eea
In particular the one form $ydx$ is:
\beq ydx(z)=\frac{\epsilon_0\left(x(z)+ e^{i\frac{\theta_0+\theta_1}{2}}\right)}{2zx(z)}dz=\epsilon_0\left(\frac{1}{2z} +\frac{1}{2(z-z_+)}-\frac{1}{2(z-z_-)}\right) \eeq
We observe again that the branchpoints are not singularities of $ydx$. Eventually we denote:
\bea z_+&=&\frac{1-\cos \frac{\theta_1-\theta_0}{2}}{\sin \frac{\theta_1-\theta_0}{2}}=\frac{\sin \frac{\theta_1-\theta_0}{2}}{1+\cos \frac{\theta_1-\theta_0}{2}} \cr
z_-&=&\frac{-1-\cos \frac{\theta_1-\theta_0}{2}}{\sin \frac{\theta_1-\theta_0}{2}}=-\frac{ \sin \frac{\theta_1-\theta_0}{2}}{1-\cos \frac{\theta_1-\theta_0}{2}}  
\eea
the two zeros of $x(z)$. Moreover we have:
\beq \label{dxsurx}\frac{dx(z)}{x(z)}=-\frac{1}{z}+\frac{1}{z-z_+}+\frac{1}{z-z-}\eeq
Eventually we observe that the factor $e^{i\frac{\theta_0+\theta_1}{2}}$ can be factorized using symplectic invariance. Indeed, if we take $\td{x}=e^{-i\frac{\theta_0+\theta_1}{2}}x$ and $\td{y}=e^{i\frac{\theta_0+\theta_1}{2}}y$ then the spectral curve reads:
\bea\label{Zhi2} \td{x}(z)&=&\cos\frac{\theta_1-\theta_0}{2}+\frac{1}{2}\sin \frac{\theta_1-\theta_0}{2}\left(z-\frac{1}{z}\right)  \cr
\td{y}(z)&=&\frac{\epsilon_0\left(\td{x}(z)+ 1\right)}{\td{x}(z)\sin \frac{\theta_1-\theta_0}{2} (z+\frac{1}{z})}
\eea
from which we can deduce that all $F^{[g]}$ will only depend on the difference $\theta_1-\theta_0$ but not directly from $(\theta_0,\theta_1)$. It would be interesting to see if this spectral curve comes from a specific integrable system.

\subsection{Computation of $F^{[-2]}$}
We want to determine $F^{[-2]}$ (noted $-F^{(0)}$ in \cite{EO}) associated to the spectral curve \eqref{Zhi}. In order to do so, we need to identify the type of singularities arising in the one-form $ydx$. Here we have $4$ singularities:
\begin{itemize}
\item $z=0$ is a pole of $x(z)$ of order $1$. In \cite{EO}, it corresponds to a type $1$ singularity
\item $z=\infty$ is a pole of $x(z)$ of order $1$. In \cite{EO}, it corresponds to a type $1$ singularity
\item $z_+$ and $z_-$ that satisfy $x(z_\pm)=0$ are simple poles of $ydx$ without being singularities of $x(z)$ nor $y(z)$. In \cite{EO}, it corresponds to a type $2$ singularity.
\end{itemize}

\medskip

\underline{Computation of the temperatures}: The first step to determine $F^{[0]}$ is to compute the so-called temperature of the singularities defined as $t_\alpha=\Res_{z\to \alpha} ydx$. We find:
$$\begin{array}{|c|c|}
\hline
\text{Singularity}&\text{Temperature}\\
\hline
t_0&\frac{\epsilon_0}{2}\\
\hline
t_\infty&-\frac{\epsilon_0}{2}\\
\hline
t_+&\frac{\epsilon_0}{2}\\
\hline
t_-&-\frac{\epsilon_0}{2}\\
\hline
\end{array}$$
For example, we get for $z_+$ the following:
\beq t_+=\Res_{z\to z_+}ydx=\Res_{z\to z_+}\frac{(x(z)+1)\epsilon_0}{2zx(z)}dz=\frac{\epsilon_0}{2}\Res_{z\to z_+}\frac{dz}{zx(z)}=\frac{\epsilon_0}{2z_+x'(z_+)}=\frac{\epsilon_0}{2}\eeq
We observe that the sum of the temperatures equals $0$ as expected from the theory.

\medskip

\underline{Computation around $z=0$}:  We now need to define some local coordinate to compute the contribution of each singularity. We start with $z=0$. It is a pole of order $1$ of $x(z)$ and therefore we need to choose in Eynard-Orantin language $z_0(z)=x(z)$. The potential $V_0(z)$ should be given by:
\beq V_0(z)=\Res_{q\to 0}y(q)dx(q)\ln\left(1-\frac{x(z)}{x(q)}\right)\eeq
But since $ydx(q)=\frac{1}{2q}+O(q)$ then we find:
\beq V_0(z)=0\eeq
Therefore we must compute:
\beq \mu_0=\int_0^K\left(ydx-dV_0+t_0\frac{d x(z)}{x(z)}\right)+V_0(K)-t_0\ln x(K)=\int_0^K\left(ydx+\frac{\epsilon_0}{2}\frac{d x(z)}{x(z)}\right)-\frac{\epsilon_0}{2}\ln x(K) \eeq
A straightforward computation from \eqref{dxsurx} and \eqref{Zhi} shows that:
\beq ydx(z)+t_0\frac{d x(z)}{x(z)}=ydx(z)+\frac{\epsilon_0}{2}\frac{d x(z)}{x(z)}=\frac{\epsilon_0}{z-z_+}\eeq
Therefore the integral gives:
\beq \mu_0=\epsilon_0\ln\left(K-z_+\right)-\epsilon_0\ln (-z_+)-\frac{\epsilon_0}{2}\ln x(K)\eeq

\medskip

\underline{Computation around $z=\infty$}: It is a pole of order $1$ of $x(z)$ and therefore we need to choose in Eynard-Orantin language $z_0(z)=x(z)$. The potential $V_0(z)$ should be given by:
\beq V_\infty(z)=\Res_{q\to \infty}y(q)dx(q)\ln\left(1-\frac{x(z)}{x(q)}\right)\eeq
But since $ydx(q)=\frac{\epsilon_0}{2q}+O(q)$ then we find:
\beq V_\infty(z)=0\eeq
A straightforward computation shows that:
\beq ydx(z)+t_\infty\frac{d x(z)}{x(z)}=ydx(z)-\frac{\epsilon_0}{2}\frac{d x(z)}{x(z)}=\frac{\epsilon_0}{z}-\frac{\epsilon_0}{z-z_-}\eeq
Therefore the integral gives:
\beq \mu_\infty=\epsilon_0\ln K-\epsilon_0\ln \left(K-z_-\right)+\frac{\epsilon_0}{2} \ln x(K)\eeq

\medskip

\underline{Computation around $z=z_+$}: This time we have to deal with a zero of $x(z)$ which is not a branchpoint. The local parameter in Eynard-Orantin formalism is $z_+(z)=\frac{1}{x(z)-x(z_+)}=\frac{1}{x(z)}$ since $x(z_+)=0$ by definition. Therefore we should compute:
\beq V_+(z)=\Res_{q\to z_+}y(q)dx(q)\ln\left(1-\frac{x(q)}{x(z)}\right)\eeq
We have $y(q)dx(q)$ of the form $\frac{\alpha}{q-z_+}$ while $\ln\left(1-\frac{x(q)}{x(z)}\right)\underset{q\to z_+}{\sim} -\frac{x'(z_+)}{x(z)}(q-z_+) +O\left((q-z_+)^2\right)$. Therefore there is no residue here since the function is regular at $q=z_+$ and we find:
\beq V_+(z)=0\eeq
Now note that $\frac{\frac{d }{dz}\left(\frac{1}{x(z)}\right)}{\frac{1}{x(z)}}=-\frac{x'(z)}{x(z)}$ so that it gives a negative sign so that the integrand of Eynard-Orantin becomes:
\beq ydx(z)+t_+\frac{d z_+(z)}{z_+(z)}=ydx(z)-\frac{\epsilon_0}{2}\frac{x'(z)}{x(z)}=\frac{\epsilon_0}{z}-\frac{\epsilon_0}{z-z_-}\eeq
and therefore we find:
\beq \mu_+=\epsilon_0\ln K-\epsilon_0\ln(K-z_-)-\epsilon_0\ln(z_+) +\epsilon_0\ln(z_+-z_-) +\frac{\epsilon_0}{2}\ln x(K)\eeq

\medskip

\underline{Computation around $z=z_-$}: The computation are very similar to the $z=z_+$ case. The local parameter is $z_-(z)=\frac{1}{x(z)}$ and as before we find:
\beq V_-(z)=\Res_{q\to z_-}y(q)dx(q)\ln\left(1-\frac{x(q)}{x(z)}\right)=0\eeq
Now since the temperature of the pole is the opposite of $z_+$, we find:
\beq ydx(z)+t_-\frac{d z_-(z)}{z_-(z)}=ydx(z)+\frac{\epsilon_0}{2}\frac{x'(z)}{x(z)}=\frac{\epsilon_0}{z-z_+}\eeq
Therefore we find:
\beq \mu_-=\epsilon_0\ln(K-z_+)-\epsilon_0\ln(z_- - z_+)-\frac{\epsilon_0}{2}\ln x(K)\eeq

\medskip

\underline{Computation of $F^{[-2]}$}: We can now insert all the results in order to compute $F^{[-2]}$. It is given in general by the formula:
\beq -F^{[-2]}=\frac{1}{2}\sum_\alpha \Res_\alpha V_\alpha ydx +\frac{1}{2}\sum_\alpha t_\alpha \mu_\alpha-\frac{1}{4i\pi}\sum_i\oint_{\mathcal{A}_i}ydx \oint_{\mathcal{B}_i}ydx \eeq
Here since the curve is of genus $0$ there are no cycles so the last term is $0$. Moreover, all potential $V_\alpha$ are $0$ so we are left with the central term. Eynard-Orantin theory proves that $F^{[-2]}$ should be independent of the choice of $K$. We can check this fact here by the explicit computation:
\beq \encadremath{F_{\theta_0,\theta_1,\epsilon_0}^{[-2]}=-\frac{\epsilon_0^2}{2}\ln\left(\frac{2}{1-\cos\frac{\theta_1-\theta_0}{2}}\right)=\epsilon_0^2\ln \left(\sin \frac{\theta_1-\theta_0}{4}\right) }\eeq
Indeed we have:
\beq t_0\mu_0+t_\infty \mu_\infty+t_+\mu_++t_-\mu_-=\frac{\epsilon_0^2}{2}\ln\left(\frac{(z_+-z_-)^2}{z_+^2}\right) =\epsilon_0^2\ln\left(\frac{2}{1-\cos\frac{\theta_1-\theta_0}{2}}\right)\eeq

\subsection{Computation of $F^{[0]}$}
In this section we compute $F^{[0]}$ (noted $-F^{(1)}$ in \cite{EO}) associated to the spectral curve \eqref{Zhi}. Since our curve is of genus $0$, we can compute $F^{[0]}$ (see \cite{EO}):
\beq \label{Correction} -F^{[0]}=-\frac{1}{2}\ln \tau_{B_x}-\frac{1}{24}\ln\prod_{a_i \text{ branchpoint} }y'(a_i) \text{ with } y'(a_i)=\frac{dy(a_i)}{dz_i(a_i)} \text{ and } z_i(p)=\frac{x(p)}{\sqrt{x(p)-x(a_i)}}\eeq
Note here a slight difference in our formula compared to definition $4.3$ of \cite{EO}. Indeed, in definition $4.3$ of \cite{EO}, the local coordinate is $z_i(p)=\sqrt{x(p)-x(a_i)}$ because the branchpoints appear traditionally in matrix models with polynomial potentials at the numerator of $y(x)$. However in our case, the branchpoints are poles of $y(z)$ (and not zeros) and the standard definition $z_i(p)=\sqrt{x(p)-x(a_i)}$ is not consistent with definition of the moduli of poles defined in section $3.4.2$ of the same paper \cite{EO}. In the case where a branchpoint is a pole of $y(z)$ the local coordinate must be taken as $z_i(p)=\frac{x(p)}{\sqrt{x(p)-x(a_i)}}$. In particular the numerator is necessary to maintain $F^{[0]}$ invariant under symplectic transformations. In our spectral curve, we have $2$ branchpoints located at $z=\pm i$. We have $x(i)=e^{i\theta_1}$ while $x(-i)=e^{i\theta_0}$ and we remind the reader that by definition of a branchpoint $x'(i)=x'(-i)=0$. It is then a straightforward computation from \eqref{Zhi} to compute $x''(i)=-\frac{1}{2}\left(e^{i\theta_1}-e^{i\theta_0}\right)$ and $x''(-i)=\frac{1}{2}\left(e^{i\theta_1}-e^{i\theta_0}\right)$. In the end we have:
\bea dz_{z=i}(z)&=&-\frac{\sqrt{2}x(i)}{\sqrt{x''(i)}(z-i)^2}+O\left(\frac{1}{z-i}\right)\cr
dz_{z=-i}(z)&=&-\frac{\sqrt{2}x(-i)}{\sqrt{x''(-i)}(z+i)^2}+O\left(\frac{1}{z+i}\right)
\eea
Moreover we have:
\beq y(z)=\frac{\epsilon_0\left(x(z)+ e^{i\frac{\theta_0+\theta_1}{2}}\right)}{2zx'(z)x(z)} \eeq
which implies:
\bea y(z)&=&\frac{\epsilon_0\left(x(i)+ e^{i\frac{\theta_0+\theta_1}{2}}\right)}{2ix''(i)x(i)(z-i)}+O(1)=\frac{\epsilon_0\left(1+ e^{-i\frac{\theta_1-\theta_0}{2}}\right)}{2ix''(i)(z-i)}+O(1)\cr
y(z)&=&-\frac{\epsilon_0\left(x(-i)+ e^{i\frac{\theta_0+\theta_1}{2}}\right)}{2ix''(-i)x(-i)(z+i)}+O(1)=-\frac{\epsilon_0\left(1+ e^{i\frac{\theta_1-\theta_0}{2}}\right)}{2ix''(-i)(z+i)}+O(1)
\eea
giving
\bea dy(z)&=&-\frac{\epsilon_0\left(1+ e^{-i\frac{\theta_1-\theta_0}{2}}\right)}{2ix''(i)(z-i)^2}+O(1)\cr
dy(z)&=&\frac{\epsilon_0\left(1+ e^{i\frac{\theta_1-\theta_0}{2}}\right)}{2ix''(-i)(z+i)^2}+O(1)
\eea
Eventually we can compute:
\bea \prod_{j} \frac{dy(a_j)}{dz_j(a_j)}&=&-\frac{\epsilon_0\left(1+ e^{-i\frac{\theta_1-\theta_0}{2}}\right)}{2\sqrt{2}x(i) i\sqrt{x''(i)}}\frac{\epsilon_0\left(1+ e^{i\frac{\theta_1-\theta_0}{2}}\right)}{2\sqrt{2}x(-i)i\sqrt{x''(-i)}}=\frac{\epsilon_0^2\left(1+ e^{-i\frac{\theta_1-\theta_0}{2}}\right)\left(1+ e^{i\frac{\theta_1-\theta_0}{2}}\right)}{8x(i)x(-i)\sqrt{x''(i)x''(-i)}}\cr
&=&-\frac{\epsilon_0^2\left(1+ \cos\frac{\theta_1-\theta_0}{2}\right)}{2i\left(e^{i\theta_1}-e^{i\theta_0}\right)}e^{-i(\theta_0+\theta_1)}=\frac{\epsilon_0^2\left(1+ \cos\frac{\theta_1-\theta_0}{2}\right)e^{-3i\frac{\theta_1+\theta_0}{2}}}{4\sin\frac{\theta_1-\theta_0}{2}}\cr
&=&\frac{\epsilon_0^2\left(1+ \cos\frac{\theta_1-\theta_0}{2}\right)e^{-3i\frac{\theta_1+\theta_0}{2}}}{4\sin\frac{\theta_1-\theta_0}{2}}=\frac{\epsilon_0^2e^{-3i\frac{\theta_1+\theta_0}{2}}}{4\tan\frac{\theta_1-\theta_0}{4}}
\eea
Since the curve is of genus $0$, the Bergmann tau-function is trivial and equals $\tau_{B_x}\propto (b-a)^\frac{1}{4}$ so that
\beq -\frac{1}{2}\ln \tau_{B_x}=-\frac{1}{8}\ln \left(\sin \frac{\theta_1-\theta_0}{2}\right)-\frac{1}{8}\ln 2 -\frac{i(\theta_0+\theta_1)}{16}\eeq
In the end we find:
\beq \label{F1t} \encadremath{-F_{\theta_0,\theta_1,\epsilon_0}^{[0]}=-\frac{1}{24}\ln 2-\frac{1}{12}\ln \epsilon_0+  \frac{1}{24}\ln\left(\tan\frac{\theta_1-\theta_0}{4}\right)-\frac{1}{8}\ln\left(\sin \frac{\theta_1-\theta_0}{2}\right) }\eeq

\subsection{Computation of $F^{[2]}$}
Computations of $F^{[2]}$ (noted $-F^{(2)}$ in \cite{EO}) and higher orders fall into the general case of the topological recursion presented in \cite{EO}. We find for the curve \eqref{Zhi}:
\beq \label{F22} F^{[2]}_{\theta_0,\theta_1,\epsilon_0}=-\frac{3\cos\left(\frac{\theta_1-\theta_0}{2}\right)-1}{128\,\epsilon_0^2\cos^2\left(\frac{\theta_1-\theta_0}{4}\right)}\eeq
In the case $\theta_0=-\theta_1$ and $\epsilon_0=1$ it reduces to:
\beq F^{[2]}_{-\theta_1,\theta_1,1}=-\frac{3\cos\left(\theta_1\right)-1}{64\left(1+\cos\theta_1\right)}\eeq

\section{\label{AppendixNew}Verifying conditions for proposition \ref{PropBorot}}
In this section, we provide details on the formalism developed by Borot, Guionnet and Kozlowski and prove that it can be applied to our situation. We want to compute the large $N$ asymptotic of the following integral:
\beq \td{Z}_{N}=\int_{A^N} \prod_{i<j} |e^{i\theta_i}-e^{i\theta_j}|^2 d\theta_1\dots d\theta_N\eeq
where $A=\underset{h=0}{\overset{g}{\bigcup}}A_h$ is a union of segments described in \eqref{Intervalsss}. In our case, the intervals (as well as how many we have) depend on a time parameter $t$ that we will consider fixed in this appendix so we omit it everywhere to simplify notations. In the formalism of \cite{BorotGuionnetKoz}, our situation corresponds to pairwise interactions, i.e. $r=2$ that can be rewritten like:
\beq \td{Z}_{N}=\int_{A^N} \left(\prod_{i<j} |\theta_i-\theta_j|^2\right) \text{exp}\left(\frac{1}{2}\sum_{1\leq i,j\leq N} T(\theta_i,\theta_j)\right)  d\theta_1\dots d\theta_N\eeq
with:
\beq T(\theta_i,\theta_j)=\ln\left(\frac{ |e^{i\theta_i}-e^{i\theta_j}|^2}{|\theta_i-\theta_j|^2}\right)=\ln\left(\sin_c^2\left(\frac{\theta_i-\theta_j}{2}\right)\right)\eeq
and the cardinal sine function is $\sin_c(x)=\frac{\sin x}{x}$. Note that in our case, the so-called $2$-linear potential $T$ is indeed symmetric and does not have any large $N$ expansion (thus only $T^{[0]}(x_1,x_2)=T(x_1,x_2)$ is present in \cite{BorotGuionnetKoz}). Moreover, we only have to deal with compact subsets of $\mathbb{R}$ since $A\subset [-\pi,\pi]$. Hence confinement issues are trivial in our case. We remind now the reader of the technical conditions to satisfy in order to have proposition \ref{PropBorot}:

\begin{proposition}\label{ConditionProperty} If the following conditions are satisfied:
\begin{itemize} \item $A$ is the union of $g+1$ disjoint segments $A_i$ not reduced to a point: $A=\underset{i=0}{\overset{g}{\bigcup}} A_i$ 
\item (Regularity) $T$ is a continuous bounded function on $A^2$
\item (Uniqueness of the minimum) The energy functional $\mathcal{E}$ defined by:
\beq \mathcal{E}[\mu]=-\int_{A^2} \left(\frac{T(x_1,x_2)}{2}+\frac{\beta}{2}\ln |x_1-x_2|\right)d\mu(x_1) d\mu(x_2)\eeq
has a unique minimum on the space of probability measure on $A$ noted $\mathcal{M}^1(A)$ in the unconstrained model. 
\item $T$ is holomorphic in a neighborhood of $A^2$
\item (Non-criticality) The equilibrium measure $\mu_{\text{eq}}$ (i.e. the unique measure minimizing $\mathcal{E}$) is off-critical in the sense that its support $S$ is a union of $g_0+1$ segments not reduced to a point and:
\beq \label{tyty}d\mu_{\text{eq}}(x)=\frac{\mathbf{1}_S(x) dx}{2\pi}M(x)\prod_{\alpha\in \partial S\setminus \partial A}|x-\alpha|^{\frac{1}{2}}\prod_{\alpha\in \partial S\cap \partial A}|x-\alpha|^{-\frac{1}{2}}\eeq
with $M(x)>0$ on $S$. 
\end{itemize}
then proposition \ref{PropBorot} is valid.
\end{proposition}

We now discuss the validity of the previous conditions in our situation:
\begin{itemize}\item First condition is obvious for times $t\notin \{2k-\epsilon,k\in \mathbb{N}^\ast\}$. Indeed except at those times we always have disjoint intervals described in \eqref{Intervalsss} that are not reduced to a single points. At times $t_k=2k-\epsilon$ with $k\in \mathbb{N}^\ast$, we have an isolated point at $\theta=\pi$ so the first condition fails.
\item The regularity condition and the holomorphic condition for the function $T(x,y)$ are obvious since $T(x,y)=\sin_c^2 \frac{x-y}{2}$
\end{itemize}
Eventually we see that only the conditions regarding the uniqueness of the minimum and the non-criticality are non trivial in our situation. In our case, the energy functional looks like:
\bea \label{energyfunct} \mathcal{E}[\mu]&=&-\frac{1}{2}\int_{A^2} \left[\ln\left(\sin_c^2\left(\frac{\theta_1-\theta_2}{2}\right)\right)+2\ln |\theta_1-\theta_2|\right]d\mu(\theta_1) d\mu(\theta_2)\cr
&=&-\frac{1}{2}\int_{A^2} \ln\left(4\sin^2\left(\frac{\theta_1-\theta_2}{2}\right)\right)d\mu(\theta_1) d\mu(\theta_2)\cr
&=&-\ln 2-\int_{A^2}\ln\left|\sin\frac{\theta_1-\theta_2}{2}\right|\,d\mu(\theta_1) d\mu(\theta_2)
\eea
Taking the Fourier transform $\hat{L}(s)$ of the function $L(z)=\ln\left|\sin \left(\frac{z}{2}\right)\right|$ gives:
\beq \int_{\mathbb{R}^2} d\mu(x)d\mu(y) L(x-y)=\int_\mathbb{R} \hat{L}(s)|\hat{\mu}(s)|^2\eeq 
Hence following \cite{AbstractLoop}, a sufficient condition to obtain uniqueness of the minimum is to show that $\hat{L}(s)$ is negative so that the interaction is strictly convex on $\mathbb{R}$ and thus also on $\mathbb{A}$. This can be done by direct computation of the Fourier transform $\hat{L}(s)=-2\pi\underset{m=1}{\overset{+\infty}{\sum}}\frac{1}{m}\left(\delta(s-\frac{m}{2\pi})+\delta(s+\frac{m}{2\pi})\right)$ or using equations $A.3$ and $A.4$ of \cite{AbstractLoop}.
We stress here that this result is valid at any time $t>0$ in our situation.\newline
Eventually the last remaining point deals with the off-criticality condition. Off-criticality is never easy to prove and in our case we expect it to fail for times of the form $t_k=2\mathbb{N}\pm \epsilon$. Indeed at those times we have:
\begin{itemize}\item Times of the form $t=2k-\epsilon$ with $k\in \mathbb{N}^\ast$. In that case, there is an isolated point at $\pm \pi$ in the domain $A=I(t)$. Hence the first condition of \ref{ConditionProperty} fails.
\item Times of the form $t=2k+\epsilon$ with $k\in \mathbb{N}^\ast$. In that case, an interval is splitting into two intervals (at $\theta=\pm \pi$). Such situations are known to be problematic and in particular we expect the equilibrium density to exhibit a singular behavior at those edge points. We expect the edge to become soft and no longer hard (the polynomial $P_d(x)$ in the spectral curve is expected to vanish there) making \eqref{tyty} fail. 
\end{itemize}
Nevertheless there are special cases where the direct computation of the spectral curve allows us to prove the off-criticality condition. In particular \textbf{when the domain $A=[\theta_0,\theta_1]$ is supported on a unique interval}, we proved in \eqref{Eqq2} (using symmetry arguments) that the equilibrium measure is supported on the whole interval $[\theta_0,\theta_1]$ and that the equilibrium measure is non-critical. Using similar arguments, we proved that \textbf{for integer times} $t\in \mathbb{N}^\ast$, the problem can be recast into several copies of the previous problem hence giving an explicit and off-critical spectral curve corresponding to an equilibrium measure supported on the whole domain $A_t$. However for general times $t>2-\epsilon$ and $t\notin \mathbb{N}$, off-criticality remains a challenge since the initial domain is supported on several intervals and no symmetry arguments can be used. On the energy functional perspective, the situation seems difficult too since moving eigenvalues from a place to another increases the energy contribution of some of the eigenvalues while it decreases the energy contribution of others. In the end, it is even unclear why a gap in one of the intervals $[a_i,b_i]$ of the domain $A_t$ may not be more beneficial to the energy than an equilibrium density supported on the whole domain $A_t$. Additionally Jensen inequalities seem difficult to use in \eqref{energyfunct} because the interaction $-\ln |\sin \frac{\Delta \theta}{2}|$ is repulsive at short distances ($\Delta \theta\leq \pi$) but attractive at long distances ($\Delta \theta\geq \pi$).

\section{A few computations with trigonometric sums}

It is known that the following identities hold \cite{Wiki}:
\beq \prod_{j=1}^{n-1} \sin \left(\frac{j\pi}{n}\right)=\frac{n}{2^{n-1}}\text{ and } \sum_{j=1}^{n-1} \cos \left(\frac{j\pi}{n}\right)=\frac{\sin\frac{\pi n}{2}}{2^{n-1}}\eeq
From them we can deduce:
\beq \prod_{j=1}^k \sin \left(\frac{j\pi}{2k+1}\right)=\frac{\sqrt{2k+1}}{2^k}\text{ and } \prod_{j=1}^{k-1}\cos\left(\frac{j\pi}{2k}\right)=\frac{\sqrt{k}}{2^{k-1}}
\eeq
and then:
\bea \label{TrigoSum}\sum_{j=1}^{2k}j\ln\left(\sin^2 \frac{j\pi}{2k+1}\right)&=&(2k+1)\ln(2k+1)-2k(2k+1)\ln 2\cr
\sum_{j=1}^{2k-1}j\ln\left(\sin^2 \frac{j\pi}{2k}\right)&=&2k\ln(2k)-2k(2k-1)\ln 2
\eea

\section{Information about numerical simulations}
Numerical simulations have been carried out on a standard laptop using Maple $11$ (or more recent versions) software and can easily be reproduced. Perhaps the most difficult step in all the simulations is to sample a random unitary matrix properly according to the Haar measure. We took procedures as described in \cite{Sampling} to sample them. Note also that exact computations for the various probabilities can directly be computed from Toeplitz determinants and turn out to be much more efficient than using matrix integrals. We can provide on demand our code to any interested reader.   

\end{appendices}
\end{document}